\documentclass[reprint,amsmath,amssymb,aps,prb,floatfix]{revtex4-2}

\usepackage{graphicx}
\usepackage{bm}
\usepackage{hyperref}
\hypersetup{hidelinks,
  pdftitle={Real-axis least-squares bath discretization for real-time
    transport in quantum-dot Josephson junctions}}

\begin{document}

\title{Real-axis least-squares bath discretization for real-time
transport in quantum-dot Josephson junctions}

\author{Ruixin Zhou}
\email{SJTU-zy-zhouruixin@sjtu.edu.cn}
\author{Bing Dong}
\author{Yiyan Wang}
\affiliation{Key Laboratory of Artificial Structures and Quantum Control Ministry of Education,Department of Physics and Astronomy, Shanghai Jiaotong University, Shanghai, China}

\date{\today}

\begin{abstract}
Efficient real-time methods are essential for resolving transient
coherence and nonlinear transport in driven superconducting
nanostructures across broad time and parameter ranges. Such
simulations remain challenging because finite representations of the
BCS continuum generate recurrences, while the damping used to
suppress them smears gap-edge structure and artificially shortens
coherent lifetimes; common finite-broadening treatments face a
related tradeoff, where regularizing the gap-edge singularity
introduces a lifetime-resolution floor. Here we extend the driven
Liouville--von Neumann approach to superconducting reservoirs---a
noninteracting quantum dot with Bogoliubov--de Gennes leads---building
an explicit-Hamiltonian framework that hosts a family of bath
discretizations (uniform and Gaussian baseline discretizations and a new real-axis
least-squares recipe, in which mode energies, spectral weights, and
mode-resolved dampings are optimized jointly under gap-aware
constraints)---making the broadening an independently calibrated
numerical parameter whose leading long-time effect extrapolates
controllably along the joint ($\gamma \to 0$, $N_b \to \infty$)
refinement path while reducing the
computational cost by more than an order of magnitude relative to the
high-resolution DLvN reference. The compressed baths recover
established current-phase relations, multiple-Andreev-reflection
structure, quasiparticle-trapping oscillations, and integer and
fractional Shapiro locking; the lifetime analysis yields the
parameter-free leading-order relation
$\kappa \simeq w_{\mathrm{bath}}\gamma_{\mathrm{eff}}$ and resolves
damping-induced decay rates below the resolution scale of a previous
finite-broadening calculation. Because the reservoir remains an explicit Hamiltonian
representation, the framework generalizes naturally to multiterminal
normal--superconducting geometries and offers a route toward
interacting impurity solvers, while finite-gap strong-drive locking
amplitudes mark its present quantitative boundary.
\end{abstract}

\maketitle

\section{Introduction}\label{sec:intro}

Superconducting quantum-dot junctions compress Cooper-pair coherence,
discrete levels, and nonequilibrium drive into a single mesoscopic
device~\cite{martinrodero2011josephson,bretheau2012superconducting}.
Their dynamics features three hallmark phenomena: long-lived coherent
oscillations of quasiparticles trapped after a bias
quench~\cite{cheng2024quasiparticle}, the multiple-Andreev-reflection
(MAR) staircase in subgap
transport~\cite{averin1995ac,cuevas1996hamiltonian,bratus1995theory},
and Shapiro locking under ac drive, including fractional
steps~\cite{shapiro1963josephson,cuevas2002subharmonic}. With the
advance of superconducting point-contact and gated quantum-dot Andreev
transport experiments~\cite{scheer1997conduction,eichler2007evenodd},
quantitative real-time simulation of these phenomena---beyond
steady-state or perturbative descriptions alone---is increasingly
needed.

The central obstacle to such simulation is the representation of the
superconducting leads, where three requirements collide. First, any
finite discrete bath suffers Poincar\'e recurrence: beyond
$t_{\mathrm{rec}} \approx \pi N_b/D$ (for approximately uniform grids) the echo of the discrete
spectrum contaminates every slow observable, so uniform
discretizations targeting a time $T$ must grow their mode number as
$N_b \propto T$---and the wall-clock cost of dense
correlation-matrix propagation then grows close to $T^4$ (the cost
accounting is quantified in Sec.~\ref{sec:lsq}). Second, the BCS
gap-edge singularity and the subgap coherence that controls Andreev
physics demand high spectral resolution exactly where discretization
is hardest. Third, damping the bath suppresses recurrence but
broadens the gap edge and artificially shortens coherent lifetimes:
a single global damping rate cannot simultaneously protect long-time
fidelity and steady-state accuracy. These competing requirements
motivate a reservoir representation that can be optimized
simultaneously in spectral and dynamical terms.

Existing routes cover complementary domains. For periodically driven
problems Floquet scattering theory is
available~\cite{cuevas1996hamiltonian,cuevas2002subharmonic}, but
standard finite-period implementations require commensurate (rational)
frequency ratios---high-denominator rationals becoming increasingly
costly---and do not apply to transients and nonperiodic protocols. The
numerical renormalization group
(NRG)~\cite{wilson1975renormalization,bulla2008numerical} excels at
statics and spectroscopy but not at long-time driven dynamics. NEGF
methods provide powerful steady-state and time-dependent
formulations~\cite{jauho1994timedependent,maciejko2006timedependent};
in real-time superconducting applications the gap edge has been
regularized by a fixed Dynes-type broadening, whose resolution scale
bounds the accessible decay rates from
below~\cite{cheng2024quasiparticle}. Numerically exact hierarchical
equations of motion (HEOM) and auxiliary-mode expansions form another
family of time-dependent routes, whose cost is governed by the number
of terms in the exponential decomposition of the bath correlation
function; structured or sharply varying spectral densities typically
drive up the decomposition
order~\cite{tanimura2020heom,croy2009propagation}, and a time-local
NEGF formulation extending this auxiliary-mode route to
superconducting leads has appeared very
recently~\cite{kawamura2026timelocal} (what is compressed and what is
retained in that route versus the present one is compared in
Sec.~\ref{sec:discussion}). Finally, the driven Liouville--von Neumann
(DLvN)
method~\cite{hod2023driven,zelovich2014state,hod2016driven,oz2023electron}
turns the finite bath into an open absorbing boundary through
level-resolved Lorentzian broadenings: early implementations used a
single global driving rate, and subsequent parameter-free variants
introduced state-dependent broadenings extracted from the reservoir
self-energy~\cite{zelovich2017parameterfree}. These developments were
formulated for normal-metal and molecular transport, where the
trade-off between recurrence suppression and lifetime fidelity
persists (too little broadening and recurrences survive, too much and
physical lifetimes are artificially shortened). Unlike these
prescribed-broadening constructions (whether global or level-by-level),
the superconducting problem addressed here requires more: a particle-hole-consistent damping structure, and
a joint optimization of bath energies, positive spectral weights, and
mode-resolved dampings against the full Nambu matrix self-energy
under gap-protection constraints.

Treating bath discretization as an optimization problem has a deep
lineage in equilibrium many-body numerics: bath parametrization in
exact-diagonalization dynamical mean-field theory
(ED-DMFT)~\cite{caffarel1994exact,liebsch2012temperature}, the
logarithmic discretization of
NRG~\cite{wilson1975renormalization,bulla2008numerical},
orthogonal-polynomial chain mappings and Markovian
embeddings~\cite{chin2010exact,woods2014mappings,devega2015thermofield},
direct studies of bath-discretization optimality for real-time
evolution~\cite{devega2015discretize},
and the auxiliary master equation approach (AMEA), which fits the
hybridization function with nondiagonal Lindblad
dissipators~\cite{arrigoni2013nonequilibrium,dorda2014auxiliary,titvinidze2015transport}.
Mathematically these are rational approximation problems: for analytic
or sufficiently smooth targets convergence is
fast~\cite{nakatsukasa2018aaa,trefethen2019approximation}, while the
branch-point singularities at the BCS gap edges are not covered by these
results; they are addressed by the gap-clean weighting and density-floor
constraints introduced below.

This work carries that idea into the damped real-time dynamics of
superconducting junctions, with contributions on two levels. On the
\emph{formal} level, we bring the DLvN construction to superconducting
(Bogoliubov--de Gennes, BdG) leads: we establish that its damping
structure is compatible with the Nambu particle-hole structure (a
structure-preservation proposition), formulate the gauge discipline
that time-dependent drive imposes on damped bath orbitals, and
delimit the scope of validity of the reference states
(Sec.~\ref{sec:dlvn}). On the \emph{numerical} level, the
discretization itself becomes the optimization problem: the
placements, weights, and mode-wise dampings $(\varepsilon_j, w_j,
\gamma_j)$ are determined jointly in one constrained nonlinear
least-squares problem (rather than prescribed externally, whether
globally or level by level), with four design constraints---gap-clean
weighting, a gap-edge density floor, dense-core-plus-wings
compression for normal leads, and $\gamma$ itself entering the
fit---whose rationale, declared hyperparameters, and measured
failure-mode provenance are given in Sec.~\ref{sec:lsq}. The
positioning of the broadening changes accordingly: $\gamma$ is
treated not as a physical parameter but as an independently
calibrated numerical broadening---tunable per mode, its dominant
long-time effect obeying the scaling relation $\kappa \simeq
w_{\mathrm{bath}}\gamma_{\mathrm{eff}}$ (to leading order in the
damping) and extrapolating controllably to $\gamma \to 0$ for the
trapped-quasiparticle decay, while the discretization as a whole
defines a joint ($\gamma \to 0$, $N_b \to \infty$) refinement path
(at fixed $N_b$, $\gamma \to 0$ alone reintroduces the discrete
spectrum and its recurrences).

The validations are organized to first establish that this
construction recovers established results---three-way cross-validated
static current-phase relations, MAR thresholds, parity, and
dispersion, trapped-quasiparticle oscillations, and integer and
fractional Shapiro locking---and then to quantify the performance of
the compression and the systematics of the damping: a
more-than-order-of-magnitude wall-clock reduction while reproducing
the high-resolution DLvN reference at the few-percent level, the
lifetime scaling law, and direct evolution at arbitrary
(including irrational) drive parameters.
Section~\ref{sec:model} presents the model, the superconducting DLvN
formulation, and the least-squares bath framework;
Sec.~\ref{sec:results} presents the four validations;
Sec.~\ref{sec:discussion} discusses the error taxonomy, the domain of
applicability, and the relation to concurrent approaches;
Sec.~\ref{sec:conclusions} concludes with the boundaries and the
outlook.

\section{Model and Methods}\label{sec:model}

\subsection{Hamiltonian model}\label{sec:modelA}

We consider a junction coupled through a single-level quantum dot: the
dot level $\varepsilon_d$ couples with hybridization strength $\Gamma$
to left and right BCS superconducting leads (gap $\Delta$, bandwidth
$D$, phases $\phi_L$, $\phi_R$); in the three-terminal calculations an
additional normal metallic probe lead is attached. The junction
Hamiltonian, written in terms of the lead modes (index $k$ a generic
lead-state label; the discretization into a specific bath is the
subject of Sec.~\ref{sec:lsq}), is
\begin{align}
H(t) &= H_{\mathrm{dot}} + \sum_\alpha H_{\mathrm{lead},\alpha}
        + H_T(t), \nonumber\\
H_{\mathrm{dot}} &= \varepsilon_d \sum_\sigma d_\sigma^\dagger
        d_\sigma, \nonumber\\
H_{\mathrm{lead},\alpha} &= \sum_k \varepsilon_{\alpha k}
        \sum_\sigma c_{\alpha k\sigma}^\dagger c_{\alpha k\sigma}
        \nonumber\\
        &\quad + \sum_k \big(\Delta_{\alpha k}\, c_{\alpha k\uparrow}^\dagger
        c_{\alpha k\downarrow}^\dagger + \mathrm{h.c.}\big),
        \nonumber\\
H_T(t) &= \sum_{\alpha k\sigma} \Big[ V_{\alpha k}\,
        e^{i\chi_\alpha(t)}\, d_\sigma^\dagger c_{\alpha k\sigma}
        + \mathrm{h.c.} \Big],
\label{eq:H}
\end{align}
where the dot carries no intrinsic pairing---the pairing
$\Delta_{\alpha k}$ lives only on the bath orbitals; a normal
(metallic) probe lead is the $\Delta_{\alpha k} = 0$ special case of
the same expressions, which closes the model definition for
normal/superconducting hybrid multi-terminal systems. We use the
convention that $\Gamma$ denotes the \emph{sum} of the two lead
hybridizations (per lead $\Gamma_\alpha = \Gamma/2$ in the
equal-hybridization case); the factor-of-2 normalization differences
with respect to the NEGF pipeline are reconciled item by item in the
Supplemental Material (SM)~\cite{supplemental}, Sec.~S10. Throughout,
$\Delta = 1$ sets the energy unit and $\hbar = e = 1$. Bias and ac drive
enter entirely through the tunneling phases $\chi_\alpha(t)$ in the
Peierls gauge ($\chi_\alpha(t) = \chi_{\alpha,0} + \int_0^t
V_\alpha\,dt'$, superconducting phase difference $\phi(t) =
2[\chi_L(t) - \chi_R(t)]$, so the static configuration is set by the
initial phases $\chi_{\alpha,0}$; the necessity of this gauge is
explained in Sec.~\ref{sec:dlvn}), and the pairing amplitudes
$\Delta_{\alpha k}$ stay real. In the Nambu basis $\Psi = (d_\uparrow,
\{c_{\alpha k\uparrow}\};\, d_\downarrow^\dagger,
\{c_{\alpha k\downarrow}^\dagger\})$ the single-particle BdG matrix is
\begin{equation}
H_{\mathrm{BdG}}(t) =
\begin{pmatrix} h(t) & \hat\Delta \\ \hat\Delta^\dagger & -h^*(t)
\end{pmatrix},
\label{eq:HBdG}
\end{equation}
where the particle block $h$ carries $+\varepsilon_d$ and
$V_{\alpha k} e^{i\chi_\alpha}$, and the hole block $-h^*$
automatically carries $-\varepsilon_d$ and the conjugate phases---the
phase orientations of the particle and hole blocks are not independent
choices but are fixed uniquely by the Nambu structure.

The dot is noninteracting throughout ($U = 0$; the hard scope
boundary of this work, revisited in Sec.~\ref{sec:discussion}).
The physical observables are the dot occupation
$n(t)$, the anomalous correlation $F(t) = \langle d_\downarrow
d_\uparrow\rangle$, and the dot-lead bond currents (particle currents,
$I_\alpha > 0$ meaning particles flow from lead $\alpha$ into the dot;
consistent with the standard bond-current definition of the
Hamiltonian approach~\cite{cuevas1996hamiltonian}),
\begin{equation}
I_\alpha(t) = -2\,\mathrm{Im} \sum_{k,\sigma} V_{\alpha k}^*(t)\,
\langle c_{\alpha k\sigma}^\dagger d_\sigma \rangle,
\label{eq:current}
\end{equation}
where the spin-$\uparrow$ component follows directly from the
particle-block correlations and the spin-$\downarrow$ component is
expressed through the hole-block correlations by the Nambu convention
(the particle-hole map flips one sign: $I_\alpha^\downarrow =
+2\,\mathrm{Im}\sum_k V_{\alpha k}^* \langle d_\downarrow
c_{\alpha k\downarrow}^\dagger\rangle$); the two components add up to
the physical particle current.

Two current combinations are used in the results. Steady transport
(the dc components of MAR and Shapiro) uses the symmetrized
combination $I \equiv (I_A - I_B)/2$; transient problems additionally
use the sum of bond currents. By continuity, in general $\sum_\alpha
I_\alpha = dn/dt$ (the sum running over all leads, probe included); in
the two-terminal quench protocol this reduces to $I_A + I_B = dn/dt$,
so that combination and the occupation $n(t)$ are
differential/integral partners carrying the same dot-charge dynamics.

\subsection{Superconducting driven Liouville--von Neumann
formulation}\label{sec:dlvn}

Adapting DLvN dynamics to superconducting reservoirs is not a purely
formal replacement of the single-particle Hamiltonian by a BdG
matrix: the damping must act consistently on particle-hole partners,
and the voltage gauge must keep the damped bath blocks compatible
with the chosen reference state. This subsection formulates the
construction and states both requirements precisely.

Since $U = 0$, the full information of the system is carried by the
single-particle (BdG) correlation matrix $C_{ij}(t) =
\langle\Psi_j^\dagger \Psi_i\rangle$. Unitary evolution of $C$ with a
finite bath is limited by the Poincar\'e
recurrence time $t_{\mathrm{rec}} \approx \pi N_b/D$ (for
approximately uniform grids; non-uniform baths are governed instead by
the memory-kernel recurrence audit of SM Sec.~S1): past
$t_{\mathrm{rec}}$, the echo of the discrete bath contaminates every
slow observable, so uniform discretizations must scale $N_b \propto T$.
The driven Liouville--von Neumann (DLvN)
method~\cite{hod2023driven,zelovich2014state,hod2016driven} lifts this
restriction by adding mode-wise damping on the bath:
\begin{equation}
\frac{dC}{dt} = -i[H(t), C]
- \tfrac{1}{2}\{\Lambda,\, C - C_{\mathrm{ref}}\},
\label{eq:dlvn}
\end{equation}
where $\{\cdot,\cdot\}$ is the matrix anticommutator,
$\Lambda = \mathrm{diag}(\gamma_j)$ acts only on the bath modes
(zero on the dot) and $C_{\mathrm{ref}}$ is the target correlation
matrix of the damping; the equation is of quadratic (Gaussian)
open-system type and closes at the level of the single-particle
correlation matrix~\cite{prosen2008third,barthel2021solving}. Each
$\gamma_j$ turns its bath mode into a Lorentzian absorber of width
$\gamma_j$: outgoing excitations are drained and the bath is
continuously ``refilled'' toward the reference state, suppressing
recurrence exponentially.

\emph{Structure preservation (proposition).} Two statements make the
superconducting use of Eq.~(\ref{eq:dlvn}) precise; both hinge on
assigning the same $\gamma_j$ to the particle and hole components of
each bath mode, as done throughout this work. (i) For any
$C_{\mathrm{ref}} = f(H_{\mathrm{ref}})$ of a BdG reference
Hamiltonian the flow preserves Hermiticity of $C$; and when the
reference occupation is particle-hole symmetric [$f(-E) = 1 - f(E)$,
as for the Fermi function at the BdG chemical potential], the
particle-hole transform of a solution solves the mirrored
equation---because the hole block of $H$ is $-h^*$ by construction and
$\Lambda$ is particle-hole symmetric---so both spin components of the
physical observables (Sec.~\ref{sec:modelA}) follow consistently from
the single propagated correlation matrix. (ii) For the decoupled
reference, $\Lambda$ is proportional to the identity within each
mode's $2\times 2$ Nambu block and therefore commutes with the lead
BdG blocks; in the lead-quasiparticle basis the dissipator takes the
canonical quadratic-Lindblad form with loss and gain rates
$\gamma_j[1 - f(\pm E_j)]$ and $\gamma_j f(\pm E_j)$, so the dynamics
is completely positive and $0 \le C \le
1$~\cite{prosen2008third,barthel2021solving}. For the coupled
reference the conservation statements of (i) still hold, while
canonical Lindblad form is not claimed, consistent with the general
DLvN literature~\cite{hod2016driven}. The derivation is given in SM
Sec.~S12.

The damping strength is squeezed from both sides: too small and
recurrence is insufficiently suppressed (criterion
$\gamma\, t_{\mathrm{rec}} \gtrsim$ a few, a uniform-grid estimate; non-uniform baths are governed by the memory-kernel recurrence audit), too large and physical
lifetimes are artificially shortened (the $\kappa$ scaling law,
Sec.~\ref{sec:quench}).

\emph{Gauge discipline.} The gauge in which time-dependent bias/drive
enters the Hamiltonian is physically equivalent without damping (we
verified numerically that the pairing gauge and the hopping-phase
gauge give currents identical point by point at the $10^{-6}$ level),
but the equivalence breaks once damping is added. The compatibility
condition for the implementation used here (static
$C_{\mathrm{ref}}$, fixed damping operators) is:
\emph{no rotating term may sit on a damped site}---not ``phases may
not enter the Hamiltonian.'' In the pairing gauge the rotating
$\Delta_j(t)$ lives on the damped superconducting bath orbitals;
damping toward a static $C_{\mathrm{ref}}$ is not self-consistent in
the rotating frame, and the DLvN boundary conditions become
inconsistent. The correct choice is the hopping-phase (Peierls) gauge:
the bias enters the tunneling amplitudes $V_j e^{i\chi_\alpha(t)}$,
the bath $\varepsilon_j$ and $\Delta_j$ are all frozen, and the static
$C_{\mathrm{ref}}$ remains valid. In short: in the mean-field BCS
representation, putting the voltage into the pairing phase makes the
damped bath blocks explicitly time-dependent, while the DLvN boundary
with a static $C_{\mathrm{ref}}$ is only compatible with frozen bath
blocks---the gauge discipline is an implementation-level compatibility
constraint, not a statement of gauge noninvariance (co-transforming
the reference state and damping operators with the Hamiltonian would
restore covariance, at the cost of time-dependent boundary data).
All calculations in this paper use the hopping gauge; the
numerical equivalence of the two gauges in the undamped limit and the
documented failure of the pairing gauge under damping are recorded in
SM Sec.~S6.

\emph{Scope of validity of the reference state.} $C_{\mathrm{ref}}$ is
chosen per scenario: the biased drive/preparation stages use the
\emph{decoupled reference} (each lead in its own equilibrium), and the
post-quench relaxation stage uses the \emph{coupled reference} (the
equilibrium correlations of the final Hamiltonian)---concretely,
validation 2 (MAR steady state) and the drive/preparation stages of
validation 3 belong to the former, and the turn-off relaxation of
validation 3 to the latter. The domain of validity must be stated
precisely: quantitative steady-transport claims (MAR, Shapiro
locking) use the completely positive decoupled-reference
construction; post-quench frequencies and damping scalings use the
non-canonical coupled-reference construction and are therefore
supported by separate positivity, bath-convergence, and recurrence
audits (SM Sec.~S1). Both live in the quasiparticle dissipation
channel. The subgap phase-locking/supercurrent channel at finite
$\Delta$ is different in kind---it is the one transport channel that
carries coherence without dissipation---and there the strong-drive
locking amplitudes deviate at the $O(1)$ level from the
frequency-domain Floquet steady state while positions and selection
rules remain robust within the resolution tested (the ``positions
accurate, amplitudes
qualitative'' split of Sec.~\ref{sec:shapiro}). The occupation pinning
of the decoupled reference (the damping drags the bath occupations
toward a reference that contains no junction coherence) is a natural
candidate contribution, but we do not claim a settled attribution of
this deviation; the boundary itself is stated as a result. Throughout,
every conclusion involving finite-$\Delta$ strong-drive amplitudes is
treated as qualitative. Since the coupled reference is not guaranteed
to generate a completely positive map, we audit positivity
empirically: along a representative coupled-reference relaxation the
eigenvalues of $C$ make only a short, small, particle-hole-symmetric
excursion outside $[0,1]$ immediately after the quench and decay
monotonically back to the integrator floor;
decoupled-reference driven runs stay at that floor throughout (the
audit numbers are collected in SM Sec.~S1).

On the positioning of broadenings: in the finite-broadening
calculation of Ref.~\cite{cheng2024quasiparticle}, the Dynes-type
$\eta$ sets a resolution scale below which decay rates cannot be
distinguished from the regularization itself. The DLvN $\gamma$ and
the Dynes $\eta$ both introduce an imaginary lifetime scale in the
retarded sector (their Keldysh/reference-state structures differ);
within this framework $\gamma$ is an
independently calibrated numerical broadening parameter: tunable
per mode, part of the optimization (Sec.~\ref{sec:lsq}), with its
dominant effect on the tested long-time observable---the ABS
coherence envelope---described to leading order by the scaling
relation $\kappa \simeq w_{\mathrm{bath}}\gamma_{\mathrm{eff}}$ whose
coefficients are given independently by a spectral calculation
(Sec.~\ref{sec:quench}), thereby supporting a controlled $\gamma \to
0$ extrapolation of that decay rate along the joint refinement path
$\gamma \to 0$, $N_b \to \infty$ (the mode density must grow as the
damping shrinks, Sec.~\ref{sec:quench}). We do not
claim that $\gamma$ is physical, only that it is calibrated---this is
what makes \emph{damping-induced} decay rates below that resolution
scale numerically resolvable, the physical statement entering through
the $\gamma \to 0$ extrapolation.

\subsection{The real-axis least-squares bath framework}\label{sec:lsq}

\begin{figure}[t]
\includegraphics[width=\columnwidth]{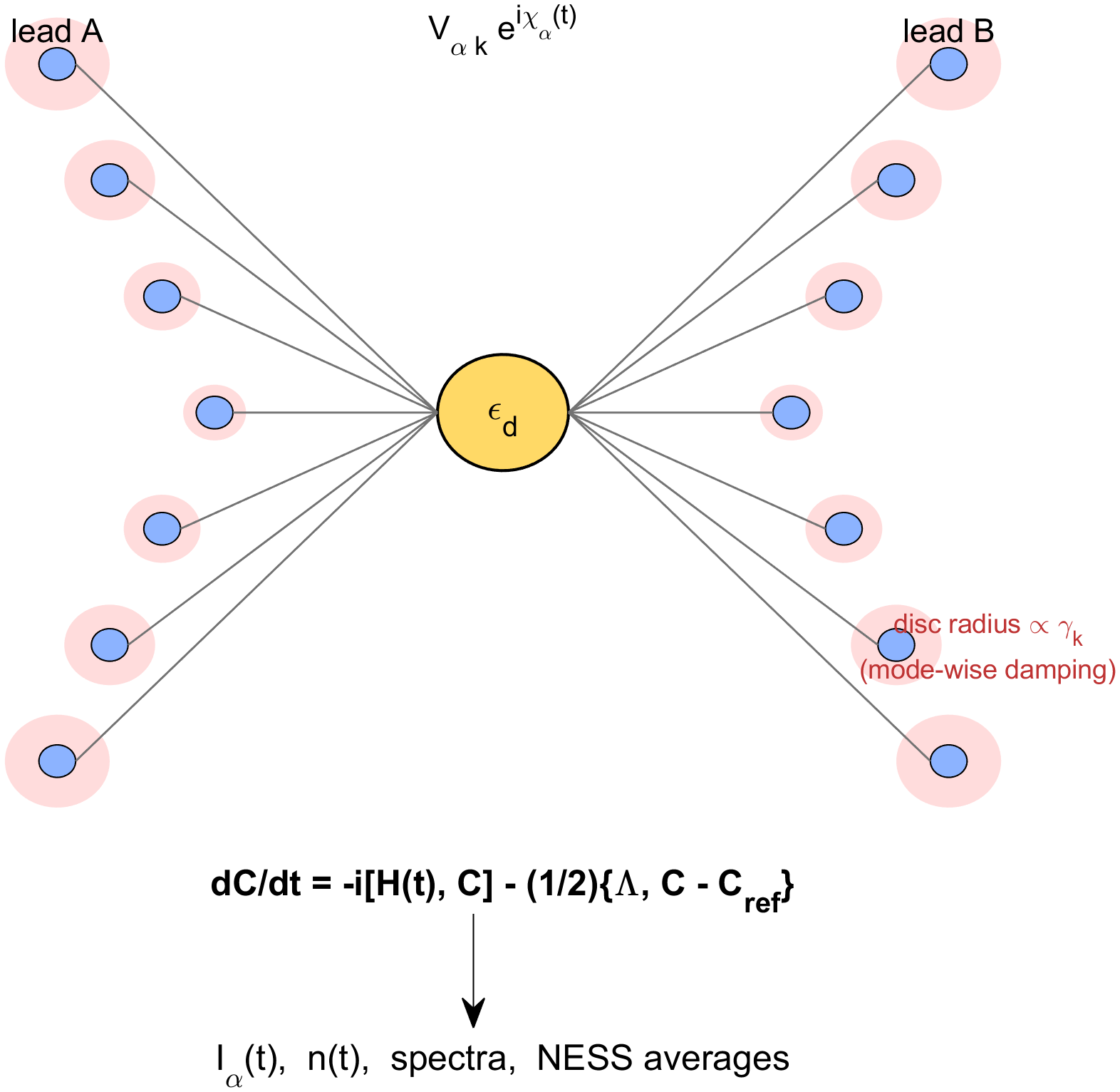}
\caption{System and workflow. A noninteracting quantum dot
($\varepsilon_d$) is coupled to two superconducting leads (A, B), each
carrying the time-dependent phase $\chi_\alpha(t)$ under bias or drive
and represented by a fan of damped bath modes whose disc radius
sketches the mode-resolved damping $\gamma_j$ (gap-edge modes take
small dampings that protect long-lived subgap physics, in-band modes
take large dampings that do the absorbing; the $\gamma_j$ structure is
detailed in SM Sec.~S3). The discretized bath is propagated with the
driven Liouville--von Neumann equation~(\ref{eq:dlvn}), and observables
(bond currents $I_\alpha(t)$, occupation $n(t)$, spectra, steady-state
averages) are read from the single-particle correlation matrix.}
\label{fig:schematic}
\end{figure}

\begin{figure*}[t]
\includegraphics[width=0.92\textwidth]{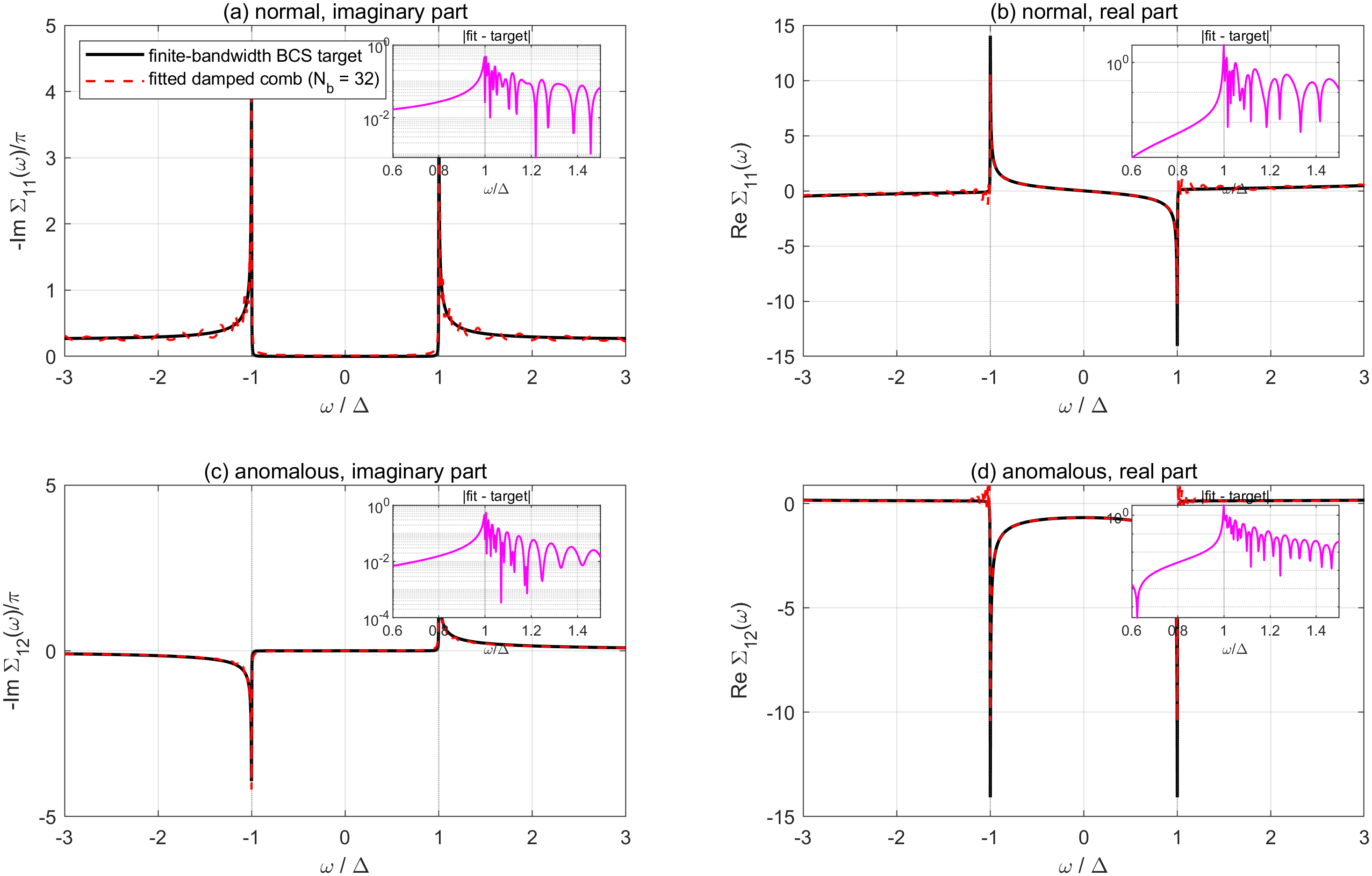}
\caption{Least-squares fit of the finite-bandwidth BCS hybridization
self-energy on the Nambu components the objective acts on
[Eq.~(\ref{eq:objective})]: (a)~$-\mathrm{Im}\,\Sigma_{11}(\omega)/\pi$
and (b)~$\mathrm{Re}\,\Sigma_{11}(\omega)$ (normal component),
(c)~$-\mathrm{Im}\,\Sigma_{12}(\omega)/\pi$ and
(d)~$\mathrm{Re}\,\Sigma_{12}(\omega)$ (anomalous component); target
(black) versus the fitted damped comb of a least-squares $N_b = 32$ bath
(red dashed). The comb is by construction a sum of single-mode
Lorentzian profiles acting jointly on the real and imaginary parts of
the three independent Nambu components $(\Sigma_{11}, \Sigma_{12},
\Sigma_{22})$ that enter the objective; the four panels display the
real and imaginary parts of the two shown here, $\Sigma_{11}$ and
$\Sigma_{12}$. Each inset shows the pointwise
error $|\Sigma_{\mathrm{fit}} - \Sigma_{\mathrm{target}}|$ near the gap
edge $\omega = \Delta$ on a logarithmic scale, where the
finite-bandwidth self-energy is near-singular; the residual ripple away
from the edge is the declared ansatz floor (Sec.~\ref{sec:lsq}, SM
Sec.~S3, where the fit convergence with mode count is also shown).}
\label{fig:selfenergy}
\end{figure*}

The task of bath discretization is to approximate the hybridization
self-energy of a continuous lead with $N_b$ discrete modes
(Fig.~\ref{fig:schematic}, fitted in Fig.~\ref{fig:selfenergy}). This section first defines the fitting
problem completely, then compares three generations of discretization
schemes and their $\gamma$ treatments, and closes with the four
constraints.

\emph{Definition of the fitting problem.} The fitting target for a
superconducting lead is the finite-bandwidth BCS hybridization
self-energy ($2\times 2$ Nambu block, retarded branch), defined by a
numerical continuum integral:
\begin{align}
\Sigma_{\mathrm{tar}}(\omega) &= \frac{\Gamma}{\pi}
\int_{-D}^{D} d\xi\, \big[(\omega + i\eta_0)\openone
- h(\xi)\big]^{-1}, \nonumber\\
h(\xi) &= \begin{pmatrix} \xi & \Delta \\ \Delta & -\xi \end{pmatrix}.
\label{eq:target}
\end{align}
Here $\Gamma$ follows the convention of Sec.~\ref{sec:modelA}, i.e.,
the sum of the two hybridizations: $\Sigma_{\mathrm{tar}}$ is the
\emph{total} embedding self-energy $\Sigma_L + \Sigma_R$ of the
equal-hybridization pair. The fit is performed once against this total
self-energy, and the resulting weights $\{w_j\}$ are split evenly
between the two leads (per lead $|V_{\alpha j}|^2 = w_j/2$, i.e.,
$\Sigma_\alpha = \tfrac{1}{2}\Sigma_{\mathrm{fit}}$). The split is
lossless: with equal hybridization the two leads are identical at
fitting time, and bias and superconducting phase enter afterwards as
Peierls phase factors multiplying each lead's $V_j$
(Sec.~\ref{sec:modelA}), never touching the weights. $D$ is the
junction bandwidth (main line $D = 4\Delta$), and $\eta_0 = 10^{-3}$
is an evaluation regulator only---it appears solely in the numerical
evaluation of the target curve and never enters the evolution; its
value equals the lower optimization bound $\gamma_{\min} = 10^{-3}$,
is smaller than every fitted mode width of the gap-clean LS-32 bath
($\min_j \gamma_j = 0.004$; the LS-64 density-floor variant parks two
far-tail anchors at the $\gamma_{\min} = 10^{-3}$ bound, SM Sec.~S1)
and than the fitting-grid resolution
($\approx 4\times 10^{-3}$), and is far below typical mode widths
($\sim$0.05). To rule out any ``hidden Dynes,'' we measured the
$\eta_0$ sensitivity directly: $\eta_0\!: 10^{-3} \to 10^{-4}$ (with
the integration resolution refined tenfold in step) moves the target
by only 0.059, below the $N_b = 32$ ansatz residual floor of 0.10--0.12;
the baths fitted under the two $\eta_0$ values differ in
cross-residuals against each other's targets by at most 0.024, with
deep-gap leakage unchanged (all numbers are dimensionless residuals in
the weighted fitting norm defined in this paragraph)---the resulting
parameter differences live in the fitting null space, and the choice
of $\eta_0$ sits below the ansatz resolution (SM Sec.~S3). The
wide-band analytic form $-\Gamma\,[\,\omega, \Delta; \Delta,
\omega\,]/\sqrt{\Delta^2 - \omega^2}$ is used only for sign/branch
checks, never as the fitting target---the three targets (wide-band
analytic, finite bandwidth, numerical integral) are inequivalent, and
this paper explicitly takes the finite-bandwidth numerical integral in
strictly the same Nambu convention as the discrete bath. The fitting
ansatz is a sum of $N_b$ damped BdG modes (mathematically a
matrix-valued rational
approximation~\cite{nakatsukasa2018aaa,trefethen2019approximation}),
\begin{align}
\Sigma_{\mathrm{fit}}(\omega) &= \sum_j w_j
\big[(\omega + i\gamma_j/2)\openone - h_j\big]^{-1}, \nonumber\\
h_j &= \begin{pmatrix} \varepsilon_j & \Delta \\
\Delta & -\varepsilon_j \end{pmatrix},
\label{eq:ansatz}
\end{align}
where $w_j \equiv |V_j|^2$, the squared mode coupling (total-$\Gamma$
convention; per lead $w_j/2$ after the split above); comparison with
the target shows the continuum measure is $(\Gamma/\pi)\,d\xi$, so a
uniform grid (spacing $h$) has weights $w_j = \Gamma h/\pi$ (per lead
$\Gamma_\alpha h/\pi$)---the entire freedom of the least squares is a
redistribution of this uniform measure along the energy axis. The
three independent Nambu components (11, 12, 22) \emph{share} one
parameter set $\{\varepsilon_j, w_j, \gamma_j\}$ (the joint-fitting
constraint, guaranteeing the constructed bath is one legitimate
physical bath rather than three mutually inconsistent spectra);
$\varepsilon_j$ is the normal-state energy $\xi$, the quasiparticle
energy is $E_j = \sqrt{\varepsilon_j^2 + \Delta^2}$, so the
inverse-square-root singularity at the gap edge is generated
automatically by $\varepsilon_j$ clustering near zero---no
hand-crafted singular basis functions. The phase is rotated globally
at bath-construction time (target co-rotated, strictly lossless). The
objective is a weighted least squares on a real-axis window,
\begin{equation}
\min \sum_\omega W(\omega)\,
\big\| \Sigma_{\mathrm{fit}}(\omega)
- \Sigma_{\mathrm{tar}}(\omega) \big\|^2,
\label{eq:objective}
\end{equation}
where $\|\cdot\|^2$ is the componentwise square sum over the three
independent Nambu components (11, 12, 22---each counted once) with
real and imaginary parts stacked at equal weight. The optimizer
minimizes this unnormalized weighted sum. Quoted residuals follow two
declared conventions: window-fit residuals (e.g., the $N_b = 32$
production value 0.152) are \emph{unweighted} relative $L_2$ values
restricted to the fit window (the norm of the in-window misfit
divided by that of the in-window target), whereas the
$\eta_0$-sensitivity and null-space numbers below are relative values
in the full \emph{weighted} objective norm of
Eq.~(\ref{eq:objective}), margins and gap weight included---the two
conventions differ, which is why the weighted-norm ansatz floor
(0.10--0.12) sits below the in-window production value. In-gap
``leakage'' is quoted separately
as the absolute $\max|\mathrm{Im}\,\Sigma|$ inside the gap.
$W(\omega)$ is a
three-segment weight: 1 inside the
physical window $[\omega_{\min}, \omega_{\max}]$; 0.1 on a soft margin
outside (extending about 0.35 window widths), preventing the fit from
dumping error at the window edge; and an extra large factor
$g_{\mathrm{gap}} \gg 1$ inside the gap ($|\omega| < 0.98\Delta$)
(gap-clean weighting, typically 30). Constraints and parametrization:
weights and dampings are log-parametrized, $w_j = e^{u_j}$, $\gamma_j
= e^{v_j}$, so positivity holds exactly by construction; $\gamma_j \in
[\gamma_{\min}, \gamma_{\max}]$ (main line $[10^{-3}, 0.25]$) guards
against undamped dead modes and overdamped smearing; $\varepsilon_j
\in [-\varepsilon_{\max}, \varepsilon_{\max}]$ stays within the
physical band for the main-line baths (the LS-64 density-floor
variant is the declared exception: its wing fit uses
$\varepsilon_{\max} = 10 > D$, and parks two auxiliary far-tail poles
at the box bound; SM Sec.~S1). The solver is trust-region nonlinear least squares with
multiple starts (uniform $\xi$ grid plus jitter) against local minima;
initial weights come from one shot of nonnegative linear least
squares; larger $N_b$ can be initialized from the smaller-bath
optimum, which reproduces the smaller-bath solution up to negligible
extra weights.

\emph{The discretization family compared.} Before the quantitative
comparison, we name the four bath constructions compared in this
work: a \emph{uniform} grid of equally spaced modes
with a single global damping; \emph{Gaussian} (gap-mapped) quadrature,
whose node energies and weights are fixed by the quadrature rule; and
the \emph{least-squares} recipe in two variants---a real-axis fit on
the transport window (the production recipe, with per-mode $\gamma_j$)
for real-time dynamics, and an imaginary-axis (Matsubara) variant
optimized for equilibrium observables. The uniform and Gaussian baths
carry an externally prescribed $\gamma$; only the least-squares
constructions optimize the dampings.

\begin{figure}[t]
\includegraphics[width=\columnwidth]{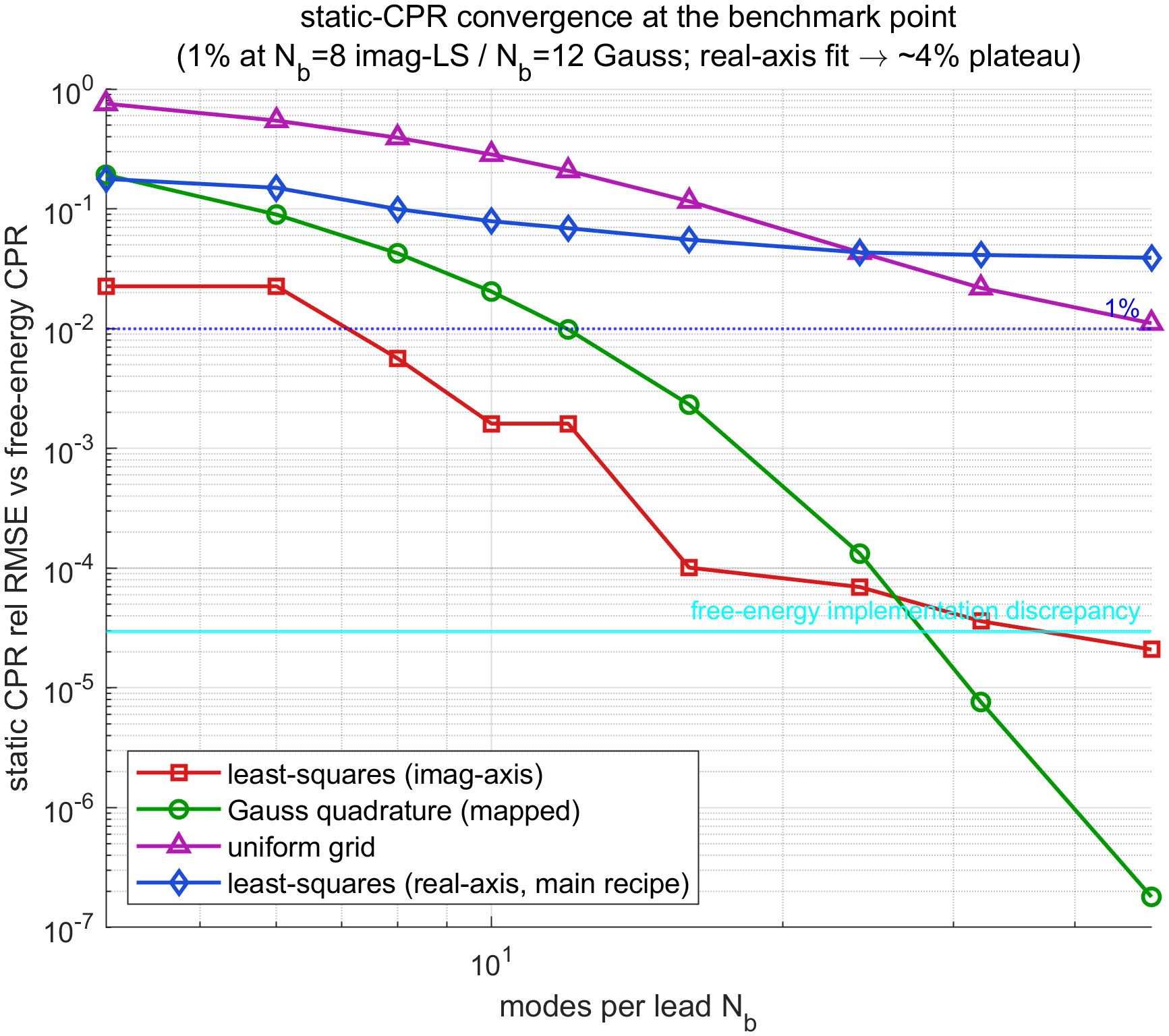}
\caption{Static-CPR convergence of the discretization schemes at the
benchmark point: 1\% at $N_b = 8$ (least squares, imaginary-axis
variant) and $N_b = 12$ (Gauss); the uniform grid and the real-axis
main recipe converge as $\sim 1/N_b$, the latter saturating near 4\%
(transport-window truncation, Sec.~\ref{sec:cpr}). The horizontal line
indicates the relative discrepancy between the two independent
free-energy-CPR implementations. Parameters in SM Sec.~S1.}
\label{fig:compare}
\end{figure}

\emph{Performance characterization: matched-budget comparison of
discretization schemes.} The following block characterizes what the
optimized bath delivers (it reports
measurements, not method definitions). The scheme-level dynamical
comparison of this block is made in a matched-budget setting (48
modes per lead, equal
total dimension) on one full MAR curve (the 117-bias-point protocol
reused verbatim) against the Gauss-110 reference
[Fig.~\ref{fig:mar}(b)].

\emph{Uniform discretization (global hand-tuned $\gamma$).} Static CPR
converges slowly (about $1/N_b$; 48 modes still 1.1\%); dynamics is
locked by recurrence, $N_b \propto T$, with total wall-clock cost
growing as $\sim T^4$ under dense correlation-matrix implementations
(the cubic per-step cost is backed by measurement: the
110$\leftrightarrow$32-mode per-step time ratio $\times 39$ matches
the cubed dimension ratio $(442/130)^3 \approx 39$). The
matched-budget comparison exposes the $\gamma$ trade-off: $\gamma =
0.045$ sits at the recurrence edge ($\gamma\, t_{\mathrm{rec}} =
1.7$), deviation 8.6\% (full 117-point grid; the SM's 59-point
subgrid convention gives 8.5\%); $\gamma = 0.09$ improves this to 5.7\%, but
$\kappa = w_{\mathrm{bath}}\gamma = 0.035$ ($w_{\mathrm{bath}} =
0.386$ is the ABS bath weight obtained by diagonalization, not a
fitted constant; Sec.~\ref{sec:quench}) damps out the
trapped-quasiparticle
ringing by $t \approx 30$, a decay rate 55 times faster than the
resolution scale of the calculation of
Ref.~\cite{cheng2024quasiparticle}---the physics of validation 3
cannot be represented in this prescription: steady-state accuracy is
bought at the cost of slow-dynamics fidelity. The two $\gamma$ values
differ by only a factor of 2 yet give materially different errors;
the outcome depends on the accidental choice of a hand-tuned
parameter.

\emph{Gaussian quadrature (global $\gamma$).} Gap-mapped Gaussian
quadrature converges exponentially for statics (1\% needs only $N_b =
12$); combined with DLvN, $N_b = 110$ per lead becomes this paper's
production reference (converged at the $10^{-7}$ level for the static
benchmark; its dynamical mode-number dispersion is measured below). But $\gamma$ remains an external global parameter,
and the quadrature rule locks the placements and weights with no room
to concede to the dynamical window: matched-budget 48 modes give
7.8\%, and one MAR curve takes 7.7~h.

\emph{Least squares ($\gamma$ in the optimization).} The solution of
the fitting problem above. Matched-budget error 3.6\%, the smallest of
the four production prescriptions and at the cross-prescription
plateau---32 modes already reach the same plateau, reproducing the
110-mode Gaussian reference at about 3.4\%. An a posteriori $\gamma$
scan of the unoptimized schemes (SM Sec.~S8) completes the fairness
picture: the Gaussian quadrature never drops below 7.8\% among the
scanned $\gamma$ values, while the uniform grid can be hand-tuned to tie the 3.6\%
on this single observable at $\gamma = 0.0675$ (confirmed on the full
117-point curve: 3.60\%, overlaid in
Fig.~\ref{fig:mar}(b))---a value that must be
located by scanning full curves against a pre-computed
high-resolution reference and that forfeits the slow-dynamics window wholesale
($\kappa = 0.026$, $46\times$ the resolution scale of validation 3's
starred configuration); the optimization delivers the same error with
per-mode dampings and no scan. The 0.2\% spread between the 48- and 32-mode
results lies within that plateau. Three observations probe the
origin of the plateau: (i) it is
insensitive to adding fit modes (24, 32,
and 48 modes all land at 3.4--3.6\%); (ii) both curves carry
$\gamma$-scale broadening of the same order, which the $\gamma$
analysis of Sec.~\ref{sec:mar} shows to dominate structure on
exactly this scale; (iii) a direct $N_b$-refinement check of the
reference itself (Gauss-96 and Gauss-140 against Gauss-110 at the
same $\gamma$, 14 representative biases) measures the reference's own
discretization dispersion at $\le 0.4\%$ across the subgap MAR window
but 1.4--1.9\% at the two gap-edge biases $V = 1.6, 2.0$ (the $n = 1$
threshold region, where the quadrature comb is sparsest against the
singular BCS edge), aggregating to 1.2\% (SM Sec.~S8; the full
117-point refinement chain is not attempted here). The plateau therefore
cannot be assigned to the fit-mode count alone: its scale is jointly set by the shared
$\gamma$ broadening, the reference's threshold-region discretization
ripple, and an unresolved residual, which
these probes cannot separate further. This result involves no
per-observable parameter scan against the reference: within the
declared prescription, $\varepsilon$, $w$, $\gamma$ come jointly
from one objective (with the declared hyperparameters of
Sec.~\ref{sec:lsq}), and the $\gamma$ trade-off is resolved by the
constrained optimization. Prescription-level design remains, openly:
which declared variant serves which observable class is itself part
of the framework (Sec.~\ref{sec:lsq} and
Sec.~\ref{sec:discussion}).
The fit itself is a one-time preprocessing step (about 20~min for the
32-mode production bath, three multistarts included; timings here and
throughout are single-node wall-clock figures, machine context in SM
Sec.~S1), amortized
across every downstream evolution. A
static by-product: the imaginary-axis variant (the Caffarel--Krauth
imaginary-axis distance function~\cite{caffarel1994exact} applied to
this system) needs only $N_b = 8$ for 1\%, whereas the real-axis recipe
itself saturates near 4\% statically (Sec.~\ref{sec:cpr})---its
objective is the transport window, not the full-band equilibrium (the
window-scan Pareto quantifying this allocation is in
Sec.~\ref{sec:cpr}).

\emph{Resulting capability.} Gauss-110 reference reproduction
(3--4\%) under $110 \to 32$ mode compression
($3.4\times$), per-step cost $1/39$, wall clock per MAR curve 7.7~h
$\to$ 20~min ($\times 23$; the end-to-end factor is smaller than the
per-step factor because fixed per-point overheads---junction
assembly, equilibrium construction, I/O---do not scale with the
dimension). Hence the $(V, \varepsilon_d)$ map
($41\times 86$), overnight Arnold-tongue maps, and direct evolution at
arbitrary (including irrational) $V_{\mathrm{dc}}$. The
speedup depends on the accounting: matched budget $\times$1.6--2.4
(most conservative) / to-floor $3.4\times$, wall clock $\times 23$ /
in the long-time small-$\gamma$ regime the uniform mode count diverges
(density $\propto 1/\gamma$) and the cubed dimension ratio gives
$\times 10^2$--$10^3$ (an arithmetic corollary).

\emph{The four constraints.} Each corresponds to a failure mode we
actually measured:
\begin{enumerate}
\item \emph{Gap-clean weighting}: Lorentzian tails of overdamped modes
leak spurious density of states into the gap, showing up in transport
as a parasitic ohmic shunt; $g_{\mathrm{gap}} \gg 1$ makes
``$\mathrm{Im}\,\Sigma \to 0$ inside the gap'' a hard target,
protecting the Andreev bound states (ABS) and the MAR threshold
structure.
\item \emph{Per-mode $\gamma$}: in-band modes take large $\gamma$ and
do the absorbing; gap-edge modes take small $\gamma$ and protect
long-lived physics; $\gamma$ is no longer global.
\item \emph{Density floor}: gap-clean fits give gap-edge modes tiny
$\gamma$, producing the a priori predictable ``edge beat'' artifact
(beats between discrete gap-edge modes and the ABS outlive the main
line and contaminate long-time spectra). Remedy: an equally spaced
dense core in the gap-edge zone (smoothly partitioned into the
least-squares wings), pushing artifact/main-peak from 1.03 to 0.09 (SM
Sec.~S4, Fig.~S3). The bath variant is chosen according to the
observable: gap-clean 32 modes for steady-state-averaged observables,
density-floor 64 modes for long-time spectroscopy.
\item \emph{Normal-lead constraint} (dense core in the transport
window + compressed wings): normal leads cannot be purely fitted.
Fermi occupations are pinned to the reference in DLvN, so the spectral
shape and occupation inside the transport window must be pointwise
correct---there is no compression budget there; purely fitted baths
produce occupation-staircase artifacts in the window. The correct
construction keeps an equally spaced dense core in the transport
window (spacing at the Gaussian-reference scale, uniform small
$\gamma$, analytic weights $\Gamma_\alpha h/\pi$) and compresses only
the wings with damped least squares (fitting the residual
hybridization after subtracting the core). All of the compression gain
comes from the wings---but the wings are the bulk of the mode count in
uniform schemes: the 77-mode hybrid bath used for three-terminal
Shapiro replaces a $\sim$400-mode purely uniform bath. Superconducting
leads are exempt: at $T \ll \Delta$ the in-band quasiparticle filling
is flat (the Fermi edge falls in the gap's zero-DOS region), so
occupation pinning is harmless---this is the spectral-structure reason
superconducting leads can be compressed more efficiently.
\end{enumerate}

\subsection{Numerical protocols and the baseline system}\label{sec:protocols}

Each headline benchmark is checked against an independent
implementation (several against two), with baselines split into
static and dynamic groups; auxiliary probes (the finite-lifetime
scaling test, bath-repair variants, high-order selection-rule floors)
are single-engine and are audited by protocol-level checks instead.

The spectral-analysis protocol is uniform throughout: signals in the
time window are demeaned and Hann-windowed before the fast Fourier
transform; slow drifts are removed by a moving-average high-pass
(window about 1.4 target periods); frequency assessments locate the
local peak within a $\pm 15\%$ \emph{search band} around the target
frequency (a line-isolation window, not an uncertainty; the measured
deviations are one to two orders of magnitude smaller); the quantitative
results (relative deviations) are quoted directly in the
corresponding figure captions.

\emph{Static baselines}: the continuum free-energy CPR benchmark (the
noninteracting junction free energy differentiated in phase; SM
Sec.~S10), evaluated by two independent implementations agreeing at
$3\times 10^{-5}$ and reproduced by the converged Gaussian reference at
$1.9\times 10^{-7}$; a physically independent real-axis Keldysh
current reproduces it at the percent level at the same point. These cross-validate at
one parameter point, with the parameter conventions (hybridization
factor of 2, current sign, temperature $\beta = 50$) aligned item by
item (SM Sec.~S10).

\emph{Dynamic baseline}: the Floquet scattering
method~\cite{cuevas1996hamiltonian,cuevas2002subharmonic} as the
independent benchmark. Two distinct error measures must be kept apart:
(i) ``vs Gaussian reference'' = reproduction error against the
Gauss-110 production reference (finite-bath DLvN)---it mainly measures
discretization/compression quality, though strictly it still contains
differences between damping parametrizations (the reference is uniform
$\gamma = 0.045$, the least-squares bath is per-mode $\gamma_j$);
(ii) ``vs Floquet'' = absolute accuracy against the broadening-free
ideal junction, where the tested DLvN bath prescriptions share a
$\gamma$-systematic layer
of about 15\% (2--3.5\% for the dominant integer ac-locking
amplitudes; larger for subharmonics, Sec.~\ref{sec:discussion}). Peak
positions, parities, frequencies, and lifetime scalings remain robust
against that layer within the numerical resolution tested here;
absolute amplitudes carry it, and wherever they appear we
report both the absolute deviation and the relative deviation
conditioned on the current magnitude.

Naming convention (uniform across text and all figure legends):
\emph{least squares} / \emph{Gauss} (Gaussian quadrature; its
110-mode configuration serves as the production Gaussian reference) /
\emph{uniform}; least-squares variants carry suffixes: $N_b = 32$
(gap-clean), $N_b = 64$ (density floor), $N_b = 77$ (normal-lead hybrid
bath, used for three-terminal Shapiro), imaginary-axis variant
(statics only). Table~\ref{tab:params} summarizes the parameters;
complete per-figure prescriptions are collected in SM Sec.~S1.

\begin{table*}
\caption{Overview of the four validations. Complete junction, drive,
and fitting parameters for every figure are collected in SM Sec.~S1;
the matched-budget protocol of Fig.~\ref{fig:mar}(b) uses four
prescriptions (LS-48 / Gauss-48 / uniform-48 at two $\gamma$ values)
at equal total dimension.}
\label{tab:params}
\begin{ruledtabular}
\begin{tabular}{p{0.14\textwidth}p{0.26\textwidth}p{0.22\textwidth}p{0.20\textwidth}l}
Validation & Physics target & Primary bath & Benchmark & Fig. \\
\hline
1 static CPR & equilibrium CPR convergence
  & LS/Gauss/uniform, $N_b{=}4$--48
  & free-energy CPR (2 impl.); Keldysh & \ref{fig:compare} \\
2 MAR & subgap staircase, parity, dispersion
  & LS-32 (gap-clean)
  & Gauss-110 reference; Floquet & \ref{fig:mar} \\
3 quench & trapped-quasiparticle ringdown, $\kappa$ law
  & Gauss-96 / LS-32 / LS-64; dense mapped
  & spectral $\kappa_{\mathrm{spec}}$ & \ref{fig:quench}, \ref{fig:quenchB} \\
4 Shapiro (eff.) & integer/fractional locking
  & LS-77 (dense core+wings)
  & Floquet & \ref{fig:shapiroA}, \ref{fig:shapiroB} \\
4 Shapiro (finite $\Delta$) & locking demonstration: resonance existence,
  selection rule
  & LS-32 (gap-clean)
  & kinematic grid (exact positions); Gauss-110 cross-check; free-energy CPR statics & \ref{fig:shapiro_fd} \\
\end{tabular}
\end{ruledtabular}
\end{table*}

\section{Results}\label{sec:results}

The validation sequence serves two purposes. First, it tests whether
the superconducting DLvN construction recovers established
equilibrium and driven-transport results: the static current-phase
relation against the continuum free-energy CPR benchmark
(Sec.~\ref{sec:cpr}); MAR thresholds, parity, and dispersion
(Sec.~\ref{sec:mar}); the trapped-quasiparticle oscillations of
Ref.~\cite{cheng2024quasiparticle} (Sec.~\ref{sec:quench}); and
Shapiro commensurability and selection rules
(Sec.~\ref{sec:shapiro}). Second, with that baseline established, the
same framework quantifies the performance of bath compression and
exposes the dynamical systematics of the damping: the
matched-budget performance and computational tradeoffs of
Sec.~\ref{sec:lsq}, the lifetime scaling
law and damping-induced decay rates below the resolution scale of a
previous finite-broadening calculation (Sec.~\ref{sec:quench}), and driven regimes that
are costly for frequency-domain methods---arbitrary (including
irrational) bias points and two-dimensional parameter maps
(Secs.~\ref{sec:mar} and \ref{sec:shapiro}).

\subsection{Validation 1: static current-phase relation, three-way
cross-validation}\label{sec:cpr}

Statics is the baseline of dynamics; this subsection establishes
consistency first. At one representative benchmark parameter point, the static-CPR
convergence curves of the discretizations are plotted together with
an independent continuum benchmark [Fig.~\ref{fig:compare}]: reaching 1\%
requires $N_b = 8$ for least squares (imaginary-axis variant), $N_b = 12$
for Gauss, and $> 48$ for uniform (about $1/N_b$); the continuum
free-energy CPR benchmark---the noninteracting junction free energy
differentiated in phase, carrying no bath discretization (SM
Sec.~S10)---is evaluated by two independent implementations that agree
at $3\times 10^{-5}$, and the converged Gaussian reference reproduces
it at $1.9\times 10^{-7}$; and a physically independent real-axis
Keldysh (Meir--Wingreen) current, evaluated at this same $D = 20$,
$\beta = 50$ point, reproduces it at the percent level ($0.3\%$ in the
wide-band limit; SM Sec.~S10). Finite Matsubara cutoffs (the
free-energy routes) and real-axis $\eta$-regularization (the Keldysh
route) make these numerically converged rather than analytically exact. The real-axis main recipe itself converges as roughly
$1/N_b$ and saturates near 4\% (17.8\% at $N_b = 4$ to 3.9\% at $N_b =
48$). This saturation is a measured resource allocation, not an
ansatz limit: refitting at fixed mode count with progressively wider
core windows traces a Pareto curve---the static error first improves
(3.7\% at window $\pm 5$ for $N_b = 32$; 3.1\% at $\pm 8$ for $N_b =
48$), then degrades as the fixed budget is diluted over the band
(9.1\%/6.1\% at the full band $\pm 20$), the optimum shifting outward
with the budget while the in-window fit residual grows monotonically
(SM Sec.~S1). The declared window deliberately spends the modes on
transport-window fidelity, which is what the dynamical validations
probe; the equilibrium CPR---a whole-band, low-frequency-weighted
observable---is served by the
imaginary-axis variant instead, the static-optimized member of the
same least-squares family. The scan shows why this division is
necessary rather than convenient: even the statics-optimal window
recovers only $\sim$3\%, three orders above the imaginary-axis
$10^{-5}$, because uniform real-axis weighting is mismatched to the
equilibrium observable. The
cross-validation is internally consistent: the two independent
free-energy-CPR implementations, the Gaussian reference, and an
independent real-axis Keldysh current agree pairwise, with the hybridization-strength
and sign conventions aligned item by item (SM Sec.~S10).

The static problem is thus solvable by all three discretizations and
is not the bottleneck; the schemes diverge in dynamics (the
matched-budget comparison of Sec.~\ref{sec:lsq}), and statics serves
here as a baseline-consistency check. We note that at the large-budget
end ($N_b \ge 32$) the Gaussian exponential convergence overtakes least
squares; the least-squares advantage lies at the dynamics-sensitive
small-budget end.

\subsection{Validation 2: multiple Andreev reflections}\label{sec:mar}

\begin{figure*}
\includegraphics[width=0.95\textwidth]{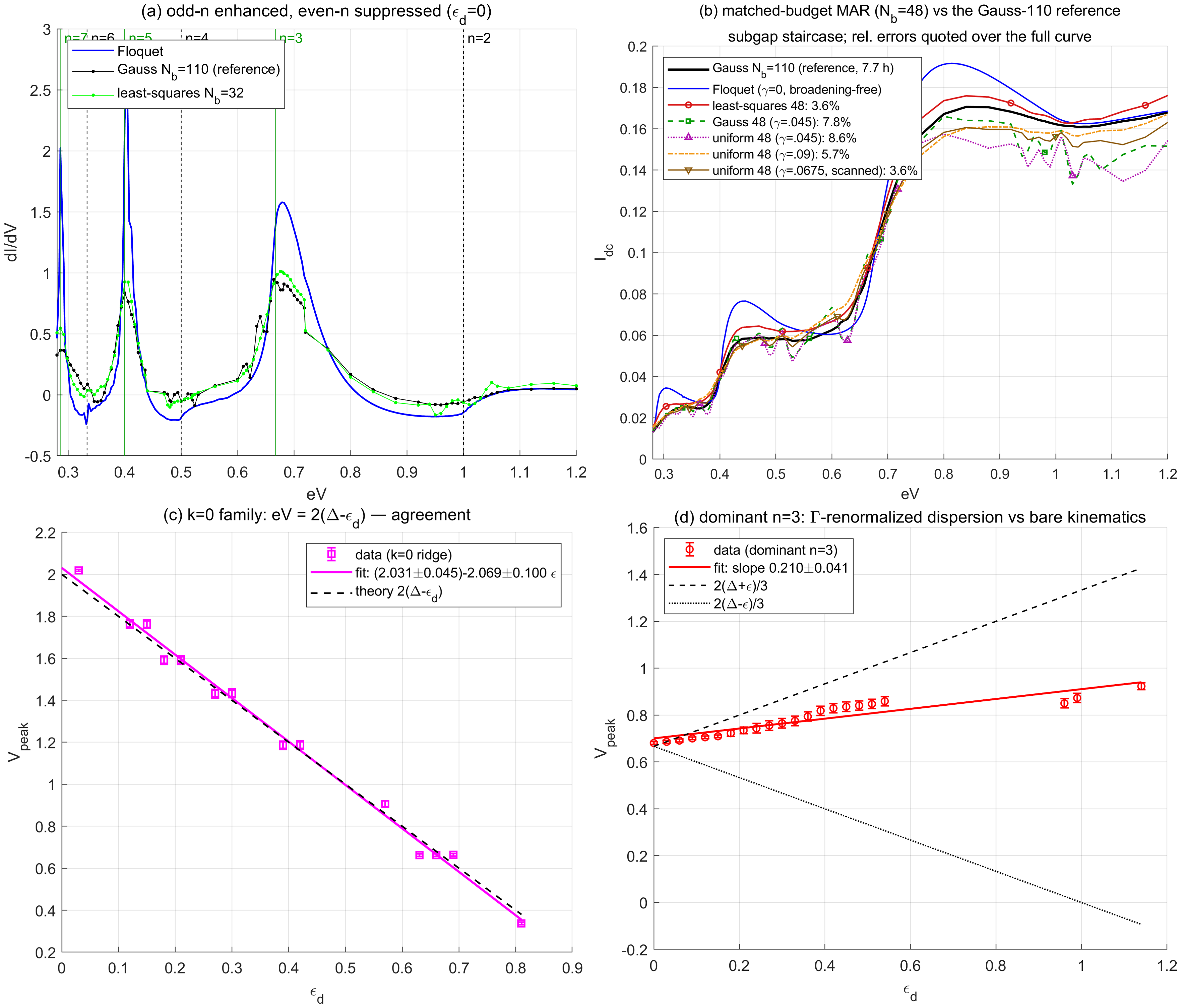}
\caption{MAR parity, matched-budget comparison, and dispersion.
(a)~$dI/dV$ at $\varepsilon_d = 0$ over the subgap staircase ($eV \le
1.2\Delta$, the multiple-Andreev regime $n \ge 2$): odd-order
enhancement; peak positions match Floquet to the bias-grid resolution.
(b)~Matched-budget MAR benchmark ($N_b = 48$ per lead) against the
Gauss-110 reference over the subgap staircase: least squares 3.6\%,
uniform 5.7--8.6\% ($\gamma$-dependent), Gauss 7.8\% (full-curve
relative errors); the a posteriori $\gamma$-scanned uniform tie point
($\gamma = 0.0675$; SM Sec.~S8) and broadening-free Floquet ($\gamma =
0$) are overlaid---least squares and Gauss-110 track Floquet to 7--10\%
over this subgap window (the finite-$\gamma$ offset shared by all
baths), while the full-curve $\approx 15\%$ is set by the high-bias
single-particle region. (c)~$k = 0$ dispersion-ridge fit, intercept and
slope inside the descriptive 95\% fit intervals of $2(\Delta -
\varepsilon_d)$. (d)~The dominant $n = 3$ resonance: renormalized
dispersion (slope $+0.210 \pm 0.041$). The production LS-32 curve
differs from the Gauss-110 reference by 3.4\%; panel~(b) is the
separate matched-budget LS-48 comparison (3.6\%). Protocols in SM Sec.~S8.}
\label{fig:mar}
\end{figure*}

\emph{Setup.} Subgap transport under dc bias is dominated by multiple
Andreev
reflections~\cite{averin1995ac,cuevas1996hamiltonian,bratus1995theory},
whose $I$-$V$ curve develops a staircase at $eV = 2\Delta/n$ (used
systematically in experiment as the fingerprint of the junction
transmission spectrum~\cite{scheer1997conduction,eichler2007evenodd});
it is a spectrally demanding steady-state-averaged test of a bath
discretization (the threshold structure requires correct gap-edge
spectral shape). Protocol: the bias enters the phase via the hopping
gauge; each bias point is evolved to the nonequilibrium steady state
(NESS) and period-averaged; the bias grid has 117 points (the
Gaussian-reference protocol reused verbatim); at the small-bias end
the DLvN evolution time grows as $t_{\max} \sim 6\cdot 2\pi/V$, so
comparison with Floquet is clipped to the common window $eV \in [0.28,
2.05]$. A dot-level scan $\varepsilon_d \in [0, 1.2]$ (step 0.03, 41
columns) yields the $(V, \varepsilon_d)$ map---exactly the kind of
two-dimensional scan enabled by the compression of Sec.~\ref{sec:lsq}
(SM Sec.~S8).

\begin{figure*}
\includegraphics[width=0.85\textwidth]{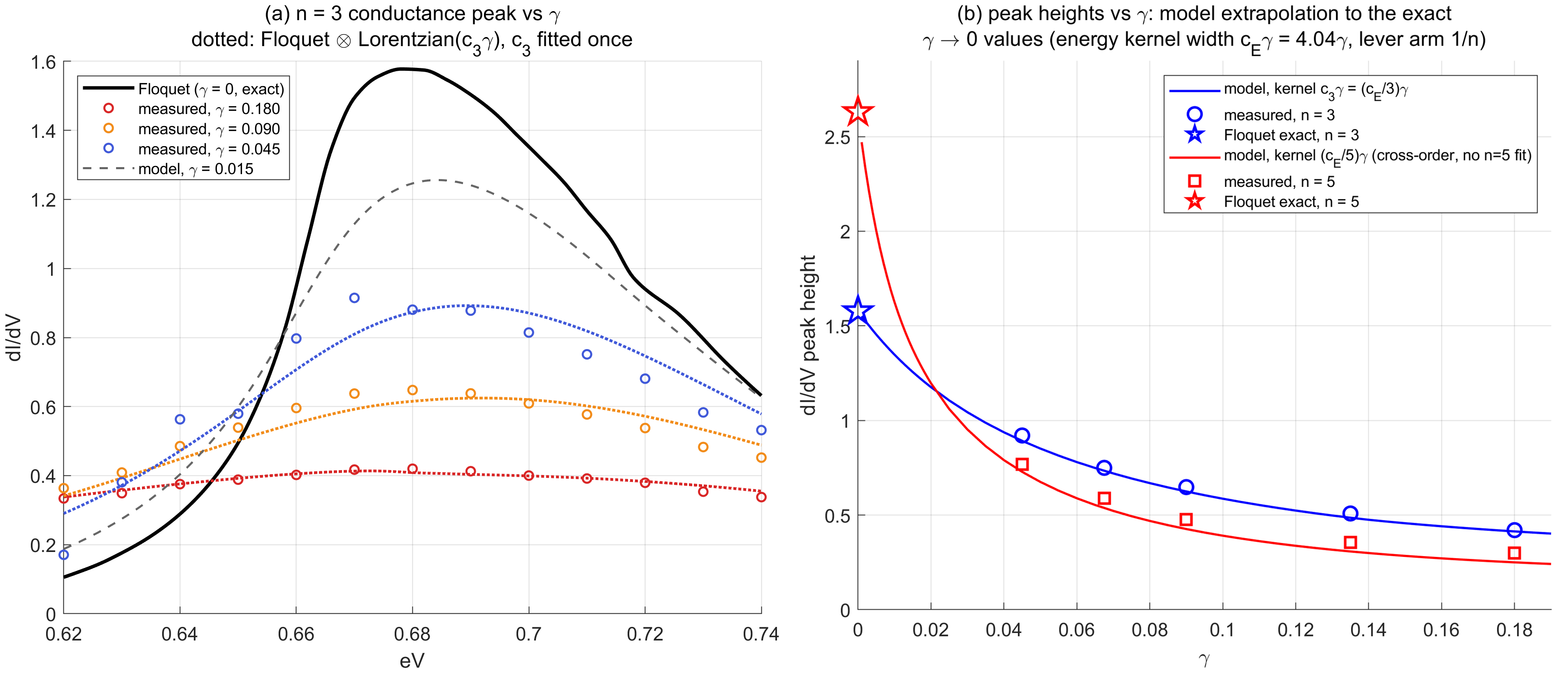}
\caption{$\gamma$ scaling of the MAR conductance peaks (the
\emph{Gaussian} reference bath, whose single global $\gamma$ provides
the knob for the extrapolation; the least-squares bath's per-mode
$\gamma_j$ has no single damping to sweep). (a)~The $n =
3$ peak: measured $dI/dV$ at three dampings (circles) against the
exact Floquet curve convolved with a truncated, renormalized
Lorentzian-core kernel of width
$c_3\gamma$ (dotted; the single linewidth coefficient $c_3 = 1.35$
fitted once on the full five-rung
family after fixing the semi-empirical locality scale $R$; kernel
convention and leave-rung-out stability check in SM
Sec.~S8); the model's small-$\gamma$ member (dashed) merges into the
exact curve. (b)~Peak heights versus $\gamma$ for the $n = 3$ and $n =
5$ peaks: measured (symbols; the $n = 5$ family spans the same five
dampings as the $n = 3$ ladder, $\gamma = 0.045$--$0.18$), model
curves with the lever-arm kernel
$c_n = c_E/n$ ($c_E = 4.0$; the $n = 5$ curve is a cross-order
prediction with no additional $n=5$-specific
fitting), and the exact Floquet heights at $\gamma = 0$ (stars).
Protocols and estimator notes in SM Sec.~S8.}
\label{fig:gammaladder}
\end{figure*}

\emph{Assessment.} The quantitative claims of this section are
threshold positions, parity, and dispersion; absolute current
amplitudes carry the $\gamma$-systematic layer of
Sec.~\ref{sec:protocols} and are not broadening-free accuracy claims.
Three levels:

(i) \emph{Curve level}: least squares with 32 modes reproduces the
Gauss-110 reference over the full curve at 3.4\% relative
deviation (the $I$-$V$ staircase itself is shown, at matched budget,
in Fig.~\ref{fig:mar}(b)); the independent Floquet benchmark has
two layers---the $\varepsilon_d = 0$ column is a full $I$-$V$
comparison against Floquet, shown at the $dI/dV$ level in
Fig.~\ref{fig:mar}(a). This comparison stratifies by bias: over the
subgap staircase ($eV \le 1.2\Delta$, the multiple-Andreev regime that
Figs.~\ref{fig:mar}(a) and~(b) display) the
least-squares current follows broadening-free Floquet to $7\%$---in
fact closer than the 110-mode Gaussian reference itself ($10\%$)---while
the full-curve deviation ($\approx 15\%$) is dominated by the high-bias
single-particle region toward the $eV = 2\Delta$ quasiparticle onset,
where finite-$\gamma$ broadening smooths the sharp threshold for the
reference and the compressed baths alike (the offset is a broadening
effect, not a discretization one). The benchmark is also examined
peak by peak in $\gamma$ in
Fig.~\ref{fig:gammaladder}. In the two-dimensional scan, every 10th
$\varepsilon_d$ column carries one control point at the fixed section
$V = 0.69$ (the bias-grid point nearest 0.7; five points total),
giving uniform
absolute deviations $\sim$0.011 and conditional relative 12\% in the
$|I| > 0.05$ domain---this is the $\gamma$-systematic layer declared
in Sec.~\ref{sec:protocols}. The relative deviation diverges as $I \to
0$ as a division artifact; we therefore report both absolute and
conditional relative deviations.

(ii) \emph{Parity level}: at $\varepsilon_d = 0$ the $dI/dV$ shows a
clean odd-even parity---odd orders ($n = 3, 5, 7$) enhanced, even
orders suppressed (a known fingerprint of resonant-junction MAR
theory~\cite{levyyeyati1997resonant,johansson1999resonant}), with peak
positions matching Floquet point by point (0.676 / 0.400 / 0.286)
[Fig.~\ref{fig:mar}(a)]. The kinematic voltages $eV = 2\Delta/n$ mark
where the $n$-th multiple-Andreev channel opens; the $dI/dV$ maximum is
a finite-width proxy for that onset and need not sit exactly on it. The
$n = 5$ and $n = 7$ maxima coincide with $2\Delta/5$ and $2\Delta/7$
within the bias resolution, while the broad, dominant $n = 3$
resonance maximum lies at most $\sim 2\%$ above $2\Delta/3$---a method-dependent
displacement of the derivative peak, within which the discretizations
and the Floquet reference scatter, rather than a shift of the
underlying MAR threshold. The residual wiggle of the finite-bath
$dI/dV$ traces in Fig.~\ref{fig:mar}(a) is the discrete-bath analog of
the discretization oscillations familiar from NRG spectral
functions~\cite{zitko2009energy} and catalogued for extended-reservoir
transport
simulations~\cite{elenewski2017markovian,wojtowicz2021dual}: it is
shared by the Gauss-110 reference and the least-squares bath alike, is
invisible at the level of the current itself, and is amplified only by
the finite-difference derivative (quantified as the $\sim$5\%
estimator scatter in SM Sec.~S8).

(iii) \emph{Dispersion level (quantitative agreement)}: away from the
symmetric point the MAR peaks disperse with $\varepsilon_d$. The
kinematic expectation is $(2k+1)\,eV = 2(\Delta \mp
\varepsilon_d)$~\cite{levyyeyati1997resonant}. Column-by-column ridge
tracking of the $k = 0$ family (continuity constraint + subgrid
parabolic refinement) and a linear fit give
\begin{equation}
V = (2.031 \pm 0.045) - (2.069 \pm 0.100)\,\varepsilon_d,
\label{eq:ridge}
\end{equation}
with intercept and slope both inside the descriptive 95\% fit
intervals (ordinary least squares on correlated comb-ripple
residuals, SM Sec.~S8) of
the theoretical $2(\Delta - \varepsilon_d)$ [Fig.~\ref{fig:mar}(c)];
error bars are the local bias-grid spacing.

\emph{Amplitude accounting: $\gamma$ scaling of the conductance
peaks.} The conductance-peak deficit against the broadening-free
Floquet benchmark is empirically captured, over the tested range, by a
semi-empirical broadening model built on the mode damping
[Fig.~\ref{fig:gammaladder}]. On a local grid around the dominant $n =
3$ peak we vary the reference-bath damping over $\gamma =
0.045$--$0.18$ (five values, the production protocol otherwise
verbatim): the measured $dI/dV$ lineshapes coincide with the exact
Floquet curve convolved with a truncated, renormalized Lorentzian-core
kernel of width $c_3\gamma$ (truncation radius $R = 0.275$, part of
the model definition: a bidirectional $R$ scan locates a fit-quality
valley coinciding with the adjacent-MAR-resonance spacing---the
locality bound of the lever-arm mapping---and the Lorentzian form
itself is the standard broadening of Lindblad-damped
modes~\cite{gruss2016landauer,dynes1978direct}; SM Sec.~S8),
with one fitted linewidth coefficient, $c_3 = 1.35$, shared across the
whole
family once the locality scale $R$ is fixed (lineshapes to 2--10\%,
peak heights to 2--4\%; $R^2 =
0.86$--$0.91$ on all five rungs; a leave-rung-out stability check (at
fixed, data-selected $R$) moves $c_3$ by only 1\%, SM Sec.~S8). Physically the
kernel is an energy-space lifetime width $c_E\gamma \approx 4\gamma$
compressed onto the voltage axis by the $n$-fold lever arm of the MAR
resonance condition $n\,eV \approx 2\Delta$: taking $c_n = c_E/n$ with
$c_E$ fixed by the $n = 3$ fit predicts the $n = 5$ peak with no
further parameters, and the measured heights follow this prediction to
12--16\% (interval-averaged estimator, the like-for-like convention
for finite-difference data; 8--16\% pointwise, SM
Sec.~S8) in a regime where the kernel exceeds the intrinsic width by
factors of 2.6 to 10, the deviation growing monotonically with the
smearing depth in both conventions. Peak positions stay pinned (0.673--0.680 across
all dampings). Since the model's $\gamma \to 0$ member is the Floquet
curve itself, the peak heights recover the exact values as the damping
is removed [Fig.~\ref{fig:gammaladder}(b)]---by construction; the
empirical content is the approach path, and it is not circular:
refitting with the anchor scale \emph{freed}, the finite-$\gamma$
data alone recover the Floquet peak to 0.5\% (blind-recovery error
budget about $\pm 5\%$; SM Sec.~S8). The recovery chain extends
across orders: with the kernel locked to the cross-order value $c_5 =
c_E/5$ and only the amplitude freed, the $n = 5$ data return the
anchor to 4\%, the per-rung amplitude approaching it monotonically
($1.16 \to 1.01$) as $\gamma$ decreases; fully freed, the $n = 5$
family alone still reaches $0.90$ of the anchor, limited by the
amplitude--width degeneracy of its much narrower peak (SM Sec.~S8).
The amplitude layer of
Sec.~\ref{sec:protocols} is thus a controlled, predictable broadening
effect. This extrapolation is carried on the Gaussian reference bath
because its single global $\gamma$ is the knob the recovery needs; the
least-squares bath, with per-mode $\gamma_j$, cannot be swept the same
way. Its fidelity to Floquet is therefore established not by its own
$\gamma \to 0$ extrapolation but through the reference: least squares
reproduces the converged Gaussian bath at the few-percent level
(Sec.~\ref{sec:lsq}, and peak by peak in
Fig.~\ref{fig:mar}), and the reference in turn recovers Floquet as the
broadening is removed here; carrying per-mode dampings of the same
scale, the least-squares bath inherits the same finite-$\gamma$
deficit, which the kernel picture quantifies. Higher-order peaks are
intrinsically sharper in energy (their
coherent MAR chains are longer), so at fixed $\gamma$ they are
proportionally more suppressed---the quantitative content of the
observable stratification discussed in Sec.~\ref{sec:discussion}. The
protocol, the estimator systematics, and a falsified $n$-independent
single-kernel hypothesis are archived in SM Sec.~S8.

\emph{Boundary.} The dispersion of the dominant $n = 3$ resonance peak
provides a counterexample: ridge slope $+0.210 \pm 0.041$, matching
neither of the bare kinematic $\pm 2/3$ [Fig.~\ref{fig:mar}(d)].
Strong hybridization ($\Gamma = 0.8$) renormalizes the dot level; the
peak follows the compressed dispersion; the linear trend of the peak
with $\varepsilon_d$ survives, but we do not assign it to the bare
kinematic relation. That the $k = 0$ family agrees while the $k = 1$
family is renormalized demonstrates, at this strong-hybridization
working point, the breakdown of the bare kinematic formula (a mapping
of the crossover would require a $\Gamma$ scan we have not performed).

\subsection{Validation 3: quasiparticle-trapping quench
dynamics}\label{sec:quench}

\begin{figure*}
\includegraphics[width=0.95\textwidth]{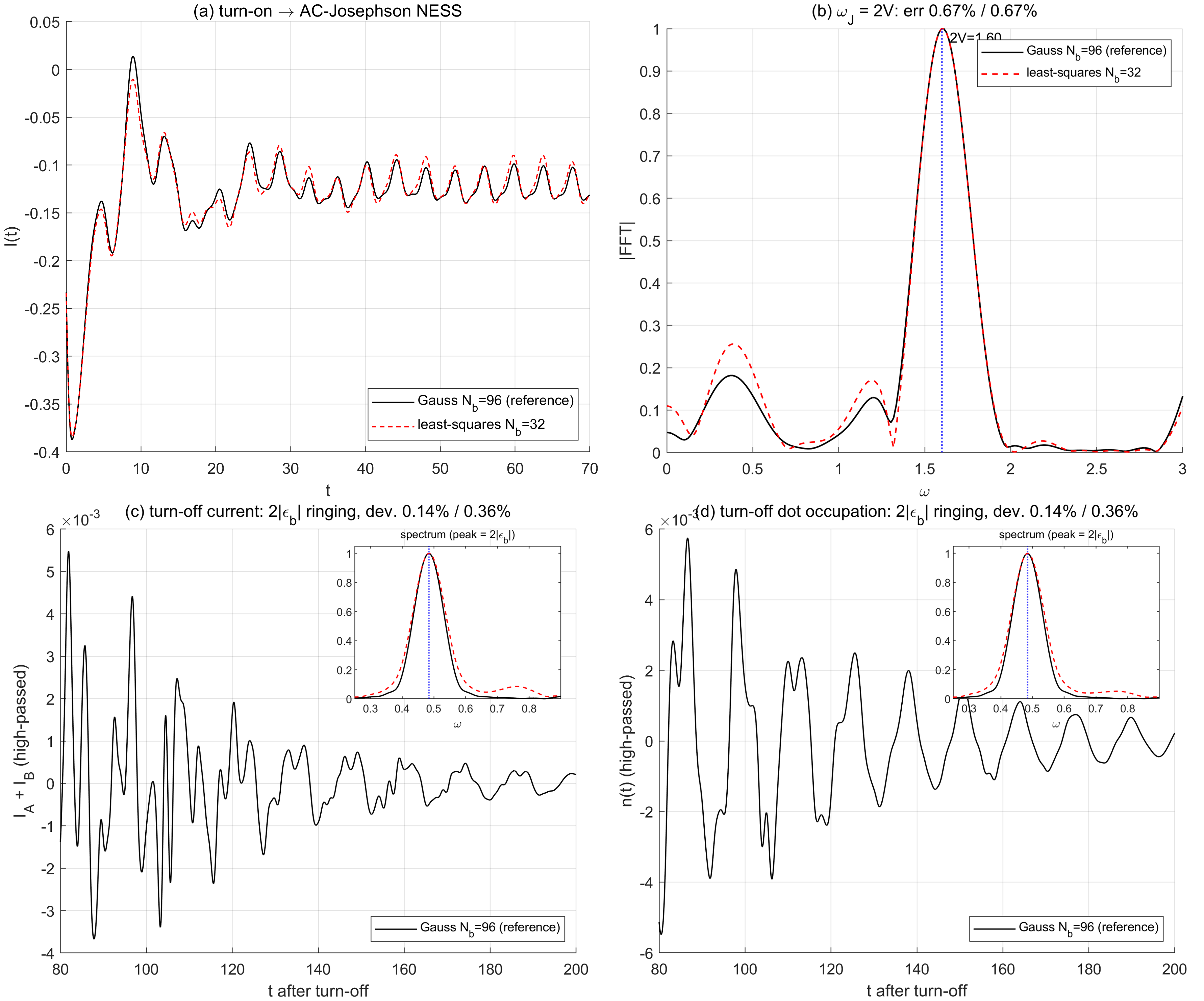}
\caption{Quench dynamics [protocol of
Ref.~\cite{cheng2024quasiparticle}]. (a,b)~Turn-on: both baths give
the Josephson peak at $\omega_J = 2V$ (0.67\%). (c,d)~Turn-off time
traces of the current $I_A + I_B$ and the occupation $n(t)$: the
trapped-quasiparticle ringing at $2|\varepsilon_b| = 0.484$,
recurrence-free over the full window (reference bath shown; each bath
decays at its own $\kappa$ by the scaling law of
Fig.~\ref{fig:quenchB}, so waveforms are not overlaid); insets:
assessment spectra of both baths, interpolated peak centers at
0.14\%/0.36\% from $2|\varepsilon_b|$ (reference/least squares;
consistent within the window resolution 0.026). Protocols in SM
Sec.~S1.}
\label{fig:quench}
\end{figure*}

\begin{figure*}
\includegraphics[width=0.95\textwidth]{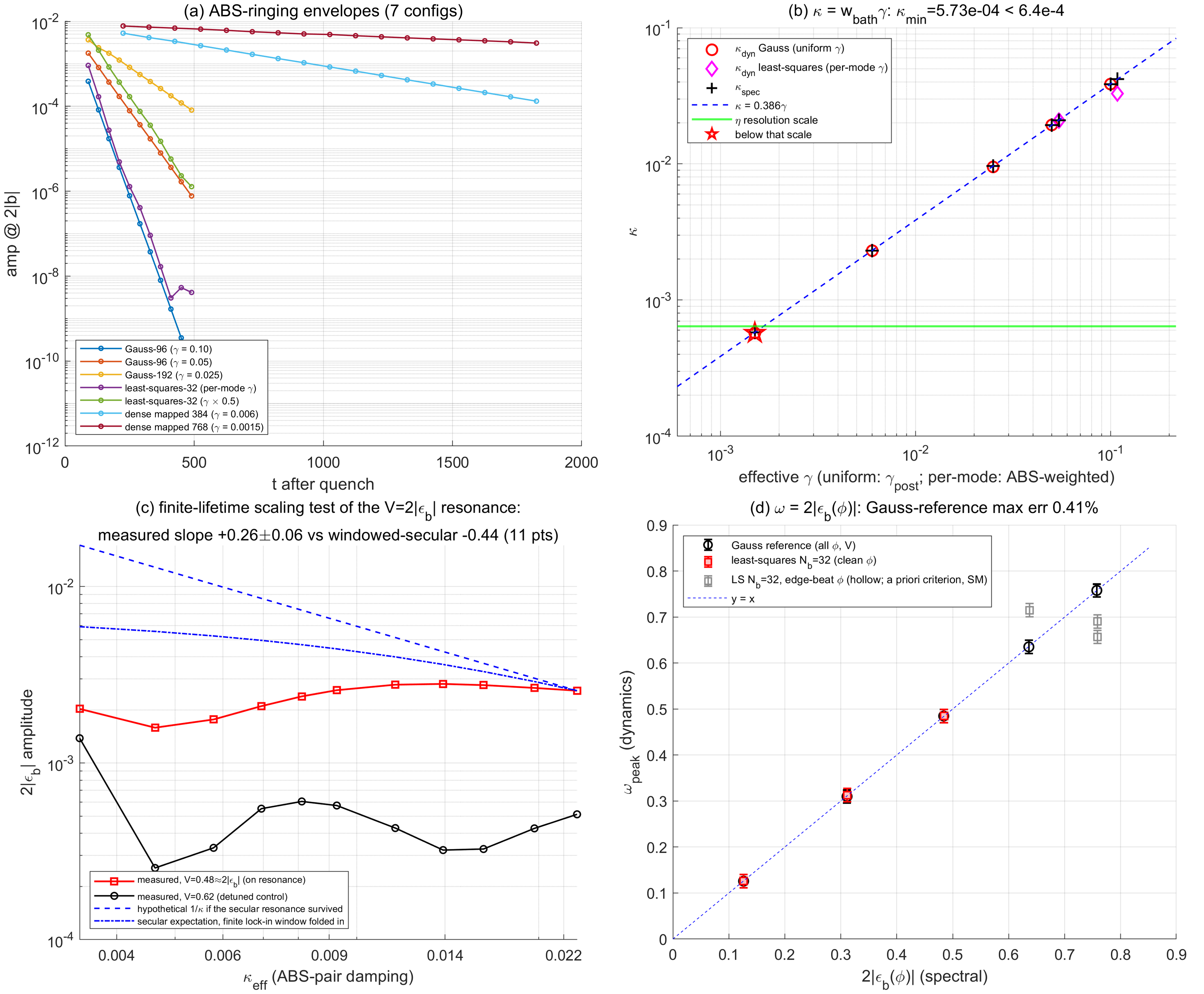}
\caption{Lifetime-scaling analysis of the quench dynamics.
(a)~Ringing envelopes across baths and nearly two decades of global
$\gamma$ ($0.0015$--$0.10$).
(b)~Scaling relation $\kappa \simeq w_{\mathrm{bath}}\gamma_{\mathrm{eff}}$
($\kappa_{\mathrm{dyn}}/\kappa_{\mathrm{spec}} = 0.986$--$1.003$ for
the six configurations whose envelope fit windows are free of
edge-beat takeover; the gap-clean LS-32 native point reads 0.78
because its fit window extends into the late-time regime where the a
priori predictable edge-beat comb overtakes the main line---over $t
\in [80, 210]$ its slope matches $\kappa_{\mathrm{spec}}$ to 4\%;
SM Sec.~S4); the starred
point [$\kappa = 5.73(10)(26)\times 10^{-4}$: descriptive OLS fit
interval and window-split systematic, SM Sec.~S1] lies below the
resolution scale of Ref.~\cite{cheng2024quasiparticle} (green line);
including both quoted uncertainty components, the upper edge
($6.09\times 10^{-4}$) remains below that scale.
(c)~Finite-lifetime scaling test of the transient resonance claimed
in Ref.~\cite{cheng2024quasiparticle}: measured amplitude (squares)
against the naive $1/\kappa$ growth (dashed) and the secular
expectation with the finite lock-in window folded in (dash-dotted);
the measured log-log trend ($+0.26 \pm 0.06$) is opposite in sign to
the windowed secular one ($-0.44$); multi-window and phasor audits
(SM Sec.~S1) bound, rather than exclude, a subdominant secular
component. (d)~Dispersion $\omega =
2|\varepsilon_b(\phi)|$ (Gauss reference max.\ 0.41\%); least-squares
points whose edge-beat comb falls a priori inside the assessment band
are shown hollow and excluded from the quantitative assessment (raw deviations 9--13\%; criterion and cure in SM
Sec.~S4). Protocols in SM Sec.~S1.}
\label{fig:quenchB}
\end{figure*}

\emph{Setup.} This section reproduces and goes beyond the quench
protocol of Ref.~\cite{cheng2024quasiparticle} (parameters as in that
work: $\varepsilon_d = 0$, $\Gamma = 0.8$, $\Delta = 1$, $D = 4$, $V =
0.8$, quench phase $\phi = 0.6\pi$): a sudden bias turn-on relaxes
into the ac Josephson NESS; driving for $t_{\mathrm{pre}} = 60$ and
then switching the bias off (turn-off) leaves trapped quasiparticles
in the final static junction oscillating at the coherent difference
frequency of the lowest ABS pair $\pm\varepsilon_b$, frequency
$2|\varepsilon_b| = 0.484$---the quasiparticle-trapping oscillation
reported in that reference. Our computed total bond current $I_A +
I_B$ and occupation $n(t)$ both show this $2\varepsilon_b$ oscillation
[Fig.~\ref{fig:quench}(c,d)], and all quantitative assessments of
frequency, dispersion, and lifetime are made on them.

\emph{Assessment: reproduction.} Turn-on: the $I(t)$ traces of the two
baths (Gauss-96 reference / least-squares-32) coincide; the NESS
spectral peak sits at $\omega_J = 2V = 1.60$ for both, deviations
0.67\% / 0.67\% [Fig.~\ref{fig:quench}(a,b)]. Turn-off: the
interpolated spectral-peak centers of the total bond current $I_A +
I_B$ land at
0.14\% / 0.36\% from $2|\varepsilon_b|$ (Gaussian reference /
least-squares-64 density-floor
variant), and the occupation $n(t)$ independently gives the same
0.14\% / 0.36\% [Fig.~\ref{fig:quench}(c,d)]. These sub-bin centers
are descriptive: the calibrated statement is that both spectra are
consistent with $2|\varepsilon_b|$ within the finite-window Fourier
resolution $\pi/T_{\mathrm{win}} = 0.026$ of this protocol's window
$t \in [80, 200]$. The dispersion relation
$\omega = 2|\varepsilon_b(\phi)|$ holds over 5 phases $\times$ 2
biases: the Gaussian reference covers all with maximum deviation
0.41\%, and the same phase at two biases gives the same frequency
($V$-independence, excluding drive residue)
[Fig.~\ref{fig:quenchB}(d)]; error bars are the frequency resolution
$\pm\pi/T_{\mathrm{win}} \approx 0.014$ of the longer
dispersion-protocol window $[80, 300]$ (SM Sec.~S1).

\emph{Assessment: the $\kappa$ scaling law and damping-induced decay
rates below the
resolution scale of Ref.~\cite{cheng2024quasiparticle}.} What controls
the decay of the post-quench ringing? Sweeping nearly two decades of
global $\gamma$ ($0.0015$--$0.10$) and several bath prescriptions
(uniform-$\gamma$ Gaussian
reference, per-mode-$\gamma$ least squares, densely mapped grids), we
measure the envelope decay rate $\kappa_{\mathrm{dyn}}$ and find a
single scaling law for this DLvN generator:
\begin{equation}
\kappa \simeq w_{\mathrm{bath}}\, \gamma_{\mathrm{eff}},
\label{eq:kappa}
\end{equation}
to leading order in the damping operator, with both factors explicitly
defined and given directly by
diagonalizing the final Hamiltonian (a damping-free, purely Hermitian
problem: both factors are fixed before any time evolution is
run)---no fitted parameters:
$w_{\mathrm{bath}} = \sum_j |\langle j|\psi_{\mathrm{ABS}}\rangle|^2$
is the total bath weight of the ABS wave function (0.386 at this
parameter point), and $\gamma_{\mathrm{eff}} = \sum_j p_j \gamma_j$
with $p_j = |\langle j|\psi_{\mathrm{ABS}}\rangle|^2 /
w_{\mathrm{bath}}$ is the mode-damping average in the ABS-weight
measure (reducing to the bare $\gamma$ for uniform baths). The product
$\kappa_{\mathrm{spec}} = \sum_j |\langle
j|\psi_{\mathrm{ABS}}\rangle|^2 \gamma_j$ is precisely the first-order
perturbative expectation of the damping operator in the ABS state---so
the scaling law is a falsifiable perturbative prediction, not a
parametrized fit; the ratio of the dynamical $\kappa_{\mathrm{dyn}}$
to the spectral $\kappa_{\mathrm{spec}}$---0.986--1.003 across the six
configurations whose envelope fit windows are free of edge-beat
takeover
[Fig.~\ref{fig:quenchB}(a,b)]---is the test of that picture. The
seventh configuration, the gap-clean LS-32 native bath, reads 0.78 on
this estimator, and the anatomy is instructive rather than damaging:
its fit window extends to $t \approx 520$, deep into the regime where
the a priori predictable edge-beat comb (decay $\sim$0.023, from the
mode table) outlives and overtakes the main line (decay 0.042). The
measured envelope slope is $\kappa_{\mathrm{spec}}$ to 4\% over $t
\in [80, 210]$ and collapses toward the beat rate at late times
($0.018$ over $[330, 490]$)---so 0.78 is a mixed slope, not an
in-window violation of the law. Halving the dampings halves the
decay-rate difference and pushes the takeover beyond the same fit
window, which is why the $\times 0.5$ variant stays clean throughout
(late-window slope 0.021 vs $\kappa_{\mathrm{spec}} = 0.021$; the
takeover bookkeeping, with both decay rates a priori from the
$\{\varepsilon_j, \gamma_j\}$ tables, is in SM Sec.~S4). The native
point is shown in
Fig.~\ref{fig:quenchB}(b) for completeness rather than as a test
point; it is the same comb that dictates the density-floor variant
for long-time observables (end of this subsection). On factor
conventions: $\kappa$ is defined as the decay rate of the bilinear
correlation (the oscillation envelope)---single-particle mode
amplitudes decay at $\gamma_j/2$, while the coherence adds the two
decay rates of the ABS pair, so the $\tfrac{1}{2}$ and the 2 cancel
exactly and no $\tfrac{1}{2}$ appears in $\kappa_{\mathrm{spec}}$
(derivation in SM Sec.~S2). The scaling relation gives a sharp
extrapolation: $\kappa \propto \gamma$, so along the
recurrence-controlled joint path ($\gamma \to 0$, $N_b \to \infty$) the
ideal persistent oscillation is recovered. A direct
measurement goes further: the densely mapped grid at $N_b = 768$,
$\gamma = 1.5\times 10^{-3}$ yields $\kappa = 5.73(10)(26)\times
10^{-4}$ [fit interval and window-split systematic],
below the resolution scale $6.4\times 10^{-4}$ associated with the
$\eta$ used in the calculation of Ref.~\cite{cheng2024quasiparticle}
[Fig.~\ref{fig:quenchB}(b), star]. This point carries its own
recurrence audit: the bath memory kernel rebuilt from the mode table
shows no revival over $t \le 2000$ (a monotonically decaying tail),
$\kappa$ refit on the window subranges $[100,600]/[600,1200]/
[1200,1900]$ drifts by only 2\%, and the independent spectral
estimator agrees at 0.8\% (SM Sec.~S1). With $\gamma$ independently
calibrated by the $\kappa$ scaling
law (Sec.~\ref{sec:dlvn}), damping-induced decay rates below that
scale become numerically resolvable and calibrated (the measured
$\kappa$ is itself damping-induced and extrapolates to zero with
$\gamma$). Scope: the mode density required in the extreme
small-$\kappa$ regime scales as $1/\gamma$, a resolution requirement
that compression cannot circumvent; the $\kappa$-minimal data point
uses the densely mapped grid rather than a compressed bath (whose
$\gamma$ floor is mode-number limited).

\emph{Boundary: the transient resonance of
Ref.~\cite{cheng2024quasiparticle}.} Figure~5 of
Ref.~\cite{cheng2024quasiparticle} claims a narrow resonance of the
quench transient amplitude at $V = 2|\varepsilon_b|$ (from a secular
$0/0$ term in perturbation theory). A minimal model fixes the
discriminating behavior: for a damped mode driven at resonance the
envelope grows as $(1 - e^{-\kappa t})/\kappa$, i.e., $\propto
1/\kappa$ once $\kappa t \gtrsim 1$; a finite measurement window must
therefore be folded into the prediction before it can discriminate.
We test this directly with $\kappa$ as the control parameter: the
$2|\varepsilon_b|$ lock-in amplitude at the resonance (detuning
folded in) and at a detuned control, over 11 values of $\kappa$
($3.5\times 10^{-3}$--$2.3\times 10^{-2}$, window $t \in [60, 160]$
of the sustained drive). Over this scan the \emph{window-folded}
secular expectation still rises by a factor 2.3 toward small $\kappa$
(log-log slope $-0.44$), while the measured amplitude shows a mildly
\emph{positive} trend ($+0.26 \pm 0.06$, descriptive OLS); on the
upper half of the scan the secular curve falls by 50\% and the
measured amplitude by 8\% [Fig.~\ref{fig:quenchB}(c)]. Two audits
qualify the conclusion (both in SM Sec.~S1): the slope is
window-dependent (a later window is statistically compatible both
with zero and with its own secular slope, but the detuned control
moves comparably there, identifying a $\kappa$-dependent background
not specific to the resonance), and the complex-phasor decomposition
into constant and window-response components is nearly collinear over
the accessible $\kappa$ range. The calibrated conclusion is therefore
a bound: over the tested lifetime range and observation windows, the
measured response does not exhibit the growth expected when the
simplest single-mode secular contribution dominates, while a smaller
secular component superposed on the background is not excluded; the
coarse ordering of that reference's Fig.~3 (larger amplitudes for $V$
near $2|\varepsilon_b|$) is reproduced. We refer to this protocol as
the finite-lifetime scaling test: measuring an observable's scaling
with $\kappa$ separates behavior that survives finite lifetimes from
behavior consistent with an $\eta \to 0$ singular limit, and bounds
the latter when the decomposition is degenerate.

A final limitation belongs to the method itself: gap-clean fitting
gives gap-edge modes tiny $\gamma$, and their difference-frequency
comb with the ABS ($E_j - |\varepsilon_b|$) outlives the main line,
contaminating long-time spectra (artifact/main peak $= 1.03$); the
contamination condition is predictable a priori from the mode table
(difference frequencies falling into the $\pm 15\%$ assessment band).
The density-floor variant (constraint 3 of Sec.~\ref{sec:lsq}) pushes
the artifact to 0.09 (SM Sec.~S4, Fig.~S3). That is why
Fig.~\ref{fig:quench}(c,d) uses least-squares-64: the bath variant is
chosen according to the observable.

\subsection{Validation 4: Shapiro locking (integer and
fractional)}\label{sec:shapiro}

\begin{figure*}
\includegraphics[width=0.85\textwidth]{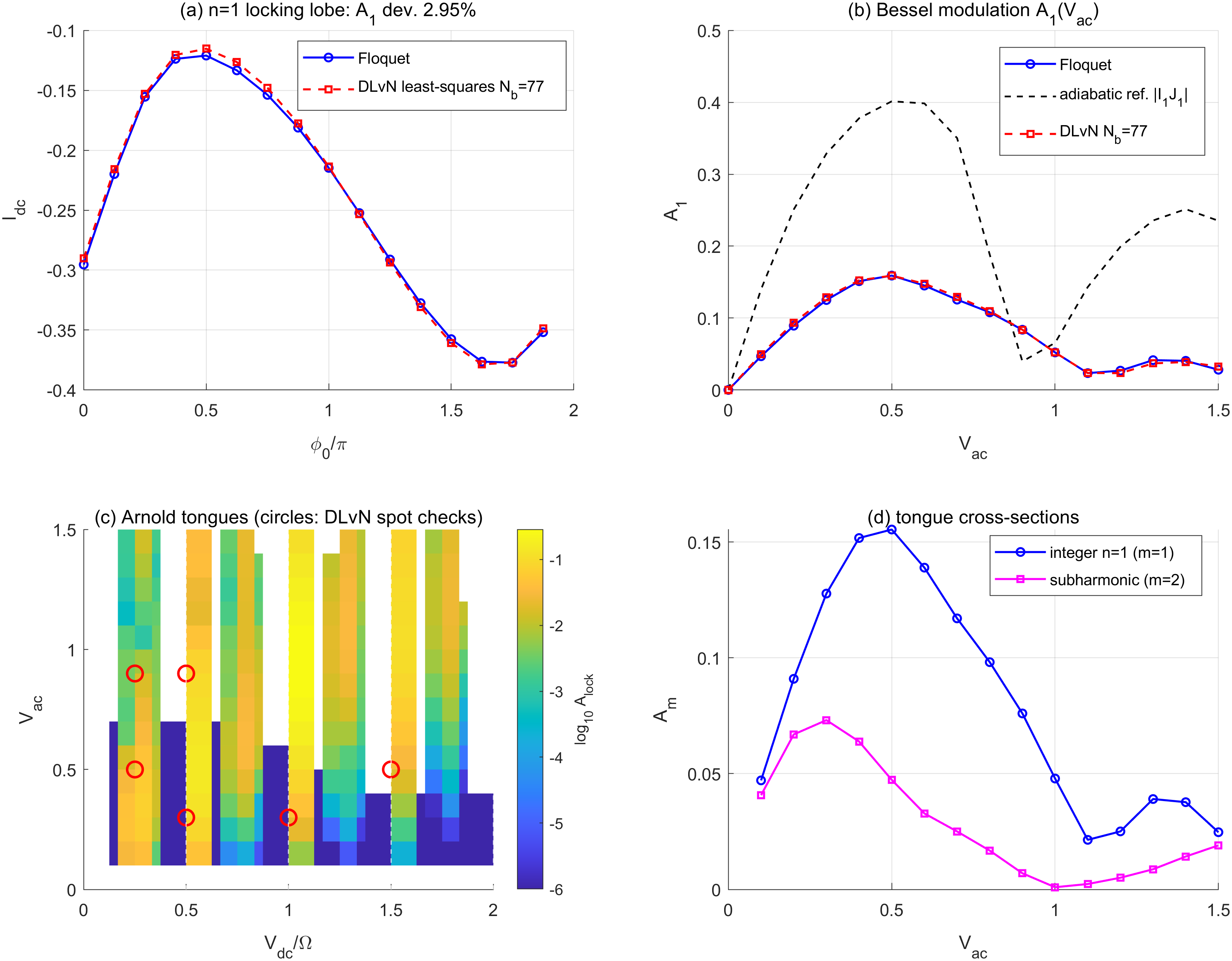}
\caption{Shapiro locking in the low-energy effective junction:
integer-step benchmark ($\Gamma_S = 1$, $\Gamma_N = 0.2$, $\Omega =
0.5$). Throughout, $A_1$ denotes the amplitude of the phase-locked
(Shapiro) first-harmonic current component---the step order parameter.
(a)~The $n = 1$ locking lobe: Floquet and DLvN (least squares
$N_b = 77$) overlap; $A_1$ deviation 2.95\%. (b)~Bessel modulation
$A_1(V_{\mathrm{ac}})$; the adiabatic reference is exact only at
$\Omega \to 0$ with sinusoidal CPR. (c,d)~Arnold-tongue map and
vertical cuts; circles mark DLvN spot checks (integer steps:
2.0--3.5\% for $n = 1, 2$; 7.5\% for $n = 3$). Protocols in SM
Sec.~S9.}
\label{fig:shapiroA}
\end{figure*}

\begin{figure*}
\includegraphics[width=0.85\textwidth]{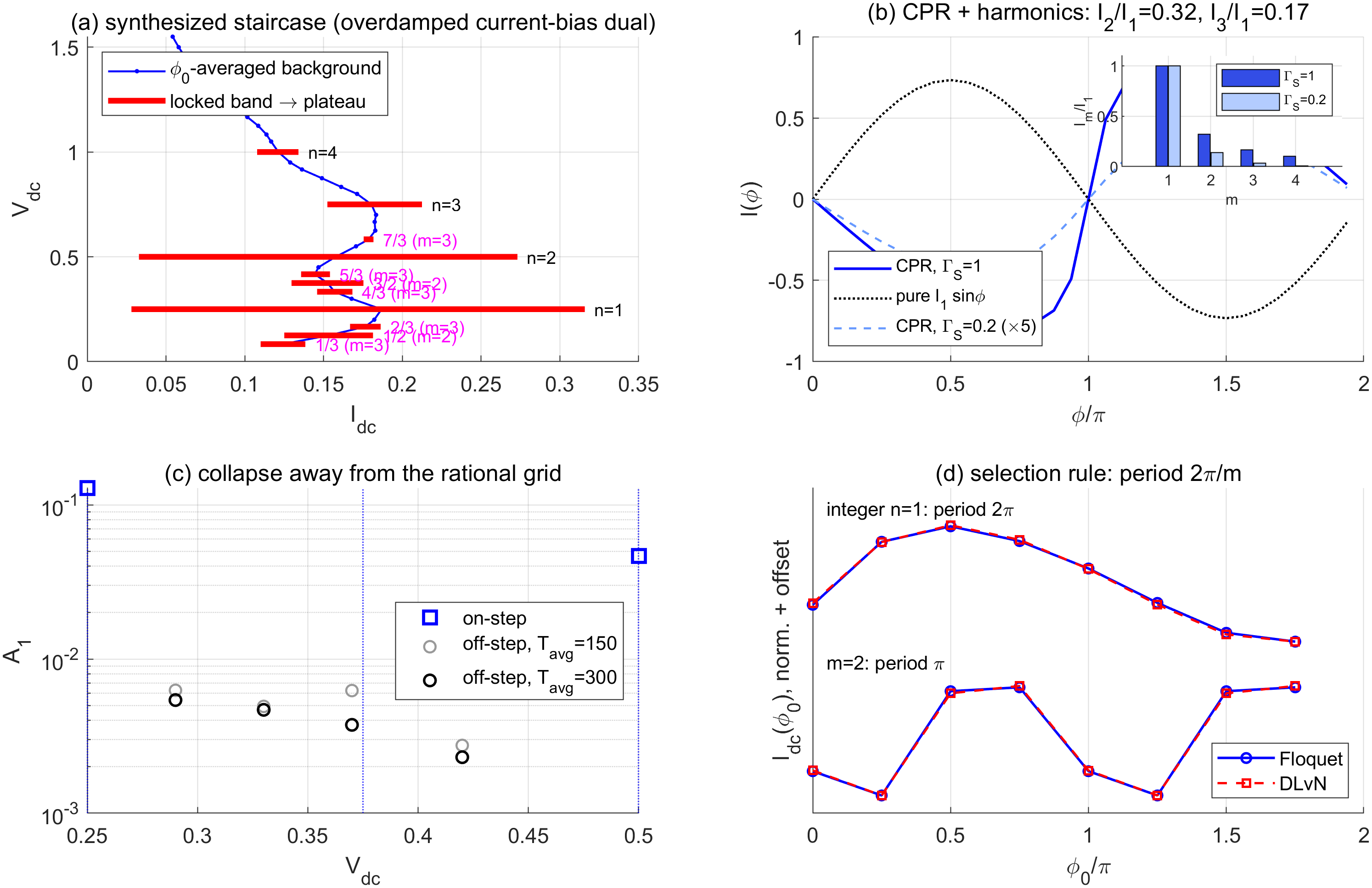}
\caption{Fractional Shapiro steps in the effective junction (same
junction as Fig.~\ref{fig:shapiroA}): application face and origin
checks. (a)~The classic staircase synthesized from the voltage-biased
data (the experimentally familiar current-biased staircase,
reconstructed from the voltage-biased simulation via the
overdamped-junction duality rather than a separate current-biased run;
data-driven plateau labels): integer and fractional plateaus. (b)~The static CPR is strongly
nonsinusoidal at the working point; inset: harmonic content $I_2/I_1
= 0.32$, $I_3/I_1 = 0.17$, collapsing to $I_{2,3,4}/I_1 =
0.14/0.03/0.01$ in the tunneling control. (c)~Away from the rational grid $A_1$ is
20--50$\times$ smaller and decreases as $1/T_{\mathrm{avg}}$.
(d)~Selection rule: the $m = 2$ lock is $\pi$-periodic in $\phi_0$;
$A_1$ on the $m = 2$ step is a numerical zero ($10^{-17}$). The
spot-check table is shown in SM Sec.~S9.}
\label{fig:shapiroB}
\end{figure*}

\begin{figure*}
\includegraphics[width=0.72\textwidth]{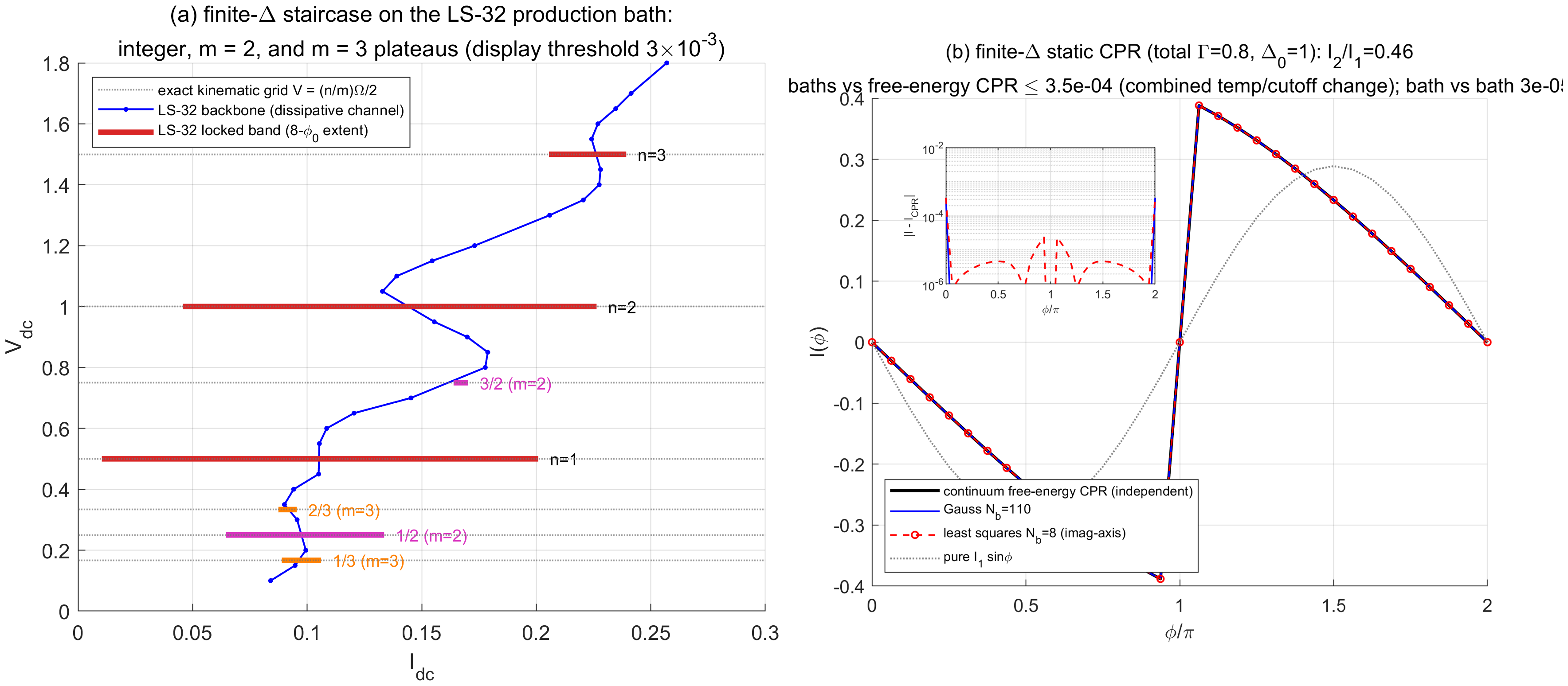}
\caption{Finite-gap Shapiro locking as a capability demonstration
(two-terminal S-QD-S; locking amplitudes qualitative). (a)~Staircase
presentation (current-bias dual) of the locking map on
the LS-32 production bath: 28-point backbone plus
\emph{every} $m \le 3$ rational resonance sampled at 8~$\phi_0$ (148
evolutions, 20~min at $1/39$ the per-step cost of the Gauss-110
recipe). Seven plateaus exceed the $3\times 10^{-3}$ display
threshold: integer $n = 1$--3, the $m = 2$ members $1/2$ and $3/2$,
and the $m = 3$ members $1/3$ and $2/3$ (orange), for which the
Floquet calculation is computationally prohibitive. Dotted lines: the exact
kinematic grid $V_{\mathrm{dc}} = (n/m)\,\Omega/2$. Locking appears
on this grid: all fifteen $m \le 3$ rationals were sampled at
8~$\phi_0$ (the $m = 3$ set extends to $11/6 \approx 1.83$, just
beyond the 1.80 backbone endpoint), with the remainder falling below
threshold. Off the grid, no plateau is resolved along the
single-$\phi_0$ backbone at the display threshold, and dedicated
8-$\phi_0$ sweeps at off-grid biases bracketing the two strongest
steps bound the locking amplitudes at $A_m \le 1.3\times 10^{-6}$,
four orders below the on-grid $A_1$ (SM Sec.~S9). The archived Gauss-110 backbone points agree at 0.1--5.4\%
inside the MAR-validated window (12.5\% at $V = 0.2$, below that
window). The commensurate resonance positions and the harmonic
selection-rule fingerprints are robust; absolute locking amplitudes,
plateau visibility relative to the fixed $3\times 10^{-3}$ display
threshold, and plateau \emph{extents} derive from
the locking amplitudes and remain qualitative
(Sec.~\ref{sec:dlvn}). (b)~Static CPR of the same junction
(total $\Gamma = 0.8$): an independent continuum free-energy CPR benchmark
(Matsubara free-energy derivative, finite band), the
Gauss-110 discretization, and the compressed least-squares $N_b = 8$
bath (imaginary-axis, statics-optimized variant)
coincide---both baths agree with the benchmark at $\le 3.5\times
10^{-4}$ (comparable to the benchmark's combined temperature/cutoff convergence change) and with each other at
$3\times 10^{-5}$; inset: pointwise deviations from the benchmark. The
CPR is more nonsinusoidal than the effective junction's ($I_2/I_1 =
0.46$), self-consistent with the wider fractional steps. Protocols in
SM Secs.~S1 and S10.}
\label{fig:shapiro_fd}
\end{figure*}

\emph{Setup and quantitative-domain declaration.} Under ac drive $V(t)
= V_{\mathrm{dc}} + V_{\mathrm{ac}}\cos(\Omega t)$, the dc current
locks at $V_{\mathrm{dc}} = (n/m)(\Omega/2)$---the commensurate
locking resonances that underlie Shapiro
steps~\cite{shapiro1963josephson}. Terminology: our simulations are
voltage-biased, so the measured objects are locking resonances and
their amplitudes; ``step/staircase/plateau'' is reserved for the
current-bias dual presentation [Fig.~\ref{fig:shapiroB}(a) and
Fig.~\ref{fig:shapiro_fd}(a)], a mapping without circuit
self-consistency. In the reduced fraction $n/m$, the
integer $m$ plays three related roles: the denominator of the step
position, the $\phi_0$ harmonic channel of the locking, and the $m$-th
coherent channel in the harmonic phase-locking picture (i.e., the
$m$-th CPR harmonic; we do not interpret at the level of microscopic
transfer events). The quantitative domain of this section is the
low-energy effective junction ($\Delta \to \infty$ integrating out the
superconducting lead, superconducting hybridization $\Gamma_S = 1$---in
that limit $\Gamma_S$ directly plays the induced pairing amplitude on
the dot; plus a normal probe lead, $\Omega = 0.5$); a finite-gap
capability demonstration follows. This approximation-domain declaration is the
scope of the section: all quantitative numbers are claimed inside it.
The finite-gap data are used only for resonance existence, position, and
selection-rule checks; they are not used as quantitative amplitude
benchmarks.

\emph{Integer-step assessment.} Three comparisons, each performed on
dense sampling grids for both methods:
the $n = 1$ locking lobe (the 16-point full curve of $I_{\mathrm{dc}}$
versus $\phi_0$) overlaps between the two methods, locking amplitude
$A_1$ deviating 2.95\% [Fig.~\ref{fig:shapiroA}(a)]; the Bessel
modulation $A_1(V_{\mathrm{ac}})$, a 16-point full curve, tracks
Floquet throughout, including the lift and right shift of the first
zero [Fig.~\ref{fig:shapiroA}(b)]---the adiabatic reference $|I_1
J_1(2V_{\mathrm{ac}}/\Omega)|$ in that panel is the
photon-assisted-tunneling-type Bessel
modulation~\cite{tien1963multiphoton}, exact only at $\Omega \to 0$
with sinusoidal CPR; both methods deviate from it together, and the
deviation itself is the physical fingerprint of ``nonadiabatic +
nonsinusoidal CPR'' (consistent with the microscopic picture in which
a microwave field strongly reshapes the
CPR~\cite{bergeret2010theory}), not an error; the Arnold-tongue map
(24 rational columns $\times$ $V_{\mathrm{ac}}$) shows the integer
tongues brightest and running the full span, with DLvN spot checks at
2.0--3.5\% on the $n = 1, 2$ integer steps and 7.5\% on the $n = 3$
step [Fig.~\ref{fig:shapiroA}(c,d); protocol, spot-check table,
and sampling details in SM Sec.~S9]. The classic staircase
synthesized from the voltage-biased data, with data-driven plateau
labels, leads the fractional-step figure
[Fig.~\ref{fig:shapiroB}(a)]; it is the overdamped current-bias
\emph{dual}---a mapping of voltage-biased locking amplitudes onto the
familiar current-biased presentation, not a current-biased simulation
with circuit self-consistency.

\emph{Fractional steps: five consistency checks.} Fractional steps
invite the suspicion of numerical artifact or probe effect
(experimentally, conventional nontopological origins of fractional or
missing steps likewise require careful
screening~\cite{raes2020fractional,dartiailh2021missing}); five
interlocking checks support an intrinsic origin within the tested
regimes:
\begin{enumerate}
\item \emph{Selection rule (exact proof)}: time-translation symmetry
forces the $I_{\mathrm{dc}}(\phi_0)$ of an $m$-th-order lock to be
$2\pi/m$ periodic---exact and nonperturbative.
Figure~\ref{fig:shapiroB}(d) shows the $m = 2$ $\pi$-periodic pattern
with the two methods agreeing point by point; the numerical
fingerprint is $A_1 =$ numerical zero (within machine precision,
$10^{-17}$) on the $m = 2$ step.
\item \emph{Mechanism (CPR harmonics)}: the static CPR at the working
point is manifestly nonsinusoidal, harmonic ratios $I_2/I_1 = 0.32$,
$I_3/I_1 = 0.17$ [Fig.~\ref{fig:shapiroB}(b)]---in the harmonic
phase-locking picture the $m$-th CPR harmonic feeds the $m$-th locking
channel directly~\cite{cuevas2002subharmonic,raes2020fractional}.
\item \emph{Control (tunneling collapse)}: the tunneling-limit control
($\Gamma_S = 0.2$) collapses the harmonics ($I_{2,3,4}/I_1$) to 0.14/0.03/0.01 and the
fractional steps vanish with them---fractional steps are suppressed in
the tunneling limit, recovering the conventional Shapiro picture as a
special case.
\item \emph{Intrinsic nature (finite $\Delta$)}: in the pure
two-terminal S-QD-S junction (no normal probe at all), the $m = 2$
lock persists ($A_2 = 3.5\times 10^{-2}$ on the $1/2$ step, with $A_1
\sim 10^{-5}$, the averaging floor, three orders below), and the first
two members of the $m = 3$ family are likewise measurable ($A_3 =
9.0\times 10^{-3}$ at $1/3$ and $4.1\times 10^{-3}$ at $2/3$, with
$A_1, A_2 \sim 10^{-5}$) [Fig.~\ref{fig:shapiro_fd}(a)]---supporting
an intrinsic origin in the coherent locking mechanism of a
nonsinusoidal CPR rather than a probe effect.
\item \emph{Collapse away from the rational grid}: on step points
$A_1$ is averaging-window independent; between steps $A_1$ is
20--50$\times$ smaller and keeps shrinking as $1/T_{\mathrm{avg}}$
with longer windows [Fig.~\ref{fig:shapiroB}(c)]---a natural
capability of time-domain propagation, unavailable to the standard
commensurate Floquet implementation used here: it requires
rational $V_{\mathrm{dc}}/\Omega$ (the commensurate base frequency
makes the dimension grow rapidly with the denominator), while DLvN
evolves any (including irrational) bias directly.
\end{enumerate}

\emph{Finite-gap capability demonstration.} At finite $\Delta$ (true
superconducting bath, run on the LS-32 production bath of
Sec.~\ref{sec:mar}; the archived Gaussian-reference points serve as a
discretization cross-check) the role of the calculation changes from
validation to demonstration, and the claims are scoped accordingly.
What is quantitative: locking resonances appear on the kinematic grid
$V_{\mathrm{dc}} = (n/m)\,\Omega/2$ and at none of the tested
off-grid controls---the position
of a Shapiro step is an exact commensurability statement, not an
engine-dependent number---with no locking plateau resolved along the
single-$\phi_0$ backbone at the display threshold, and the locking
amplitudes themselves bounded four orders below the on-grid values by
the four dedicated off-grid 8-$\phi_0$ sweeps (a bound at the tested
controls, not a continuum scan; a single-$\phi_0$ backbone point by
itself cannot bound a locking amplitude, SM Sec.~S9)
[Fig.~\ref{fig:shapiro_fd}(a)]; the
selection rule holds to machine precision in the Floquet control
($A_1 \sim 10^{-17}$ on the $m = 2$ step; the real-time average
reproduces it to its $10^{-5}$ averaging floor); and the statics of the same junction
are certified against an independent continuum free-energy CPR benchmark---the
Gauss-110 discretization and the compressed $N_b = 8$
least-squares bath (the imaginary-axis, statics-optimized variant)
both match it at $\le 3.5\times 10^{-4}$ (comparable to its combined
temperature/cutoff convergence change), and match each other at $3\times 10^{-5}$
[Fig.~\ref{fig:shapiro_fd}(b)]. What is not: the locking amplitudes
(integer step $\sim$35\% from Floquet, $m = 2$ amplitude $\sim
3\times$) reside in the dissipation-free coherent locking channel
discussed in Sec.~\ref{sec:dlvn} and are reported as qualitative,
without a settled attribution. The $m = 3$ Floquet control is
computationally prohibitive (the commensurate base frequency makes
the Floquet dimension explode) and is replaced by the exact symmetry
selection rule, satisfied numerically to the DLvN averaging floor
($\sim 10^{-5}$), as an internal consistency check;
the $m = 3$ locking features of Fig.~\ref{fig:shapiro_fd}(a) are thus a
demonstration that, at this working point, the real-time route
remains computationally accessible within our implementation and
budget where the Floquet control is not. The finite-$\Delta$ CPR is more nonsinusoidal ($I_2/I_1 =
0.46$), self-consistent with its wider fractional steps
[Fig.~\ref{fig:shapiro_fd}(b)].

\section{Discussion}\label{sec:discussion}

\emph{Observable hierarchy.} The most transferable outcome of the
validation program is a hierarchy of observables. Peak positions,
parities, oscillation frequencies, selection rules, and lifetime
scaling laws remain robust against the finite-broadening layer within
the tested parameter ranges and numerical resolution (for the
lifetime law, after the a priori mode-table screening of assessment
windows contaminated by discrete edge beats, SM Sec.~S4); absolute
amplitudes carry it, and wherever they appear we report both absolute
and conditional relative deviations. Observables also stratify by
tolerance: MAR (steady-state averaged) is the most forgiving of
coarse discretization, long-time coherence spectra the
harshest---``passing'' on any single observable (e.g., uniform
discretization's 5.7\% on MAR) cannot be extrapolated to global
capability, which is why the matched-budget comparison uses
multi-observable stratified assessment.

\emph{Error taxonomy.} Quantitatively, all results are organized by
two error measures. ``Vs Gaussian reference'' (reproduction error
against the Gauss-110 production reference, Sec.~\ref{sec:protocols})
measures discretization/compression quality: this layer falls with
added modes to a plateau of about 3--4\% relative to the Gauss-110
DLvN reference (reached
already at 32 modes), consistent with the reference's
own broadening, its measured threshold-region discretization ripple
(1--2\% at the gap-edge biases, Sec.~\ref{sec:lsq}), and
damping-parametrization differences---increasing the fitted mode
count from 24 to 48 does not reduce it. ``Vs Floquet'' (against the
ideal broadening-free
junction) contains the $\gamma$-systematic layer shared by the
tested DLvN bath prescriptions: about 15\% in dc amplitudes; in ac locking amplitudes
2--3.5\% for the dominant $n = 1, 2$ integer steps, 7.5\% at the
tested $n = 3$ point, and 11--28\% for the small $m = 2$
subharmonics (Table~S4 of~\cite{supplemental}).

\emph{Domain of applicability.} The hard boundary of the method is $U
= 0$ (single-particle BdG level). Soft boundaries: the construction is
not tied to two-terminal geometry and straightforwardly generalizes to
mixed normal/superconducting multi-terminal setups (demonstrated here
up to three terminals), with finite and infinite gaps; the mode
density required in the extreme small-$\kappa$ regime scales as
$1/\gamma$---compression optimizes coefficients but cannot circumvent
this resolution requirement; the finite-$\Delta$ subgap
locking/supercurrent channel---the one transport channel that carries
coherence without dissipation---deviates at the $O(1)$ level in
strong-drive locking amplitudes (the $\sim$35\% of finite-$\Delta$
Shapiro; Sec.~\ref{sec:dlvn}), whence those amplitudes are reported
as qualitative. The LS-64 density-floor bath is a windowed auxiliary
rational representation validated only for the stated in-gap
spectroscopic observables---its far-tail anchors spoil the
high-frequency spectral moments (SM Secs.~S1 and S4), so high-frequency
observables lie outside its scope. The bath choice is
observable-dependent:
gap-clean for steady-state averages, density floor for long-time
spectra, dense core plus wings for normal leads.

\emph{Relation to auxiliary-mode routes.} The time-local NEGF
formulation of Ref.~\cite{kawamura2026timelocal}, which extends the
auxiliary-mode route to superconducting leads, is a closely related
concurrent approach: the BCS self-energy is decomposed by
Pad\'e-plus-AAA rational approximation (with a phenomenological Dynes
broadening regularizing the gap edge), and the periodic steady state
after a voltage quench is validated against Floquet Green functions.
The two routes differ in what is compressed and what is retained:
Ref.~\cite{kawamura2026timelocal} propagates auxiliary current
matrices built on general complex poles, retaining the non-Markovian
memory of the continuum, whereas the present construction compresses
the lead into an explicit positive-weight bath \emph{Hamiltonian}
with mode-resolved Markovian dampings---a representation that plugs
directly into correlation-matrix (and, prospectively, many-body
impurity-solver) machinery, keeps the fitted spectral weights
positive by construction, and makes the broadening itself an
optimization variable rather than a fixed regularizer. The gap-edge
broadening parameter, notably, appears in that auxiliary-mode route as
well; here it is made part of the optimization and its dynamical
systematics are quantified by scaling laws (Secs.~\ref{sec:lsq} and
\ref{sec:mar}), at the price of a fitting residual that we track
observable by observable (this section and SM Sec.~S3).

\emph{Methodological implications.} Three elements carry beyond this
work: (i) the finite-lifetime scaling test---measuring an observable's
scaling with $\kappa$ diagnoses, for infinite-lifetime-dependent
literature claims, whether the response is consistent with
finite-lifetime robustness, and provides a bound rather than a unique
decomposition when the components are degenerate
(Sec.~\ref{sec:quench});
(ii) the gauge discipline---no rotating terms on damped sites (the
compatibility condition of the static-reference implementation); all
time-dependent drive goes through the hopping gauge; (iii) the
matched-budget comparison + stratified assessment---a fair comparison
paradigm for discretization schemes (fixed total dimension, one
protocol, multiple observables stratified). More broadly, the
workflow is a bath-\emph{design} framework: a declared prescription
is selected per observable class (transport window, density floor,
normal core, imaginary axis), with the selection rules and their
costs stated up front, rather than one discretization claimed
optimal for all observables at once.

\section{Conclusions}\label{sec:conclusions}

We have formulated a superconductivity-aware driven Liouville--von
Neumann framework---structure preservation in Nambu (BdG) space,
gauge discipline for damped bath orbitals, and delimited reference
states---and, on top of it, a real-axis least-squares
bath-discretization framework for the damped dynamics of
superconducting quantum-dot junctions. Its numerical contribution is
condensed into four design constraints---$\gamma$ enters a
constrained optimization, differentiated in/out-of-gap weighting, a
gap-edge density floor, and dense-core-plus-wings compression for
normal leads---which together change the status of the damping from
an externally prescribed broadening to a jointly optimized,
independently calibrated parameter whose leading effect on the
trapped-quasiparticle decay extrapolates along a joint ($\gamma \to
0$, $N_b \to \infty$) refinement path.

The validations expose one common methodological pattern. The
framework first recovers established results---static current-phase
relations against the continuum free-energy CPR benchmark; MAR thresholds,
parity, and dispersion; trapped-quasiparticle oscillations; Shapiro
commensurability and selection rules---and the recovered baseline
then licenses the new capabilities. The most demanding dynamical
compression benchmark is the MAR reproduction: 32 modes per lead
reproduce the 110-mode Gaussian reference
to 3--4\%, with no further reduction from 24 to 48 fitted modes, at
$1/39$ of the per-step cost, reducing one $I$-$V$ curve from 7.7~h to
20~min and making two-dimensional parameter scans feasible on
workstation-scale hardware. All results obey a two-measure error
taxonomy: positions, parities, frequencies, selection rules, and the
parameter-free scaling relation $\kappa \simeq
w_{\mathrm{bath}}\gamma_{\mathrm{eff}}$ remain robust against the
$\gamma$-systematic layer within the tested parameter ranges and
numerical resolution (the lifetime law after the a priori mode-table
screening of edge-beat-contaminated assessment windows); absolute
amplitudes carry it and are reported
with both absolute and conditional relative errors.

The boundaries are stated as results and map directly onto the next
steps. (i) Finite-gap strong-drive locking amplitudes---the
dissipation-free coherent subgap channel---remain qualitative;
whether this can be closed within the Lindblad bath class, for
instance by reference-state or dissipator-structure improvements
toward auxiliary-master-equation-type
constructions~\cite{arrigoni2013nonequilibrium,dorda2014auxiliary,titvinidze2015transport},
or whether it reflects a deeper difference between real-time
relaxation and the frequency-domain periodic steady state in a
dissipation-free coherent channel, is under active investigation.
(ii) The construction is demonstrated here on a single-level dot with
up to three terminals; extending it to multilevel and multidot
junctions and richer normal--superconducting geometries does not
alter the underlying DLvN structure, although matrix-valued bath
fitting may be required. (iii) The
least-squares bath layer decouples from the on-dot physics: the
explicit Hamiltonian bath is a natural input for correlated
finite-reservoir methods such as AMEA, exact diagonalization, or
tensor-network impurity solvers. Extending the propagation to $U \ne
0$ remains future work, for which the three-way equilibrium
cross-validation of this paper (Sec.~\ref{sec:cpr}) provides
benchmarks and calibrated conventions. (iv) In the extreme
small-$\kappa$ regime the required mode density scales as
$1/\gamma$---a resolution requirement that adaptive bath design can
optimize but not circumvent.

\begin{acknowledgments}
We thank Hui Jiang for useful discussions.
This research did not receive any specific grant from funding agencies
in the public, commercial, or not-for-profit sectors.
\end{acknowledgments}

\section*{Data availability}
The bath parameter tables (mode energies, weights, and dampings for
all least-squares baths), the processed data underlying
Figs.~\ref{fig:schematic}--\ref{fig:shapiro_fd} and the Supplemental
Material figures, the DLvN-side fitting and propagation code, and
representative reproduction demos (bath refit, static CPR, quench
decay-rate scaling, and one full MAR $I$-$V$ curve) will be deposited
in a public repository with a citable DOI upon publication. The Floquet reference engine is
not included in the package; its outputs used in the figures are
part of the archived data, and the protocols to regenerate the raw
time traces are specified in the Supplemental Material.


\begin{thebibliography}{47}%
\makeatletter
\providecommand \@ifxundefined [1]{%
 \@ifx{#1\undefined}
}%
\providecommand \@ifnum [1]{%
 \ifnum #1\expandafter \@firstoftwo
 \else \expandafter \@secondoftwo
 \fi
}%
\providecommand \@ifx [1]{%
 \ifx #1\expandafter \@firstoftwo
 \else \expandafter \@secondoftwo
 \fi
}%
\providecommand \natexlab [1]{#1}%
\providecommand \enquote  [1]{``#1''}%
\providecommand \bibnamefont  [1]{#1}%
\providecommand \bibfnamefont [1]{#1}%
\providecommand \citenamefont [1]{#1}%
\providecommand \href@noop [0]{\@secondoftwo}%
\providecommand \href [0]{\begingroup \@sanitize@url \@href}%
\providecommand \@href[1]{\@@startlink{#1}\@@href}%
\providecommand \@@href[1]{\endgroup#1\@@endlink}%
\providecommand \@sanitize@url [0]{\catcode `\\12\catcode `\$12\catcode
  `\&12\catcode `\#12\catcode `\^12\catcode `\_12\catcode `\%12\relax}%
\providecommand \@@startlink[1]{}%
\providecommand \@@endlink[0]{}%
\providecommand \url  [0]{\begingroup\@sanitize@url \@url }%
\providecommand \@url [1]{\endgroup\@href {#1}{\urlprefix }}%
\providecommand \urlprefix  [0]{URL }%
\providecommand \Eprint [0]{\href }%
\providecommand \doibase [0]{https://doi.org/}%
\providecommand \selectlanguage [0]{\@gobble}%
\providecommand \bibinfo  [0]{\@secondoftwo}%
\providecommand \bibfield  [0]{\@secondoftwo}%
\providecommand \translation [1]{[#1]}%
\providecommand \BibitemOpen [0]{}%
\providecommand \bibitemStop [0]{}%
\providecommand \bibitemNoStop [0]{.\EOS\space}%
\providecommand \EOS [0]{\spacefactor3000\relax}%
\providecommand \BibitemShut  [1]{\csname bibitem#1\endcsname}%
\let\auto@bib@innerbib\@empty
\bibitem [{\citenamefont {Mart{\'i}n-Rodero}\ and\ \citenamefont
  {Levy~Yeyati}(2011)}]{martinrodero2011josephson}%
  \BibitemOpen
  \bibfield  {author} {\bibinfo {author} {\bibfnamefont {A.}~\bibnamefont
  {Mart{\'i}n-Rodero}}\ and\ \bibinfo {author} {\bibfnamefont {A.}~\bibnamefont
  {Levy~Yeyati}},\ }\bibfield  {title} {\bibinfo {title} {Josephson and
  {Andreev} transport through quantum dots},\ }\href@noop {} {\bibfield
  {journal} {\bibinfo  {journal} {Adv. Phys.}\ }\textbf {\bibinfo {volume}
  {60}},\ \bibinfo {pages} {899} (\bibinfo {year} {2011})}\BibitemShut
  {NoStop}%
\bibitem [{\citenamefont {Bretheau}\ \emph {et~al.}(2012)\citenamefont
  {Bretheau}, \citenamefont {Girit}, \citenamefont {Tosi}, \citenamefont
  {Goffman}, \citenamefont {Joyez}, \citenamefont {Pothier}, \citenamefont
  {Esteve},\ and\ \citenamefont {Urbina}}]{bretheau2012superconducting}%
  \BibitemOpen
  \bibfield  {author} {\bibinfo {author} {\bibfnamefont {L.}~\bibnamefont
  {Bretheau}}, \bibinfo {author} {\bibfnamefont {{\c C}.}~\bibnamefont
  {Girit}}, \bibinfo {author} {\bibfnamefont {L.}~\bibnamefont {Tosi}},
  \bibinfo {author} {\bibfnamefont {M.}~\bibnamefont {Goffman}}, \bibinfo
  {author} {\bibfnamefont {P.}~\bibnamefont {Joyez}}, \bibinfo {author}
  {\bibfnamefont {H.}~\bibnamefont {Pothier}}, \bibinfo {author} {\bibfnamefont
  {D.}~\bibnamefont {Esteve}},\ and\ \bibinfo {author} {\bibfnamefont
  {C.}~\bibnamefont {Urbina}},\ }\bibfield  {title} {\bibinfo {title}
  {Superconducting quantum point contacts},\ }\href@noop {} {\bibfield
  {journal} {\bibinfo  {journal} {C. R. Physique}\ }\textbf {\bibinfo {volume}
  {13}},\ \bibinfo {pages} {89} (\bibinfo {year} {2012})}\BibitemShut {NoStop}%
\bibitem [{\citenamefont {Cheng}\ \emph {et~al.}(2024)\citenamefont {Cheng},
  \citenamefont {Zuo}, \citenamefont {Wang},\ and\ \citenamefont
  {Xing}}]{cheng2024quasiparticle}%
  \BibitemOpen
  \bibfield  {author} {\bibinfo {author} {\bibfnamefont {J.}~\bibnamefont
  {Cheng}}, \bibinfo {author} {\bibfnamefont {X.}~\bibnamefont {Zuo}}, \bibinfo
  {author} {\bibfnamefont {J.}~\bibnamefont {Wang}},\ and\ \bibinfo {author}
  {\bibfnamefont {Y.}~\bibnamefont {Xing}},\ }\bibfield  {title} {\bibinfo
  {title} {Quasiparticle trapping in quench dynamics of superconductor/quantum
  dot/superconductor {J}osephson junctions},\ }\href@noop {} {\bibfield
  {journal} {\bibinfo  {journal} {Phys. Rev. B}\ }\textbf {\bibinfo {volume}
  {110}},\ \bibinfo {pages} {125417} (\bibinfo {year} {2024})}\BibitemShut
  {NoStop}%
\bibitem [{\citenamefont {Averin}\ and\ \citenamefont
  {Bardas}(1995)}]{averin1995ac}%
  \BibitemOpen
  \bibfield  {author} {\bibinfo {author} {\bibfnamefont {D.}~\bibnamefont
  {Averin}}\ and\ \bibinfo {author} {\bibfnamefont {A.}~\bibnamefont
  {Bardas}},\ }\bibfield  {title} {\bibinfo {title} {ac {J}osephson effect in a
  single quantum channel},\ }\href@noop {} {\bibfield  {journal} {\bibinfo
  {journal} {Phys. Rev. Lett.}\ }\textbf {\bibinfo {volume} {75}},\ \bibinfo
  {pages} {1831} (\bibinfo {year} {1995})}\BibitemShut {NoStop}%
\bibitem [{\citenamefont {Cuevas}\ \emph {et~al.}(1996)\citenamefont {Cuevas},
  \citenamefont {Mart{\'i}n-Rodero},\ and\ \citenamefont
  {Levy~Yeyati}}]{cuevas1996hamiltonian}%
  \BibitemOpen
  \bibfield  {author} {\bibinfo {author} {\bibfnamefont {J.~C.}\ \bibnamefont
  {Cuevas}}, \bibinfo {author} {\bibfnamefont {A.}~\bibnamefont
  {Mart{\'i}n-Rodero}},\ and\ \bibinfo {author} {\bibfnamefont
  {A.}~\bibnamefont {Levy~Yeyati}},\ }\bibfield  {title} {\bibinfo {title}
  {Hamiltonian approach to the transport properties of superconducting quantum
  point contacts},\ }\href@noop {} {\bibfield  {journal} {\bibinfo  {journal}
  {Phys. Rev. B}\ }\textbf {\bibinfo {volume} {54}},\ \bibinfo {pages} {7366}
  (\bibinfo {year} {1996})}\BibitemShut {NoStop}%
\bibitem [{\citenamefont {Bratus'}\ \emph {et~al.}(1995)\citenamefont
  {Bratus'}, \citenamefont {Shumeiko},\ and\ \citenamefont
  {Wendin}}]{bratus1995theory}%
  \BibitemOpen
  \bibfield  {author} {\bibinfo {author} {\bibfnamefont {E.~N.}\ \bibnamefont
  {Bratus'}}, \bibinfo {author} {\bibfnamefont {V.~S.}\ \bibnamefont
  {Shumeiko}},\ and\ \bibinfo {author} {\bibfnamefont {G.}~\bibnamefont
  {Wendin}},\ }\bibfield  {title} {\bibinfo {title} {Theory of subharmonic gap
  structure in superconducting mesoscopic tunnel contacts},\ }\href@noop {}
  {\bibfield  {journal} {\bibinfo  {journal} {Phys. Rev. Lett.}\ }\textbf
  {\bibinfo {volume} {74}},\ \bibinfo {pages} {2110} (\bibinfo {year}
  {1995})}\BibitemShut {NoStop}%
\bibitem [{\citenamefont {Shapiro}(1963)}]{shapiro1963josephson}%
  \BibitemOpen
  \bibfield  {author} {\bibinfo {author} {\bibfnamefont {S.}~\bibnamefont
  {Shapiro}},\ }\bibfield  {title} {\bibinfo {title} {Josephson currents in
  superconducting tunneling: the effect of microwaves and other observations},\
  }\href@noop {} {\bibfield  {journal} {\bibinfo  {journal} {Phys. Rev. Lett.}\
  }\textbf {\bibinfo {volume} {11}},\ \bibinfo {pages} {80} (\bibinfo {year}
  {1963})}\BibitemShut {NoStop}%
\bibitem [{\citenamefont {Cuevas}\ \emph {et~al.}(2002)\citenamefont {Cuevas},
  \citenamefont {Heurich}, \citenamefont {Mart{\'i}n-Rodero}, \citenamefont
  {Levy~Yeyati},\ and\ \citenamefont {Sch{\"o}n}}]{cuevas2002subharmonic}%
  \BibitemOpen
  \bibfield  {author} {\bibinfo {author} {\bibfnamefont {J.~C.}\ \bibnamefont
  {Cuevas}}, \bibinfo {author} {\bibfnamefont {J.}~\bibnamefont {Heurich}},
  \bibinfo {author} {\bibfnamefont {A.}~\bibnamefont {Mart{\'i}n-Rodero}},
  \bibinfo {author} {\bibfnamefont {A.}~\bibnamefont {Levy~Yeyati}},\ and\
  \bibinfo {author} {\bibfnamefont {G.}~\bibnamefont {Sch{\"o}n}},\ }\bibfield
  {title} {\bibinfo {title} {Subharmonic {S}hapiro steps and assisted tunneling
  in superconducting point contacts},\ }\href@noop {} {\bibfield  {journal}
  {\bibinfo  {journal} {Phys. Rev. Lett.}\ }\textbf {\bibinfo {volume} {88}},\
  \bibinfo {pages} {157001} (\bibinfo {year} {2002})}\BibitemShut {NoStop}%
\bibitem [{\citenamefont {Scheer}\ \emph {et~al.}(1997)\citenamefont {Scheer},
  \citenamefont {Joyez}, \citenamefont {Esteve}, \citenamefont {Urbina},\ and\
  \citenamefont {Devoret}}]{scheer1997conduction}%
  \BibitemOpen
  \bibfield  {author} {\bibinfo {author} {\bibfnamefont {E.}~\bibnamefont
  {Scheer}}, \bibinfo {author} {\bibfnamefont {P.}~\bibnamefont {Joyez}},
  \bibinfo {author} {\bibfnamefont {D.}~\bibnamefont {Esteve}}, \bibinfo
  {author} {\bibfnamefont {C.}~\bibnamefont {Urbina}},\ and\ \bibinfo {author}
  {\bibfnamefont {M.~H.}\ \bibnamefont {Devoret}},\ }\bibfield  {title}
  {\bibinfo {title} {Conduction channel transmissions of atomic-size aluminum
  contacts},\ }\href@noop {} {\bibfield  {journal} {\bibinfo  {journal} {Phys.
  Rev. Lett.}\ }\textbf {\bibinfo {volume} {78}},\ \bibinfo {pages} {3535}
  (\bibinfo {year} {1997})}\BibitemShut {NoStop}%
\bibitem [{\citenamefont {Eichler}\ \emph {et~al.}(2007)\citenamefont
  {Eichler}, \citenamefont {Weiss}, \citenamefont {Oberholzer}, \citenamefont
  {Sch{\"o}nenberger}, \citenamefont {Levy~Yeyati}, \citenamefont {Cuevas},\
  and\ \citenamefont {Mart{\'i}n-Rodero}}]{eichler2007evenodd}%
  \BibitemOpen
  \bibfield  {author} {\bibinfo {author} {\bibfnamefont {A.}~\bibnamefont
  {Eichler}}, \bibinfo {author} {\bibfnamefont {M.}~\bibnamefont {Weiss}},
  \bibinfo {author} {\bibfnamefont {S.}~\bibnamefont {Oberholzer}}, \bibinfo
  {author} {\bibfnamefont {C.}~\bibnamefont {Sch{\"o}nenberger}}, \bibinfo
  {author} {\bibfnamefont {A.}~\bibnamefont {Levy~Yeyati}}, \bibinfo {author}
  {\bibfnamefont {J.~C.}\ \bibnamefont {Cuevas}},\ and\ \bibinfo {author}
  {\bibfnamefont {A.}~\bibnamefont {Mart{\'i}n-Rodero}},\ }\bibfield  {title}
  {\bibinfo {title} {Even-odd effect in {A}ndreev transport through a carbon
  nanotube quantum dot},\ }\href@noop {} {\bibfield  {journal} {\bibinfo
  {journal} {Phys. Rev. Lett.}\ }\textbf {\bibinfo {volume} {99}},\ \bibinfo
  {pages} {126602} (\bibinfo {year} {2007})}\BibitemShut {NoStop}%
\bibitem [{\citenamefont {Wilson}(1975)}]{wilson1975renormalization}%
  \BibitemOpen
  \bibfield  {author} {\bibinfo {author} {\bibfnamefont {K.~G.}\ \bibnamefont
  {Wilson}},\ }\bibfield  {title} {\bibinfo {title} {The renormalization group:
  critical phenomena and the {K}ondo problem},\ }\href@noop {} {\bibfield
  {journal} {\bibinfo  {journal} {Rev. Mod. Phys.}\ }\textbf {\bibinfo {volume}
  {47}},\ \bibinfo {pages} {773} (\bibinfo {year} {1975})}\BibitemShut
  {NoStop}%
\bibitem [{\citenamefont {Bulla}\ \emph {et~al.}(2008)\citenamefont {Bulla},
  \citenamefont {Costi},\ and\ \citenamefont {Pruschke}}]{bulla2008numerical}%
  \BibitemOpen
  \bibfield  {author} {\bibinfo {author} {\bibfnamefont {R.}~\bibnamefont
  {Bulla}}, \bibinfo {author} {\bibfnamefont {T.~A.}\ \bibnamefont {Costi}},\
  and\ \bibinfo {author} {\bibfnamefont {T.}~\bibnamefont {Pruschke}},\
  }\bibfield  {title} {\bibinfo {title} {Numerical renormalization group method
  for quantum impurity systems},\ }\href@noop {} {\bibfield  {journal}
  {\bibinfo  {journal} {Rev. Mod. Phys.}\ }\textbf {\bibinfo {volume} {80}},\
  \bibinfo {pages} {395} (\bibinfo {year} {2008})}\BibitemShut {NoStop}%
\bibitem [{\citenamefont {Jauho}\ \emph {et~al.}(1994)\citenamefont {Jauho},
  \citenamefont {Wingreen},\ and\ \citenamefont
  {Meir}}]{jauho1994timedependent}%
  \BibitemOpen
  \bibfield  {author} {\bibinfo {author} {\bibfnamefont {A.-P.}\ \bibnamefont
  {Jauho}}, \bibinfo {author} {\bibfnamefont {N.~S.}\ \bibnamefont
  {Wingreen}},\ and\ \bibinfo {author} {\bibfnamefont {Y.}~\bibnamefont
  {Meir}},\ }\bibfield  {title} {\bibinfo {title} {Time-dependent transport in
  interacting and noninteracting resonant-tunneling systems},\ }\href@noop {}
  {\bibfield  {journal} {\bibinfo  {journal} {Phys. Rev. B}\ }\textbf {\bibinfo
  {volume} {50}},\ \bibinfo {pages} {5528} (\bibinfo {year}
  {1994})}\BibitemShut {NoStop}%
\bibitem [{\citenamefont {Maciejko}\ \emph {et~al.}(2006)\citenamefont
  {Maciejko}, \citenamefont {Wang},\ and\ \citenamefont
  {Guo}}]{maciejko2006timedependent}%
  \BibitemOpen
  \bibfield  {author} {\bibinfo {author} {\bibfnamefont {J.}~\bibnamefont
  {Maciejko}}, \bibinfo {author} {\bibfnamefont {J.}~\bibnamefont {Wang}},\
  and\ \bibinfo {author} {\bibfnamefont {H.}~\bibnamefont {Guo}},\ }\bibfield
  {title} {\bibinfo {title} {Time-dependent quantum transport far from
  equilibrium: An exact nonlinear response theory},\ }\href@noop {} {\bibfield
  {journal} {\bibinfo  {journal} {Phys. Rev. B}\ }\textbf {\bibinfo {volume}
  {74}},\ \bibinfo {pages} {085324} (\bibinfo {year} {2006})}\BibitemShut
  {NoStop}%
\bibitem [{\citenamefont {Tanimura}(2020)}]{tanimura2020heom}%
  \BibitemOpen
  \bibfield  {author} {\bibinfo {author} {\bibfnamefont {Y.}~\bibnamefont
  {Tanimura}},\ }\bibfield  {title} {\bibinfo {title} {Numerically ``exact''
  approach to open quantum dynamics: The hierarchical equations of motion
  ({HEOM})},\ }\href@noop {} {\bibfield  {journal} {\bibinfo  {journal} {J.
  Chem. Phys.}\ }\textbf {\bibinfo {volume} {153}},\ \bibinfo {pages} {020901}
  (\bibinfo {year} {2020})}\BibitemShut {NoStop}%
\bibitem [{\citenamefont {Croy}\ and\ \citenamefont
  {Saalmann}(2009)}]{croy2009propagation}%
  \BibitemOpen
  \bibfield  {author} {\bibinfo {author} {\bibfnamefont {A.}~\bibnamefont
  {Croy}}\ and\ \bibinfo {author} {\bibfnamefont {U.}~\bibnamefont
  {Saalmann}},\ }\bibfield  {title} {\bibinfo {title} {Propagation scheme for
  nonequilibrium dynamics of electron transport in nanoscale devices},\
  }\href@noop {} {\bibfield  {journal} {\bibinfo  {journal} {Phys. Rev. B}\
  }\textbf {\bibinfo {volume} {80}},\ \bibinfo {pages} {245311} (\bibinfo
  {year} {2009})}\BibitemShut {NoStop}%
\bibitem [{\citenamefont {Kawamura}(2026)}]{kawamura2026timelocal}%
  \BibitemOpen
  \bibfield  {author} {\bibinfo {author} {\bibfnamefont {T.}~\bibnamefont
  {Kawamura}},\ }\href@noop {} {\bibinfo {title} {Time-local nonequilibrium
  {G}reen's function method for real-time dynamics in quantum systems coupled
  to superconducting leads}} (\bibinfo {year} {2026}),\ \Eprint
  {https://arxiv.org/abs/2606.29266} {arXiv:2606.29266} \BibitemShut {NoStop}%
\bibitem [{\citenamefont {Hod}\ and\ \citenamefont
  {Kronik}(2023)}]{hod2023driven}%
  \BibitemOpen
  \bibfield  {author} {\bibinfo {author} {\bibfnamefont {O.}~\bibnamefont
  {Hod}}\ and\ \bibinfo {author} {\bibfnamefont {L.}~\bibnamefont {Kronik}},\
  }\bibfield  {title} {\bibinfo {title} {The driven {L}iouville von {N}eumann
  approach to electron dynamics in open quantum systems},\ }\href@noop {}
  {\bibfield  {journal} {\bibinfo  {journal} {Isr. J. Chem.}\ }\textbf
  {\bibinfo {volume} {63}},\ \bibinfo {pages} {e202300058} (\bibinfo {year}
  {2023})}\BibitemShut {NoStop}%
\bibitem [{\citenamefont {Zelovich}\ \emph {et~al.}(2014)\citenamefont
  {Zelovich}, \citenamefont {Kronik},\ and\ \citenamefont
  {Hod}}]{zelovich2014state}%
  \BibitemOpen
  \bibfield  {author} {\bibinfo {author} {\bibfnamefont {T.}~\bibnamefont
  {Zelovich}}, \bibinfo {author} {\bibfnamefont {L.}~\bibnamefont {Kronik}},\
  and\ \bibinfo {author} {\bibfnamefont {O.}~\bibnamefont {Hod}},\ }\bibfield
  {title} {\bibinfo {title} {State representation approach for atomistic
  time-dependent transport calculations in molecular junctions},\ }\href@noop
  {} {\bibfield  {journal} {\bibinfo  {journal} {J. Chem. Theory Comput.}\
  }\textbf {\bibinfo {volume} {10}},\ \bibinfo {pages} {2927} (\bibinfo {year}
  {2014})}\BibitemShut {NoStop}%
\bibitem [{\citenamefont {Hod}\ \emph {et~al.}(2016)\citenamefont {Hod},
  \citenamefont {Rodr{\'i}guez-Rosario}, \citenamefont {Zelovich},\ and\
  \citenamefont {Frauenheim}}]{hod2016driven}%
  \BibitemOpen
  \bibfield  {author} {\bibinfo {author} {\bibfnamefont {O.}~\bibnamefont
  {Hod}}, \bibinfo {author} {\bibfnamefont {C.~A.}\ \bibnamefont
  {Rodr{\'i}guez-Rosario}}, \bibinfo {author} {\bibfnamefont {T.}~\bibnamefont
  {Zelovich}},\ and\ \bibinfo {author} {\bibfnamefont {T.}~\bibnamefont
  {Frauenheim}},\ }\bibfield  {title} {\bibinfo {title} {Driven {L}iouville von
  {N}eumann equation in {L}indblad form},\ }\href@noop {} {\bibfield  {journal}
  {\bibinfo  {journal} {J. Phys. Chem. A}\ }\textbf {\bibinfo {volume} {120}},\
  \bibinfo {pages} {3278} (\bibinfo {year} {2016})}\BibitemShut {NoStop}%
\bibitem [{\citenamefont {Oz}\ \emph {et~al.}(2023)\citenamefont {Oz},
  \citenamefont {Nitzan}, \citenamefont {Hod},\ and\ \citenamefont
  {Peralta}}]{oz2023electron}%
  \BibitemOpen
  \bibfield  {author} {\bibinfo {author} {\bibfnamefont {A.}~\bibnamefont
  {Oz}}, \bibinfo {author} {\bibfnamefont {A.}~\bibnamefont {Nitzan}}, \bibinfo
  {author} {\bibfnamefont {O.}~\bibnamefont {Hod}},\ and\ \bibinfo {author}
  {\bibfnamefont {J.~E.}\ \bibnamefont {Peralta}},\ }\bibfield  {title}
  {\bibinfo {title} {Electron dynamics in open quantum systems: The driven
  {L}iouville-von {N}eumann methodology within time-dependent density
  functional theory},\ }\href@noop {} {\bibfield  {journal} {\bibinfo
  {journal} {J. Chem. Theory Comput.}\ }\textbf {\bibinfo {volume} {19}},\
  \bibinfo {pages} {7496} (\bibinfo {year} {2023})}\BibitemShut {NoStop}%
\bibitem [{\citenamefont {Zelovich}\ \emph {et~al.}(2017)\citenamefont
  {Zelovich}, \citenamefont {Hansen}, \citenamefont {Liu}, \citenamefont
  {Neaton}, \citenamefont {Kronik},\ and\ \citenamefont
  {Hod}}]{zelovich2017parameterfree}%
  \BibitemOpen
  \bibfield  {author} {\bibinfo {author} {\bibfnamefont {T.}~\bibnamefont
  {Zelovich}}, \bibinfo {author} {\bibfnamefont {T.}~\bibnamefont {Hansen}},
  \bibinfo {author} {\bibfnamefont {Z.-F.}\ \bibnamefont {Liu}}, \bibinfo
  {author} {\bibfnamefont {J.~B.}\ \bibnamefont {Neaton}}, \bibinfo {author}
  {\bibfnamefont {L.}~\bibnamefont {Kronik}},\ and\ \bibinfo {author}
  {\bibfnamefont {O.}~\bibnamefont {Hod}},\ }\bibfield  {title} {\bibinfo
  {title} {Parameter-free driven {L}iouville-von {N}eumann approach for
  time-dependent electronic transport simulations in open quantum systems},\
  }\href@noop {} {\bibfield  {journal} {\bibinfo  {journal} {J. Chem. Phys.}\
  }\textbf {\bibinfo {volume} {146}},\ \bibinfo {pages} {092331} (\bibinfo
  {year} {2017})}\BibitemShut {NoStop}%
\bibitem [{\citenamefont {Caffarel}\ and\ \citenamefont
  {Krauth}(1994)}]{caffarel1994exact}%
  \BibitemOpen
  \bibfield  {author} {\bibinfo {author} {\bibfnamefont {M.}~\bibnamefont
  {Caffarel}}\ and\ \bibinfo {author} {\bibfnamefont {W.}~\bibnamefont
  {Krauth}},\ }\bibfield  {title} {\bibinfo {title} {Exact diagonalization
  approach to correlated fermions in infinite dimensions: {M}ott transition and
  superconductivity},\ }\href@noop {} {\bibfield  {journal} {\bibinfo
  {journal} {Phys. Rev. Lett.}\ }\textbf {\bibinfo {volume} {72}},\ \bibinfo
  {pages} {1545} (\bibinfo {year} {1994})}\BibitemShut {NoStop}%
\bibitem [{\citenamefont {Liebsch}\ and\ \citenamefont
  {Ishida}(2012)}]{liebsch2012temperature}%
  \BibitemOpen
  \bibfield  {author} {\bibinfo {author} {\bibfnamefont {A.}~\bibnamefont
  {Liebsch}}\ and\ \bibinfo {author} {\bibfnamefont {H.}~\bibnamefont
  {Ishida}},\ }\bibfield  {title} {\bibinfo {title} {Temperature and bath size
  in exact diagonalization dynamical mean field theory},\ }\href@noop {}
  {\bibfield  {journal} {\bibinfo  {journal} {J. Phys.: Condens. Matter}\
  }\textbf {\bibinfo {volume} {24}},\ \bibinfo {pages} {053201} (\bibinfo
  {year} {2012})}\BibitemShut {NoStop}%
\bibitem [{\citenamefont {Chin}\ \emph {et~al.}(2010)\citenamefont {Chin},
  \citenamefont {Rivas}, \citenamefont {Huelga},\ and\ \citenamefont
  {Plenio}}]{chin2010exact}%
  \BibitemOpen
  \bibfield  {author} {\bibinfo {author} {\bibfnamefont {A.~W.}\ \bibnamefont
  {Chin}}, \bibinfo {author} {\bibfnamefont {{\'A}.}~\bibnamefont {Rivas}},
  \bibinfo {author} {\bibfnamefont {S.~F.}\ \bibnamefont {Huelga}},\ and\
  \bibinfo {author} {\bibfnamefont {M.~B.}\ \bibnamefont {Plenio}},\ }\bibfield
   {title} {\bibinfo {title} {Exact mapping between system-reservoir quantum
  models and semi-infinite discrete chains using orthogonal polynomials},\
  }\href@noop {} {\bibfield  {journal} {\bibinfo  {journal} {J. Math. Phys.}\
  }\textbf {\bibinfo {volume} {51}},\ \bibinfo {pages} {092109} (\bibinfo
  {year} {2010})}\BibitemShut {NoStop}%
\bibitem [{\citenamefont {Woods}\ \emph {et~al.}(2014)\citenamefont {Woods},
  \citenamefont {Groux}, \citenamefont {Chin}, \citenamefont {Huelga},\ and\
  \citenamefont {Plenio}}]{woods2014mappings}%
  \BibitemOpen
  \bibfield  {author} {\bibinfo {author} {\bibfnamefont {M.~P.}\ \bibnamefont
  {Woods}}, \bibinfo {author} {\bibfnamefont {R.}~\bibnamefont {Groux}},
  \bibinfo {author} {\bibfnamefont {A.~W.}\ \bibnamefont {Chin}}, \bibinfo
  {author} {\bibfnamefont {S.~F.}\ \bibnamefont {Huelga}},\ and\ \bibinfo
  {author} {\bibfnamefont {M.~B.}\ \bibnamefont {Plenio}},\ }\bibfield  {title}
  {\bibinfo {title} {Mappings of open quantum systems onto chain
  representations and {M}arkovian embeddings},\ }\href@noop {} {\bibfield
  {journal} {\bibinfo  {journal} {J. Math. Phys.}\ }\textbf {\bibinfo {volume}
  {55}},\ \bibinfo {pages} {032101} (\bibinfo {year} {2014})}\BibitemShut
  {NoStop}%
\bibitem [{\citenamefont {de~Vega}\ and\ \citenamefont
  {Ba{\~n}uls}(2015)}]{devega2015thermofield}%
  \BibitemOpen
  \bibfield  {author} {\bibinfo {author} {\bibfnamefont {I.}~\bibnamefont
  {de~Vega}}\ and\ \bibinfo {author} {\bibfnamefont {M.-C.}\ \bibnamefont
  {Ba{\~n}uls}},\ }\bibfield  {title} {\bibinfo {title} {Thermofield-based
  chain-mapping approach for open quantum systems},\ }\href@noop {} {\bibfield
  {journal} {\bibinfo  {journal} {Phys. Rev. A}\ }\textbf {\bibinfo {volume}
  {92}},\ \bibinfo {pages} {052116} (\bibinfo {year} {2015})}\BibitemShut
  {NoStop}%
\bibitem [{\citenamefont {de~Vega}\ \emph {et~al.}(2015)\citenamefont
  {de~Vega}, \citenamefont {Schollw{\"o}ck},\ and\ \citenamefont
  {Wolf}}]{devega2015discretize}%
  \BibitemOpen
  \bibfield  {author} {\bibinfo {author} {\bibfnamefont {I.}~\bibnamefont
  {de~Vega}}, \bibinfo {author} {\bibfnamefont {U.}~\bibnamefont
  {Schollw{\"o}ck}},\ and\ \bibinfo {author} {\bibfnamefont {F.~A.}\
  \bibnamefont {Wolf}},\ }\bibfield  {title} {\bibinfo {title} {How to
  discretize a quantum bath for real-time evolution},\ }\href@noop {}
  {\bibfield  {journal} {\bibinfo  {journal} {Phys. Rev. B}\ }\textbf {\bibinfo
  {volume} {92}},\ \bibinfo {pages} {155126} (\bibinfo {year}
  {2015})}\BibitemShut {NoStop}%
\bibitem [{\citenamefont {Arrigoni}\ \emph {et~al.}(2013)\citenamefont
  {Arrigoni}, \citenamefont {Knap},\ and\ \citenamefont {von~der
  Linden}}]{arrigoni2013nonequilibrium}%
  \BibitemOpen
  \bibfield  {author} {\bibinfo {author} {\bibfnamefont {E.}~\bibnamefont
  {Arrigoni}}, \bibinfo {author} {\bibfnamefont {M.}~\bibnamefont {Knap}},\
  and\ \bibinfo {author} {\bibfnamefont {W.}~\bibnamefont {von~der Linden}},\
  }\bibfield  {title} {\bibinfo {title} {Nonequilibrium dynamical mean-field
  theory: An auxiliary quantum master equation approach},\ }\href@noop {}
  {\bibfield  {journal} {\bibinfo  {journal} {Phys. Rev. Lett.}\ }\textbf
  {\bibinfo {volume} {110}},\ \bibinfo {pages} {086403} (\bibinfo {year}
  {2013})}\BibitemShut {NoStop}%
\bibitem [{\citenamefont {Dorda}\ \emph {et~al.}(2014)\citenamefont {Dorda},
  \citenamefont {Nuss}, \citenamefont {von~der Linden},\ and\ \citenamefont
  {Arrigoni}}]{dorda2014auxiliary}%
  \BibitemOpen
  \bibfield  {author} {\bibinfo {author} {\bibfnamefont {A.}~\bibnamefont
  {Dorda}}, \bibinfo {author} {\bibfnamefont {M.}~\bibnamefont {Nuss}},
  \bibinfo {author} {\bibfnamefont {W.}~\bibnamefont {von~der Linden}},\ and\
  \bibinfo {author} {\bibfnamefont {E.}~\bibnamefont {Arrigoni}},\ }\bibfield
  {title} {\bibinfo {title} {Auxiliary master equation approach to
  nonequilibrium correlated impurities},\ }\href@noop {} {\bibfield  {journal}
  {\bibinfo  {journal} {Phys. Rev. B}\ }\textbf {\bibinfo {volume} {89}},\
  \bibinfo {pages} {165105} (\bibinfo {year} {2014})}\BibitemShut {NoStop}%
\bibitem [{\citenamefont {Titvinidze}\ \emph {et~al.}(2015)\citenamefont
  {Titvinidze}, \citenamefont {Dorda}, \citenamefont {von~der Linden},\ and\
  \citenamefont {Arrigoni}}]{titvinidze2015transport}%
  \BibitemOpen
  \bibfield  {author} {\bibinfo {author} {\bibfnamefont {I.}~\bibnamefont
  {Titvinidze}}, \bibinfo {author} {\bibfnamefont {A.}~\bibnamefont {Dorda}},
  \bibinfo {author} {\bibfnamefont {W.}~\bibnamefont {von~der Linden}},\ and\
  \bibinfo {author} {\bibfnamefont {E.}~\bibnamefont {Arrigoni}},\ }\bibfield
  {title} {\bibinfo {title} {Transport through a correlated interface:
  Auxiliary master equation approach},\ }\href@noop {} {\bibfield  {journal}
  {\bibinfo  {journal} {Phys. Rev. B}\ }\textbf {\bibinfo {volume} {92}},\
  \bibinfo {pages} {245125} (\bibinfo {year} {2015})}\BibitemShut {NoStop}%
\bibitem [{\citenamefont {Nakatsukasa}\ \emph {et~al.}(2018)\citenamefont
  {Nakatsukasa}, \citenamefont {S{\`e}te},\ and\ \citenamefont
  {Trefethen}}]{nakatsukasa2018aaa}%
  \BibitemOpen
  \bibfield  {author} {\bibinfo {author} {\bibfnamefont {Y.}~\bibnamefont
  {Nakatsukasa}}, \bibinfo {author} {\bibfnamefont {O.}~\bibnamefont
  {S{\`e}te}},\ and\ \bibinfo {author} {\bibfnamefont {L.~N.}\ \bibnamefont
  {Trefethen}},\ }\bibfield  {title} {\bibinfo {title} {The {AAA} algorithm for
  rational approximation},\ }\href@noop {} {\bibfield  {journal} {\bibinfo
  {journal} {SIAM J. Sci. Comput.}\ }\textbf {\bibinfo {volume} {40}},\
  \bibinfo {pages} {A1494} (\bibinfo {year} {2018})}\BibitemShut {NoStop}%
\bibitem [{\citenamefont {Trefethen}(2019)}]{trefethen2019approximation}%
  \BibitemOpen
  \bibfield  {author} {\bibinfo {author} {\bibfnamefont {L.~N.}\ \bibnamefont
  {Trefethen}},\ }\href@noop {} {\emph {\bibinfo {title} {Approximation Theory
  and Approximation Practice, Extended Edition}}}\ (\bibinfo  {publisher}
  {SIAM},\ \bibinfo {address} {Philadelphia},\ \bibinfo {year}
  {2019})\BibitemShut {NoStop}%
\bibitem [{sup()}]{supplemental}%
  \BibitemOpen
  \href@noop {} {}\bibinfo {note} {See Supplemental Material at [URL will be
  inserted by publisher] for complete parameter prescriptions, convention
  bookkeeping, validation details and numerical consistency checks, and the
  derivations referenced in the text.}\BibitemShut {Stop}%
\bibitem [{\citenamefont {Prosen}(2008)}]{prosen2008third}%
  \BibitemOpen
  \bibfield  {author} {\bibinfo {author} {\bibfnamefont {T.}~\bibnamefont
  {Prosen}},\ }\bibfield  {title} {\bibinfo {title} {Third quantization: a
  general method to solve master equations for quadratic open {F}ermi
  systems},\ }\href@noop {} {\bibfield  {journal} {\bibinfo  {journal} {New J.
  Phys.}\ }\textbf {\bibinfo {volume} {10}},\ \bibinfo {pages} {043026}
  (\bibinfo {year} {2008})}\BibitemShut {NoStop}%
\bibitem [{\citenamefont {Barthel}\ and\ \citenamefont
  {Zhang}(2022)}]{barthel2021solving}%
  \BibitemOpen
  \bibfield  {author} {\bibinfo {author} {\bibfnamefont {T.}~\bibnamefont
  {Barthel}}\ and\ \bibinfo {author} {\bibfnamefont {Y.}~\bibnamefont
  {Zhang}},\ }\bibfield  {title} {\bibinfo {title} {Solving quasi-free and
  quadratic {L}indblad master equations for open fermionic and bosonic
  systems},\ }\href@noop {} {\bibfield  {journal} {\bibinfo  {journal} {J.
  Stat. Mech.}\ }\textbf {\bibinfo {volume} {2022}},\ \bibinfo {pages} {113101}
  (\bibinfo {year} {2022})}\BibitemShut {NoStop}%
\bibitem [{\citenamefont {Levy~Yeyati}\ \emph {et~al.}(1997)\citenamefont
  {Levy~Yeyati}, \citenamefont {Cuevas}, \citenamefont
  {L{\'o}pez-D{\'a}valos},\ and\ \citenamefont
  {Mart{\'i}n-Rodero}}]{levyyeyati1997resonant}%
  \BibitemOpen
  \bibfield  {author} {\bibinfo {author} {\bibfnamefont {A.}~\bibnamefont
  {Levy~Yeyati}}, \bibinfo {author} {\bibfnamefont {J.~C.}\ \bibnamefont
  {Cuevas}}, \bibinfo {author} {\bibfnamefont {A.}~\bibnamefont
  {L{\'o}pez-D{\'a}valos}},\ and\ \bibinfo {author} {\bibfnamefont
  {A.}~\bibnamefont {Mart{\'i}n-Rodero}},\ }\bibfield  {title} {\bibinfo
  {title} {Resonant tunneling through a small quantum dot coupled to
  superconducting leads},\ }\href@noop {} {\bibfield  {journal} {\bibinfo
  {journal} {Phys. Rev. B}\ }\textbf {\bibinfo {volume} {55}},\ \bibinfo
  {pages} {R6137} (\bibinfo {year} {1997})}\BibitemShut {NoStop}%
\bibitem [{\citenamefont {Johansson}\ \emph {et~al.}(1999)\citenamefont
  {Johansson}, \citenamefont {Bratus}, \citenamefont {Shumeiko},\ and\
  \citenamefont {Wendin}}]{johansson1999resonant}%
  \BibitemOpen
  \bibfield  {author} {\bibinfo {author} {\bibfnamefont {G.}~\bibnamefont
  {Johansson}}, \bibinfo {author} {\bibfnamefont {E.~N.}\ \bibnamefont
  {Bratus}}, \bibinfo {author} {\bibfnamefont {V.~S.}\ \bibnamefont
  {Shumeiko}},\ and\ \bibinfo {author} {\bibfnamefont {G.}~\bibnamefont
  {Wendin}},\ }\bibfield  {title} {\bibinfo {title} {Resonant multiple
  {A}ndreev reflections in mesoscopic superconducting junctions},\ }\href@noop
  {} {\bibfield  {journal} {\bibinfo  {journal} {Phys. Rev. B}\ }\textbf
  {\bibinfo {volume} {60}},\ \bibinfo {pages} {1382} (\bibinfo {year}
  {1999})}\BibitemShut {NoStop}%
\bibitem [{\citenamefont {{\v Z}itko}\ and\ \citenamefont
  {Pruschke}(2009)}]{zitko2009energy}%
  \BibitemOpen
  \bibfield  {author} {\bibinfo {author} {\bibfnamefont {R.}~\bibnamefont {{\v
  Z}itko}}\ and\ \bibinfo {author} {\bibfnamefont {T.}~\bibnamefont
  {Pruschke}},\ }\bibfield  {title} {\bibinfo {title} {Energy resolution and
  discretization artifacts in the numerical renormalization group},\
  }\href@noop {} {\bibfield  {journal} {\bibinfo  {journal} {Phys. Rev. B}\
  }\textbf {\bibinfo {volume} {79}},\ \bibinfo {pages} {085106} (\bibinfo
  {year} {2009})}\BibitemShut {NoStop}%
\bibitem [{\citenamefont {Elenewski}\ \emph {et~al.}(2017)\citenamefont
  {Elenewski}, \citenamefont {Gruss},\ and\ \citenamefont
  {Zwolak}}]{elenewski2017markovian}%
  \BibitemOpen
  \bibfield  {author} {\bibinfo {author} {\bibfnamefont {J.~E.}\ \bibnamefont
  {Elenewski}}, \bibinfo {author} {\bibfnamefont {D.}~\bibnamefont {Gruss}},\
  and\ \bibinfo {author} {\bibfnamefont {M.}~\bibnamefont {Zwolak}},\
  }\bibfield  {title} {\bibinfo {title} {Communication: Master equations for
  electron transport: The limits of the {M}arkovian limit},\ }\href@noop {}
  {\bibfield  {journal} {\bibinfo  {journal} {J. Chem. Phys.}\ }\textbf
  {\bibinfo {volume} {147}},\ \bibinfo {pages} {151101} (\bibinfo {year}
  {2017})}\BibitemShut {NoStop}%
\bibitem [{\citenamefont {W{\'o}jtowicz}\ \emph {et~al.}(2021)\citenamefont
  {W{\'o}jtowicz}, \citenamefont {Elenewski}, \citenamefont {Rams},\ and\
  \citenamefont {Zwolak}}]{wojtowicz2021dual}%
  \BibitemOpen
  \bibfield  {author} {\bibinfo {author} {\bibfnamefont {G.}~\bibnamefont
  {W{\'o}jtowicz}}, \bibinfo {author} {\bibfnamefont {J.~E.}\ \bibnamefont
  {Elenewski}}, \bibinfo {author} {\bibfnamefont {M.~M.}\ \bibnamefont
  {Rams}},\ and\ \bibinfo {author} {\bibfnamefont {M.}~\bibnamefont {Zwolak}},\
  }\bibfield  {title} {\bibinfo {title} {Dual current anomalies and quantum
  transport within extended reservoir simulations},\ }\href@noop {} {\bibfield
  {journal} {\bibinfo  {journal} {Phys. Rev. B}\ }\textbf {\bibinfo {volume}
  {104}},\ \bibinfo {pages} {165131} (\bibinfo {year} {2021})}\BibitemShut
  {NoStop}%
\bibitem [{\citenamefont {Gruss}\ \emph {et~al.}(2016)\citenamefont {Gruss},
  \citenamefont {Velizhanin},\ and\ \citenamefont
  {Zwolak}}]{gruss2016landauer}%
  \BibitemOpen
  \bibfield  {author} {\bibinfo {author} {\bibfnamefont {D.}~\bibnamefont
  {Gruss}}, \bibinfo {author} {\bibfnamefont {K.~A.}\ \bibnamefont
  {Velizhanin}},\ and\ \bibinfo {author} {\bibfnamefont {M.}~\bibnamefont
  {Zwolak}},\ }\bibfield  {title} {\bibinfo {title} {Landauer's formula with
  finite-time relaxation: {K}ramers' crossover in electronic transport},\
  }\href@noop {} {\bibfield  {journal} {\bibinfo  {journal} {Sci. Rep.}\
  }\textbf {\bibinfo {volume} {6}},\ \bibinfo {pages} {24514} (\bibinfo {year}
  {2016})}\BibitemShut {NoStop}%
\bibitem [{\citenamefont {Dynes}\ \emph {et~al.}(1978)\citenamefont {Dynes},
  \citenamefont {Narayanamurti},\ and\ \citenamefont
  {Garno}}]{dynes1978direct}%
  \BibitemOpen
  \bibfield  {author} {\bibinfo {author} {\bibfnamefont {R.~C.}\ \bibnamefont
  {Dynes}}, \bibinfo {author} {\bibfnamefont {V.}~\bibnamefont
  {Narayanamurti}},\ and\ \bibinfo {author} {\bibfnamefont {J.~P.}\
  \bibnamefont {Garno}},\ }\bibfield  {title} {\bibinfo {title} {Direct
  measurement of quasiparticle-lifetime broadening in a strong-coupled
  superconductor},\ }\href@noop {} {\bibfield  {journal} {\bibinfo  {journal}
  {Phys. Rev. Lett.}\ }\textbf {\bibinfo {volume} {41}},\ \bibinfo {pages}
  {1509} (\bibinfo {year} {1978})}\BibitemShut {NoStop}%
\bibitem [{\citenamefont {Tien}\ and\ \citenamefont
  {Gordon}(1963)}]{tien1963multiphoton}%
  \BibitemOpen
  \bibfield  {author} {\bibinfo {author} {\bibfnamefont {P.~K.}\ \bibnamefont
  {Tien}}\ and\ \bibinfo {author} {\bibfnamefont {J.~P.}\ \bibnamefont
  {Gordon}},\ }\bibfield  {title} {\bibinfo {title} {Multiphoton process
  observed in the interaction of microwave fields with the tunneling between
  superconductor films},\ }\href@noop {} {\bibfield  {journal} {\bibinfo
  {journal} {Phys. Rev.}\ }\textbf {\bibinfo {volume} {129}},\ \bibinfo {pages}
  {647} (\bibinfo {year} {1963})}\BibitemShut {NoStop}%
\bibitem [{\citenamefont {Bergeret}\ \emph {et~al.}(2010)\citenamefont
  {Bergeret}, \citenamefont {Virtanen}, \citenamefont {Heikkil{\"a}},\ and\
  \citenamefont {Cuevas}}]{bergeret2010theory}%
  \BibitemOpen
  \bibfield  {author} {\bibinfo {author} {\bibfnamefont {F.~S.}\ \bibnamefont
  {Bergeret}}, \bibinfo {author} {\bibfnamefont {P.}~\bibnamefont {Virtanen}},
  \bibinfo {author} {\bibfnamefont {T.~T.}\ \bibnamefont {Heikkil{\"a}}},\ and\
  \bibinfo {author} {\bibfnamefont {J.~C.}\ \bibnamefont {Cuevas}},\ }\bibfield
   {title} {\bibinfo {title} {Theory of microwave-assisted supercurrent in
  quantum point contacts},\ }\href@noop {} {\bibfield  {journal} {\bibinfo
  {journal} {Phys. Rev. Lett.}\ }\textbf {\bibinfo {volume} {105}},\ \bibinfo
  {pages} {117001} (\bibinfo {year} {2010})}\BibitemShut {NoStop}%
\bibitem [{\citenamefont {Raes}\ \emph {et~al.}(2020)\citenamefont {Raes},
  \citenamefont {Tubsrinuan}, \citenamefont {Sreedhar}, \citenamefont {Guala},
  \citenamefont {Panghotra}, \citenamefont {Dausy}, \citenamefont
  {de~Souza~Silva},\ and\ \citenamefont {Van~de Vondel}}]{raes2020fractional}%
  \BibitemOpen
  \bibfield  {author} {\bibinfo {author} {\bibfnamefont {B.}~\bibnamefont
  {Raes}}, \bibinfo {author} {\bibfnamefont {N.}~\bibnamefont {Tubsrinuan}},
  \bibinfo {author} {\bibfnamefont {R.}~\bibnamefont {Sreedhar}}, \bibinfo
  {author} {\bibfnamefont {D.~S.}\ \bibnamefont {Guala}}, \bibinfo {author}
  {\bibfnamefont {R.}~\bibnamefont {Panghotra}}, \bibinfo {author}
  {\bibfnamefont {H.}~\bibnamefont {Dausy}}, \bibinfo {author} {\bibfnamefont
  {C.~C.}\ \bibnamefont {de~Souza~Silva}},\ and\ \bibinfo {author}
  {\bibfnamefont {J.}~\bibnamefont {Van~de Vondel}},\ }\bibfield  {title}
  {\bibinfo {title} {Fractional {S}hapiro steps in resistively shunted
  {J}osephson junctions as a fingerprint of a skewed current-phase
  relationship},\ }\href@noop {} {\bibfield  {journal} {\bibinfo  {journal}
  {Phys. Rev. B}\ }\textbf {\bibinfo {volume} {102}},\ \bibinfo {pages}
  {054507} (\bibinfo {year} {2020})}\BibitemShut {NoStop}%
\bibitem [{\citenamefont {Dartiailh}\ \emph {et~al.}(2021)\citenamefont
  {Dartiailh}, \citenamefont {Cuozzo}, \citenamefont {Elfeky}, \citenamefont
  {Mayer}, \citenamefont {Yuan}, \citenamefont {Wickramasinghe}, \citenamefont
  {Rossi},\ and\ \citenamefont {Shabani}}]{dartiailh2021missing}%
  \BibitemOpen
  \bibfield  {author} {\bibinfo {author} {\bibfnamefont {M.~C.}\ \bibnamefont
  {Dartiailh}}, \bibinfo {author} {\bibfnamefont {J.~J.}\ \bibnamefont
  {Cuozzo}}, \bibinfo {author} {\bibfnamefont {B.~H.}\ \bibnamefont {Elfeky}},
  \bibinfo {author} {\bibfnamefont {W.}~\bibnamefont {Mayer}}, \bibinfo
  {author} {\bibfnamefont {J.}~\bibnamefont {Yuan}}, \bibinfo {author}
  {\bibfnamefont {K.~S.}\ \bibnamefont {Wickramasinghe}}, \bibinfo {author}
  {\bibfnamefont {E.}~\bibnamefont {Rossi}},\ and\ \bibinfo {author}
  {\bibfnamefont {J.}~\bibnamefont {Shabani}},\ }\bibfield  {title} {\bibinfo
  {title} {Missing {S}hapiro steps in topologically trivial {J}osephson
  junction on {InAs} quantum well},\ }\href@noop {} {\bibfield  {journal}
  {\bibinfo  {journal} {Nat. Commun.}\ }\textbf {\bibinfo {volume} {12}},\
  \bibinfo {pages} {78} (\bibinfo {year} {2021})}\BibitemShut {NoStop}%
\end{thebibliography}
\end{document}


\title{Supplemental Material for ``Real-axis least-squares bath
discretization for real-time transport in quantum-dot Josephson
junctions''}

\author{Ruixin Zhou}
\author{Bing Dong}
\author{Yiyan Wang}
\affiliation{Key Laboratory of Artificial Structures and Quantum Control Ministry of Education,Department of Physics and Astronomy, Shanghai Jiaotong University, Shanghai, China}

\date{\today}

\maketitle

\setcounter{section}{-1}

\section{Reading guide: map to the main text}\label{sec:s0}

Table~\ref{tab:guide} maps each section of this Supplemental Material
(SM) to the main-text statement it supports. Sections are cited from
the main text as ``SM Sec.~Sx''.

\begin{table}[b]
\caption{\label{tab:guide}Map between SM sections and the main text.}
\begin{ruledtabular}
\begin{tabular}{lp{0.48\columnwidth}p{0.26\columnwidth}}
SM & Content & Main text \\
\hline
S1 & full parameter prescriptions & Table I; setup paragraphs \\
S2 & $\kappa$ scaling-law factor derivation & Sec.~III.C \\
S3 & $\eta_0$ sensitivity test & Sec.~II.C \\
S4 & edge-beat artifact; density floor & Sec.~II.C (constraint 3) \\
S5 & normal-lead pinning; hybrid bath & Sec.~II.C (constraint 4) \\
S6 & gauge-choice numerical archive & Sec.~II.B \\
S7 & protocol verification and error correction & methods credibility \\
S8 & MAR map and ridge-tracking details & Sec.~III.B \\
S9 & Shapiro protocol and sampling details & Sec.~III.D \\
S10 & static-baseline bookkeeping & Secs.~II.A, II.D, III.A \\
S11 & naming map (paper $\leftrightarrow$ archive) & all \\
S12 & structure-preservation proposition & Sec.~II.B \\
\end{tabular}
\end{ruledtabular}
\end{table}

\section{Full parameter prescriptions (expansion of Table I)}\label{sec:s1}

Each row of main-text Table~I is expanded here into a reproducible
prescription; Table~\ref{tab:prescriptions} collects the same
information in tabular form. Common conventions: $\Delta = 1$, $\hbar
= e = 1$; the spectral-analysis protocol is uniform (Hann window plus
a moving-average high-pass of width about 1.4 target periods,
$2\pi/(0.7\,\omega_{\mathrm{target}})$ scale); frequency verdicts use
the local peak in the $\pm 15\%$ band around the target. All
dynamical runs are at zero temperature: the initial condition is the
exact $T = 0$ coupled equilibrium of $H(0)$, and the DLvN damping
targets are the corresponding $T = 0$ equilibria (decoupled or
coupled reference, as declared per validation). Bias enters through
the hopping Peierls gauge, $\chi_\alpha(t) = \chi_{\alpha,0} +
\int_0^t V_\alpha(t')\,dt'$ (the initial phases $\chi_{\alpha,0}$ set
the static configuration), switched on suddenly at $t = 0$; the assignment
is symmetric, $V_{L,R} = \pm V/2$, for MAR and finite-$\Delta$
Shapiro, and single-sided, $V_L = V$, $V_R = 0$, for the quench and
scaling-test protocols. The ac drive is a cosine, $V_\alpha(t) =
V_{\alpha,\mathrm{dc}} + V_{\alpha,\mathrm{ac}}\cos\Omega t$, i.e.,
$\chi_\alpha(t) = \chi_{\alpha,0} + V_{\alpha,\mathrm{dc}}\,t +
(V_{\alpha,\mathrm{ac}}/\Omega)\sin\Omega t$. Deviation metrics: curve-level
percentages (e.g., the MAR 3.4\%/3.6\%) are relative $\ell_2$ norms
$\lVert I - I_{\mathrm{ref}}\rVert_2 / \lVert I_{\mathrm{ref}}
\rVert_2$ over the stated bias grid; the ``conditional relative''
deviation of the main-text Floquet control points is the mean
per-point $|\Delta I|/|I|$ over the control points with $|I| > 0.05$; locking amplitudes are
$A_m = 2|\hat c_m|$ with $\hat c_m = N_\phi^{-1} \sum_j
I_{\mathrm{dc}}(\phi_{0,j})\, e^{-\mathrm{i} m \phi_{0,j}}$ on the
uniform $N_\phi$-point $\phi_0$ grid.

\emph{Validation 1 (static CPR, main Fig.~3).} Benchmark point
$\varepsilon_d = 0$, $\Gamma = 1$ (per lead $\Gamma_\alpha = 0.5$),
$\Delta = 1$, $D = 20$, $\beta = 50$; the three discretizations at $N_b
= 4$--48 (parameter-passing conventions in Sec.~\ref{sec:s11}). Real-axis main-recipe static supplement
(\texttt{run\_fig5\_realaxis.m}): the main-line fitting prescription
verbatim (window $[-3,3]$, gap weight 30, $\gamma \in [10^{-3},
0.25]$, 3 starts), target $D_{\mathrm{int}} = 20$, $N_{\mathrm{int}} =
100001$, static readout through the same downstream path as the
imaginary-axis variant; 17.8\% at $N_b = 4$ to 3.9\% at $N_b = 48$,
$\sim 1/N_b$ then saturation near 4\%.

\emph{Window-scan Pareto (attribution of the static saturation).}
Whether the 4\% saturation is transport-window truncation was tested
directly: the same prescription refit with core windows $\pm 3 / \pm
5 / \pm 8 / \pm 12 / \pm 20$ (full band) at $N_b = 32$ and $48$,
everything else fixed ($\varepsilon_{\max}$ capped at the band edge
$D$ for the wide windows; the $\pm 3$ baseline refit reproduces the
production numbers, $4.08\%$ vs $4.2\%$). Static CPR error, $N_b = 32$:
$4.08 / 3.67 / 4.85 / 7.38 / 9.06\%$; $N_b = 48$: $3.93 / 3.32 / 3.06 /
4.30 / 6.12\%$. Both mode counts trace a Pareto valley---widening
first helps (the wings enter the core-weight region), then hurts as
the fixed budget is diluted over 40 energy units, with the optimum
shifting outward with the budget ($\pm 5$ at $N_b = 32$, $\pm 8$ at $N_b
= 48$) while the in-window fit residual grows monotonically ($0.18
\to 0.68$ at $N_b = 32$; same unweighted in-window convention as
Sec.~\ref{sec:s3}---the scan's three-start refits are rougher than
the production fit, whose value at the production window is 0.152). Two conclusions, stated in main Sec.~III.A:
the window is a resource-allocation dial, and widening it cannot
substitute for the imaginary-axis variant---even the statics-optimal
window recovers only $\sim$3\%, three orders above the
imaginary-axis $10^{-5}$, because uniform real-axis weighting is
mismatched to the low-frequency-weighted equilibrium observable.
(Caveat: the wide-window fit landscapes are rougher and the
three-start multistart is finite; the scan values are descriptive.)

\emph{Validation 2 (MAR, main Fig.~4).} $\varepsilon_d \in [0, 1.2]$
in steps of 0.03 (41 columns); 117 bias points (the
Gaussian-reference protocol reused verbatim; the color map is thinned
to 86); evolution $dt = 0.05$, $t_{\max} = \min(260,\;
t_{\mathrm{settle}} + N_{\mathrm{per}} \cdot 2\pi/V)$ with
$t_{\mathrm{settle}} = 75$ the discarded initial transient and
$N_{\mathrm{per}} = 6$ the number of averaging blocks of length
$2\pi/V$ (each an integer number---two---of Josephson current periods
$\pi/V$; $\omega_J = 2V$) in the
steady-state averaging window; NESS period averaging; errors computed
over the common window $eV \in [0.28, 2.05]$ (the full curve), while
main Figs.~4(a) and~(b) display the subgap staircase $eV \le
1.2\Delta$ for clarity. The Floquet comparison has two layers: the
$\varepsilon_d = 0$ column is a full comparison against Floquet [the
current staircase and differential-conductance curve in main Figs.~4(b)
and~4(a), respectively], which stratifies by bias---least squares follows Floquet to
$7\%$ over the subgap staircase ($eV \le 1.2\Delta$, closer than the
Gauss-110 reference at $10\%$), whereas the full-curve deviation
($\approx 15\%$) is set by the high-bias single-particle region toward
$eV = 2\Delta$, where finite-$\gamma$ broadening smooths the sharp
threshold for every bath alike---and in the two-dimensional scan every 10th
$\varepsilon_d$ column carries one control point at the fixed section
$V = 0.69$, the bias-grid point nearest 0.7 (five points; deviations
reported in the main text). Baths: LS-32 (gap-clean,
prescription at the end of this section) versus the Gauss-110
reference ($\gamma = 0.045$).

\emph{Validation 3 (quench, main Figs.~6--7).} $\varepsilon_d = 0$,
$\Gamma = 0.8$, $\Delta = 1$, $D = 4$, $V = 0.8$; turn-on $0 \to 70$,
spectral window $[35, 70]$; turn-off: pre-drive $t_{\mathrm{pre}} =
60$ ($\gamma_{\mathrm{pre}} = 0.12$), quench phase $\phi = 0.6\pi$,
relaxation $t_{\mathrm{post}} = 200$ ($\gamma_{\mathrm{post}} =
0.05$), spectral window $[80, 200]$; $2|\varepsilon_b| = 0.484$.
Envelope family: Gauss-96 ($\gamma = 0.10$ and $0.05$) and Gauss-192
($\gamma = 0.025$; the smaller damping requires the denser
quadrature comb), least-squares-32
native and $\times 0.5$, densely mapped $N_b = 384$ ($\gamma =
0.006$) and 768 ($\gamma = 1.5\times 10^{-3}$, the
$\kappa$-minimal configuration). The $\kappa$-minimal point has its
own long-trace protocol: total
trace $t = 2000$, envelope from 17 sliding lock-in windows of length
250 anchored at $t_0 = 100$--$1700$, over which the amplitude decays
by 60\%; a linear fit of the log envelope gives
$\kappa_{\mathrm{dyn}} = 5.73(10)(26)\times 10^{-4}$ [first bracket:
OLS fit interval, quoted as descriptive since the sliding windows
overlap; second bracket: window-split systematic, half the
front/back-half difference $6.19/5.66\times 10^{-4}$; log-envelope
residual rms $1.0\times 10^{-2}$], versus $\kappa_{\mathrm{spec}} =
5.78\times 10^{-4}$ (0.8\%, an estimator with no time stepping at
all). \emph{Recurrence audit} (both dense configurations): refitting
$\kappa$ on the window-center subranges $[100,600] / [600,1200] /
[1200,1900]$ gives $5.64/5.64/5.53\times 10^{-4}$ (2\% spread; $N_b =
384$: 3.4\% spread)---no systematic late-time drift; the bath memory
kernel $K(t) = \sum_j |V_j|^2 e^{-\mathrm{i}E_j t - \gamma t/2}$
rebuilt from the mode table decays with no revival over $t \le 2000$
(max $|K(t)|/|K(0)| = 0.14$ for $t \ge 200$ and $0.09$ for $t \ge
500$, a smooth gap-edge tail rather than a recurrence spike; $N_b =
384$: bound $0.10$); the minimal gap-edge level spacings put pairwise
beat periods at $452$ and above, and the sliding-window drift test
above is the direct check that none of them contaminates the fit.
$\eta \to 0$ discrimination: $V = 0.48$
(resonant) and 0.62 (detuned), \emph{sustained} drive with lock-in
window $t \in [60, 160]$, 11 values of $\kappa \in [3.5\times
10^{-3}, 2.3\times 10^{-2}]$ (window-end $\kappa t = 0.56$--$3.7$);
the window-folded secular expectation drawn in main
Fig.~7(c) is $A(\kappa) \propto \kappa^{-1}\{1 - [e^{-60\kappa} -
e^{-160\kappa}]/(100\kappa)\}$, normalized at the largest $\kappa$
(detuning $\delta = V - 2|\varepsilon_b| = -0.0048$ folded into the
fitted phasor model below shifts its log-log slope from $-0.45$ to
$-0.44$). \emph{Multi-window and phasor audit} (upgrading the
amplitude analysis to the complex lock-in phasor $Z$): the windows
$[40,120] / [60,160] / [100,160]$ give measured resonant slopes
$-1.22 \pm 0.10$ / $+0.26 \pm 0.06$ / $-0.42 \pm 0.22$ (descriptive
OLS) against the corresponding window-folded secular slopes
$-0.35 / -0.44 / -0.49$, with detuned-control slopes $-0.81 / -0.25 /
-0.57$: the early window is transient-dominated (both biases steeply
negative), and in the late window the resonant and detuned biases
move together---a $\kappa$-dependent background that is not specific
to the resonance. A joint complex fit $Z(\kappa) = Z_0 +
B\,F_{\mathrm{win}}(\kappa,\delta)$ is ill-conditioned over the
accessible $\kappa$ range ($Z_0$ and $F_{\mathrm{win}}$ nearly
collinear; relative fit residual 0.09--0.32 across windows), so the
protocol bounds rather than isolates the secular amplitude $B$; the
production window's opposite-sign slope carries the quoted conclusion,
hedged accordingly in the main text. The drive amplitude is not a
separate knob in this protocol (the dc bias is the drive), so
linear-response order was not independently varied---declared.

\emph{Time step and positivity.} All production runs use RK4 with
$dt = 0.05$. Halving and quartering the step shifts NESS observables
by $\le 3.2\times 10^{-4}$ (relative; five MAR bias points,
LS-32) and a Shapiro locked point by $3\times 10^{-8}$---far below
every error quoted in the paper. Positivity audit (the coupled
reference is not guaranteed Lindblad-form): along a representative
coupled-reference relaxation (phase quench $0 \to 0.6\pi$, LS-32
per-mode $\gamma$ and Gauss-96 $\gamma = 0.05$), the eigenvalues of
$C$ make a particle-hole-symmetric excursion outside $[0,1]$ of at
most $1.3\times 10^{-3}$ (LS-32) / $4.4\times 10^{-4}$ (Gauss-96)
immediately after the quench, decaying monotonically to $3\times
10^{-8}$ by $t = 200$; decoupled-reference driven runs stay within
$10^{-8}$ (integrator floor) throughout.

\emph{Preparation-length audit.} The production turn-off inherits
$t_{\mathrm{pre}} = 60$ verbatim from
Ref.~\cite{cheng2024quasiparticle}, although the preparation study of
Sec.~\ref{sec:s7} shows St\"uckelberg-type fringes that die out only
around $t_{\mathrm{pre}} = 150$. Rerunning the production Gauss-96
turn-off with $t_{\mathrm{pre}} = 150$ (quench phase pinned at $\phi
= 0.6\pi$ for both preparations) leaves all three assessed
observables unchanged: the interpolated peak frequency is identical
at the $10^{-6}$ level (both $-0.008\%$ from $2|\varepsilon_b|$ under
this audit's occupation-channel estimator), the
lock-in amplitude moves by $-0.13\%$, and the envelope decay rate by
$0.01\%$ ($\kappa_{\mathrm{env}} = 1.947\times 10^{-2}$ vs the
scaling-law $0.386\,\gamma = 1.93\times 10^{-2}$)---the fringes live
in the prepared state, not in the frequency/amplitude/$\kappa$
estimators. (The same
protocol on a denser Gauss-110 configuration gives the same verdict:
frequency shift at the $10^{-5}$ level, amplitude $-0.09\%$, decay
rate $0.02\%$.)

\emph{Audit coverage extension.} The LS-64
density-floor coupled-reference turn-off (the production 0.36\%
variant) rerun at $dt = 0.05$ vs $0.025$ shifts the peak frequency by
$1.4\times 10^{-8}$, the lock-in amplitude by $4\times 10^{-5}$
(relative), and $\kappa_{\mathrm{env}}$ by $5\times 10^{-5}$
(relative); its positivity excursion is $\le 1.16\times 10^{-3}$
(peaked at $t \approx 2$, decaying), matching the LS-32 audit above.
The $\kappa$-minimal dense-768 configuration is audited separately:
its post-quench stage is \emph{exact eigen-propagation} (no time
stepping enters the decay-rate estimate; $dt$ affects only the
prepared state, and the $dt$-free spectral estimator agrees at
0.8\%), and its positivity, sampled across the full $t \le 2000$
span of the coupled-reference propagation, stays within $2.6\times
10^{-6}$ of $[0, 1]$---at this small $\gamma$ the map is
positivity-preserving to near machine relevance.

\emph{Validation 4 (Shapiro, main Figs.~8--10).} Effective junction
$\Gamma_S = 1$, $\Gamma_N = 0.2$, $\Omega = 0.5$, dense 16-$\phi_0$
sweeps; Arnold tongues on 24 rational columns (denominators $q \in
\{1,2,3,4,6,8\}$) $\times$ $V_{\mathrm{ac}} = 0.10:0.10:1.5$, with
the fast Floquet kernel (row-wise Green-function solves; validation
gate $2.5\times 10^{-14}$; $N_E = 1401/801$; dimension cap 700, cells
beyond the cap left blank and marked gray). Finite $\Delta$:
$\varepsilon_d = 0$, $\Gamma = 0.8$, $\Omega = 1$, $V_{\mathrm{ac}} =
1$; production run on the LS-32 production bath (per-lead $w_j/2$
split, per-mode $\gamma_j$, exactly the MAR pipeline): 28-point
backbone at a single $\phi_0$ ($V = 0.10$--$1.80$, step 0.05, minus
the rational set) plus every $m \le 3$ rational resonance at
8~$\phi_0$; per-point settledness monitored (late-window vs
full-window average, drift $\le 5\times 10^{-4}$). A single-$\phi_0$
backbone point cannot by itself bound a locking amplitude, so the
off-grid claim is closed by dedicated 8-$\phi_0$ sweeps at $V = 0.45,
0.55, 0.95, 1.05$ (bracketing the two strongest integer steps): $A_1
\le 1.3\times 10^{-6}$, $A_2 \le 7.9\times 10^{-7}$, $A_3 \le
3.4\times 10^{-7}$ at all four biases---four orders of magnitude
below the on-grid $A_1 \approx 2\times 10^{-2}$ and three and a half
below the $3\times 10^{-3}$ display threshold. Archived
cross-checks: Gaussian-reference recipe ($N_b = 110$, $\gamma =
0.045$) backbone points and Floquet resonance bands
(Sec.~\ref{sec:s9}).

\emph{Superconducting-lead least-squares prescription (main line).}
Target = finite-bandwidth numerical integral ($D_{\mathrm{int}} = D$,
$N_{\mathrm{int}} = 40001$, $\eta_0 = 10^{-3}$); window $[-3, 3]$;
grid $N_\omega = 2400$, soft margin 0.35 window widths at weight 0.1;
gap weight 30 ($|\omega| < 0.98\Delta$); $\gamma \in [10^{-3},
0.25]$; multistart with 3 starts (uniform $\xi$ grid plus jitter),
initial weights from one NNLS step; larger $N_b$ warm-started from
the smaller-bath optimum. Density-floor variant (LS-64): gap-edge
equal-$\xi$ dense core $h = 0.05$, $\gamma = 0.015$, smooth partition
$s(\xi) = [1 - \tanh((|\xi| - 0.85)/0.15)]/2$, core weights
$(\Gamma/\pi)h\,s$, wings fitted to the $\rho(1 - s)$ residual with
position box $|\varepsilon| \le 10$. \emph{Far Re-tail anchors
(disclosure)}: the wing optimizer parks two modes at the box bound
$\varepsilon = \pm 10$ ($w \approx 3.6$, $\gamma$ at the lower bound
$10^{-3}$; rows 49 and 64 of Table~\ref{tab:bath64}) and leaves the eight
intermediate wing slots at numerically zero weight ($\le 4\times
10^{-12}$). These anchors lie outside the physical band and act
purely as a rational-approximation device for the smooth
Kramers--Kronig tail of $\mathrm{Re}\,\Sigma$ from outside the fit
window: their in-window spectral density is $\le 1.5\times 10^{-5}$
(relative $10^{-5}$ of the comb maximum), their in-window
$\mathrm{Re}\,\Sigma$ contribution is $\pm 0.23$ (5\% of the full
range), and removing them shifts the static CPR by 7.3\% (the Re-tail
weight they carry). Dynamically they are auxiliary far-tail poles
that encode this low-energy dispersive contribution; they do not
represent physical states of the finite band ($D = 4$), and their
weights break the global high-frequency sum rule by construction:
$\Sigma(\omega \to \infty) \sim (\sum_j w_j)/\omega$ gives $\sum_j
w_j \approx 7.24$ from the two anchors alone against the physical
moment $2D\Gamma/\pi \approx 2.04$. The LS-64 bath is therefore a
windowed rational approximation of $\Sigma(\omega)$, valid inside
the fit window, not a global discretization of the physical spectral
density---which is exactly how it is used (long-time in-gap
spectra). An anchors-removed control run bounds this construction
dynamically: the production turn-off quench protocol was rerun
verbatim on the 62-mode bath with the two anchors deleted, with three
findings. (i) The static role is confirmed: removing the anchors
shifts the bath's ABS assignment from $2|\varepsilon_b| = 0.485$ to
$0.453$ (the dispersive weight they carry), and each bath's
post-quench spectral line lands on its \emph{own} static assignment
to within $0.14$ of the window resolution in both channels. The full
bath's assignment is moreover anchored to the continuum, not merely
internally consistent: solving the ABS pole of the
continuum-self-energy junction at the same working point
[$\det\{\omega\openone - \Sigma_L(\omega) - \Sigma_R(\omega)\} = 0$
with per-lead finite-band ($D = 4$) numerical-integral self-energies
at phases $\pm\phi/2$, $\phi = 0.6\pi$, dot level $\varepsilon_d =
0$; the unique in-gap root located on $(0, \Delta)$] gives
$2|\varepsilon_b|_{\mathrm{cont}} = 0.4839$ (stable under regulator
$10^{-6}$--$10^{-3}$ and integration-resolution doubling), within
$0.23\%$ of the full bath and $6\%$ from the control---the anchors
supply precisely the dispersive weight the continuum junction
requires. (ii) The
anchors are nevertheless dynamically excited by the protocol: the
full bath's bond-current spectrum carries a line cluster at $\omega
\approx 10.0$--$10.8$---the anchor quasiparticle energy scale
$\sqrt{\varepsilon^2 + \Delta^2} \approx 10.05$---reaching
$2.8\times$ the ABS-line amplitude ($0.13\times$ in the occupation
channel), absent from the control ($\le 0.01$); the ringing decays at
$\approx 0.023$, the anchor--bath coherence rate set by the partner
mode dampings. (iii) Inside the in-gap assessment band the two baths
agree at the comb-background level, and the lock-in decay rate of the
ABS coherence is bath-independent to 3\% (channel ratios
1.03/0.98). The windowed-approximation statement is thus both
confirmed and bounded by measurement: the bath is valid for the
in-gap spectral observables it is used for (main Fig.~6(c,d): line
within $0.36\%$ of the reference assignment), while spectral content
at the anchor scale $\omega \approx 10$ is unphysical by construction
and measurably populated---observables with support there must not
be read from this bath. A sum-rule-preserving density-floor refit is
a natural next iteration of the prescription. The
main-line LS-32 bath is fitted inside the window $[-3, 3]$ and has no
such anchors.
$\Gamma$-convention note (to prevent misreading): in the
superconducting-lead main-line prescription $\Gamma$ always denotes
the main-text Sec.~II.C total embedding self-energy convention
($\Gamma \equiv \Gamma_L + \Gamma_R$); the fit is performed once and
the weights are split evenly between the two leads (per lead
$|V_{\alpha j}|^2 = w_j/2$). In the normal-lead prescription
(Sec.~\ref{sec:s5}) $\Gamma$ denotes that normal lead's own
$\Gamma_\alpha$. The resulting mode tables $\{\varepsilon_j, w_j,
\gamma_j\}$ of the production baths are listed in
Tables~\ref{tab:bath32}--\ref{tab:bath77}.

\emph{Optimizer and implementation conditions (reproducibility;
verified against \texttt{fb\_fit\_bath\_bdg}).} Trust-region
reflective \texttt{lsqnonlin}, \texttt{FunctionTolerance} =
\texttt{StepTolerance} = $10^{-12}$, \texttt{MaxIterations} = 3000,
\texttt{MaxFunctionEvaluations} = $6\times 10^4$; fitting grid
$N_\omega = 2400$ (window plus 0.35-window soft margin); multistart
with 3 starts (first start = deterministic uniform $\xi$ grid, the
rest jittered); initial weights from one NNLS step; larger $N_b$
warm-started from the smaller-bath optimum. Machine context for all
wall-clock numbers: a single workstation with an Intel Core Ultra 9
275HX (24 cores), 192~GB memory, one MATLAB process (default
multithreaded BLAS), RK4 step $dt = 0.05$.

\begin{table*}
\caption{\label{tab:prescriptions}Complete numerical prescriptions of
the four validations: junction parameters, drive protocol, bath
configurations, and benchmarks. Main-text Table~I is the compressed
version; the surrounding text of Sec.~\ref{sec:s1} carries the same
content with additional commentary.}
\begin{ruledtabular}
\begin{tabular}{p{0.10\textwidth}p{0.22\textwidth}p{0.28\textwidth}p{0.18\textwidth}p{0.14\textwidth}}
Validation & Junction & Drive / protocol & Baths & Benchmark \\
\hline
1 static CPR &
$\varepsilon_d{=}0$, $\Gamma{=}1$ ($\Gamma_\alpha{=}0.5$),
$\Delta{=}1$, $D{=}20$, $\beta{=}50$ &
equilibrium CPR; real-axis static run: $D_{\mathrm{int}}{=}20$,
$N_{\mathrm{int}}{=}100001$ &
LS / Gauss / uniform, $N_b{=}4$--48; imag-axis variant &
free-energy CPR (2 impl.); Keldysh \\
2 MAR &
$\varepsilon_d\in[0,1.2]$ step 0.03 (41 col.), $\Gamma{=}0.8$,
$\Delta{=}1$, $D{=}4$ &
$V_{L,R}{=}{\pm}V/2$; 117 bias pts; $dt{=}0.05$;
$t_{\max}{=}\min(260, 75 + 6\cdot 2\pi/V)$;
common window $eV\in[0.28,2.05]$ &
LS-32 (gap-clean); Gauss-110 ($\gamma{=}0.045$) &
Gauss-110; Floquet ($\varepsilon_d{=}0$ + 5 control pts) \\
3 quench &
$\varepsilon_d{=}0$, $\Gamma{=}0.8$, $\Delta{=}1$, $D{=}4$, $V{=}0.8$ &
$V_L{=}V$, $V_R{=}0$; turn-on $0\to70$ (window $[35,70]$); turn-off
$t_{\mathrm{pre}}{=}60$,
$\phi{=}0.6\pi$, $t_{\mathrm{post}}{=}200$ (window $[80,200]$;
dispersion protocol $t_{\mathrm{post}}{=}300$, window $[80,300]$) &
Gauss $\gamma{=}0.10/0.05/0.025$; LS-32/64; dense mapped
$N_b{=}384/768$ &
spectral $\kappa_{\mathrm{spec}}$; protocol of
Ref.~\cite{cheng2024quasiparticle} \\
4 Shapiro &
eff.: $\Gamma_S{=}1$, $\Gamma_N{=}0.2$, $\Omega{=}0.5$;
finite $\Delta$: $\varepsilon_d{=}0$, $\Gamma{=}0.8$, $\Delta{=}1$,
$D{=}4$, $\Omega{=}1$, $V_{\mathrm{ac}}{=}1$ &
dc ${\pm}V/2$, ac ${\pm}V_{\mathrm{ac}}/2$ cosine (initial superconducting
phase difference $\phi_0 = 2(\chi_{L,0} - \chi_{R,0})$, symmetric
$\chi_{L,0} = {+}\phi_0/4$, $\chi_{R,0} = {-}\phi_0/4$, swept); 16-$\phi_0$
sweeps; tongues 24 rational col.\ $\times$
$V_{\mathrm{ac}}{=}0.1{:}0.1{:}1.5$; 8-$\phi_0$ subharmonic rescan &
eff.: LS-77 hybrid (Sec.~\ref{sec:s5});
finite $\Delta$: LS-32 (MAR production bath); Gauss-110 recipe
cross-check &
Floquet fast kernel (gate $2.5{\times}10^{-14}$); kinematic grid \\
\end{tabular}
\end{ruledtabular}
\end{table*}

\section{Factor bookkeeping of the \texorpdfstring{$\kappa$}{kappa}
scaling law}\label{sec:s2}

Linearized mode analysis of the DLvN equation $dC/dt = -i[H, C] -
\tfrac{1}{2}\{\Lambda, C - C_{\mathrm{ref}}\}$: the amplitude of a
single-particle eigenmode $|i\rangle$ decays at $\Gamma_i \simeq
\tfrac{1}{2}\sum_j |\langle j|i\rangle|^2 \gamma_j$ (first-order
perturbation theory in the damping; the $\tfrac{1}{2}$ comes from the
anticommutator). The observed $2|\varepsilon_b|$ oscillation is a
\emph{bilinear} coherence of the correlation matrix, $C_{12}(t)
\propto e^{-i(E_1 - E_2)t}\, e^{-(\Gamma_1 + \Gamma_2)t}$.
Particle-hole symmetry gives $|\langle j|\psi_1\rangle|^2 = |\langle
j|\psi_2\rangle|^2$, hence
\begin{equation}
\kappa = \Gamma_1 + \Gamma_2
\simeq 2 \cdot \tfrac{1}{2} \sum_j |\langle j|\psi_{\mathrm{ABS}}\rangle|^2
\gamma_j
= w_{\mathrm{bath}}\, \gamma_{\mathrm{eff}}.
\label{eq:kappafactor}
\end{equation}
The $\tfrac{1}{2}$ (single-particle amplitude) and the 2 (two branches
of the bilinear coherence) cancel exactly at this leading order---the
absence of a $\tfrac{1}{2}$ in $\kappa_{\mathrm{spec}}$ is a derived
result, not a convention. For the six configurations whose envelope fit windows are
free of edge-beat takeover (the a priori criterion of
Sec.~\ref{sec:s4}), the measured
$\kappa_{\mathrm{dyn}}/\kappa_{\mathrm{spec}} = 0.986$--$1.003$
across nearly two decades in global $\gamma$ and a broader range in
mode-resolved dampings---the test of this first-order picture. The
gap-clean LS-32 native bath reads 0.783 because its fit window
extends into the beat-dominated late-time regime: its envelope slope
matches $\kappa_{\mathrm{spec}}$ to 4\% over $t \in [80, 210]$ and
collapses toward the beat rate at late times (the measured anatomy is
in Sec.~\ref{sec:s4}), so 0.783 is a mixed slope, not an in-window
violation of the law; the point is retained as an estimator-boundary
diagnostic rather than included in the scaling-law test.
Abscissa bookkeeping: uniform baths may use the bare
$\gamma_{\mathrm{post}}$, but per-mode baths must use the ABS-weighted
average $\gamma_{\mathrm{eff}} = \kappa_{\mathrm{spec}} /
w_{\mathrm{bath}}$---other proxies (e.g., the median mode damping)
deviate systematically from the line, a bookkeeping artifact rather
than a physical deviation.

\section{\texorpdfstring{$\eta_0$}{eta0} sensitivity: the
target-evaluation regulator is not
a hidden Dynes broadening}\label{sec:s3}

This section expands the $\eta_0$ note of main-text Sec.~II.C (the
probe scripts are preserved with the data archive).

\begin{figure}[tb]
\includegraphics[width=\columnwidth]{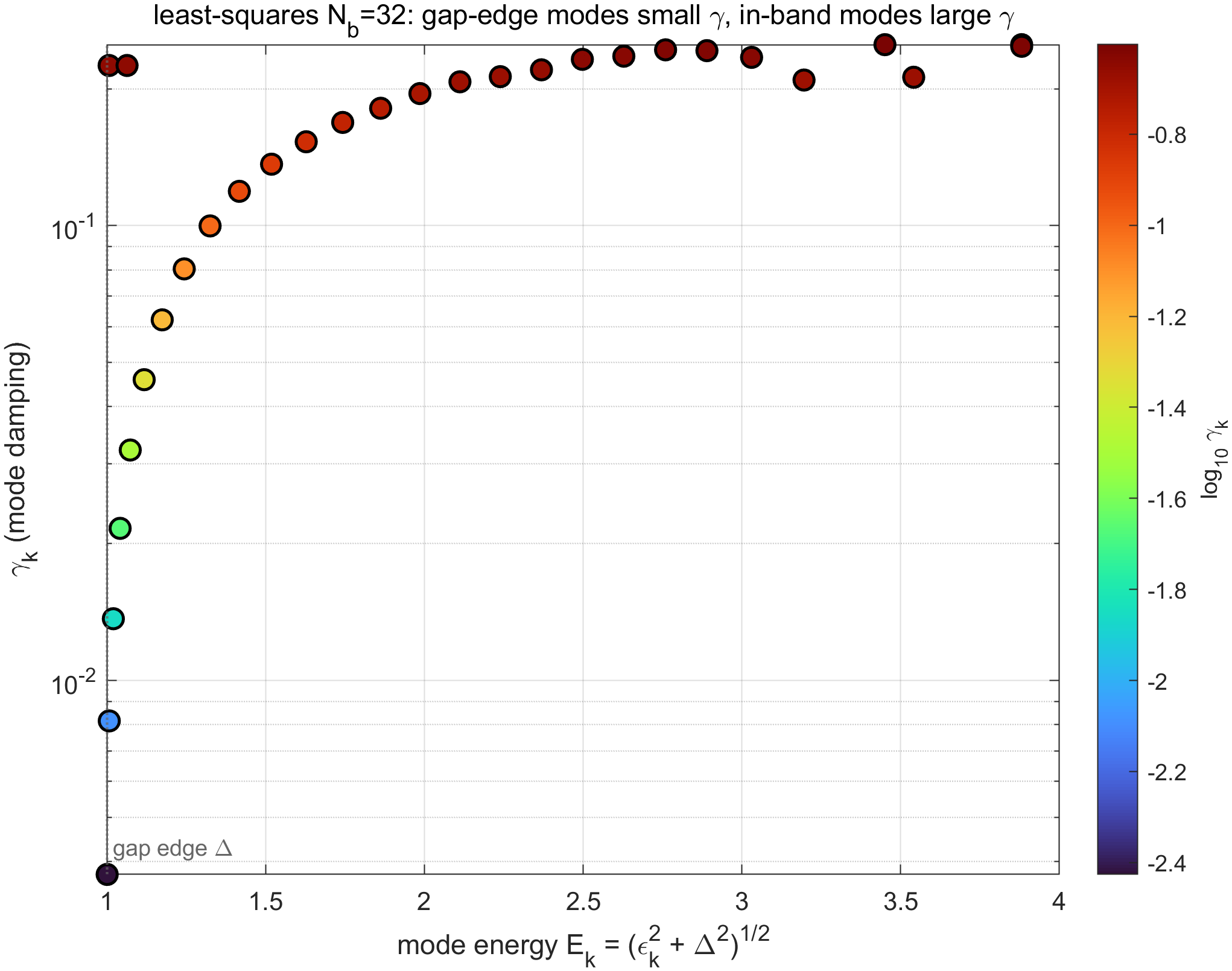}
\caption{Per-mode damping $\gamma_j$ versus mode energy $E_j$ for the
least-squares $N_b = 32$ bath: gap-edge modes take small $\gamma$,
in-band modes large $\gamma$.}
\label{fig:sm_gamma}
\end{figure}

\begin{figure}[tb]
\includegraphics[width=\columnwidth]{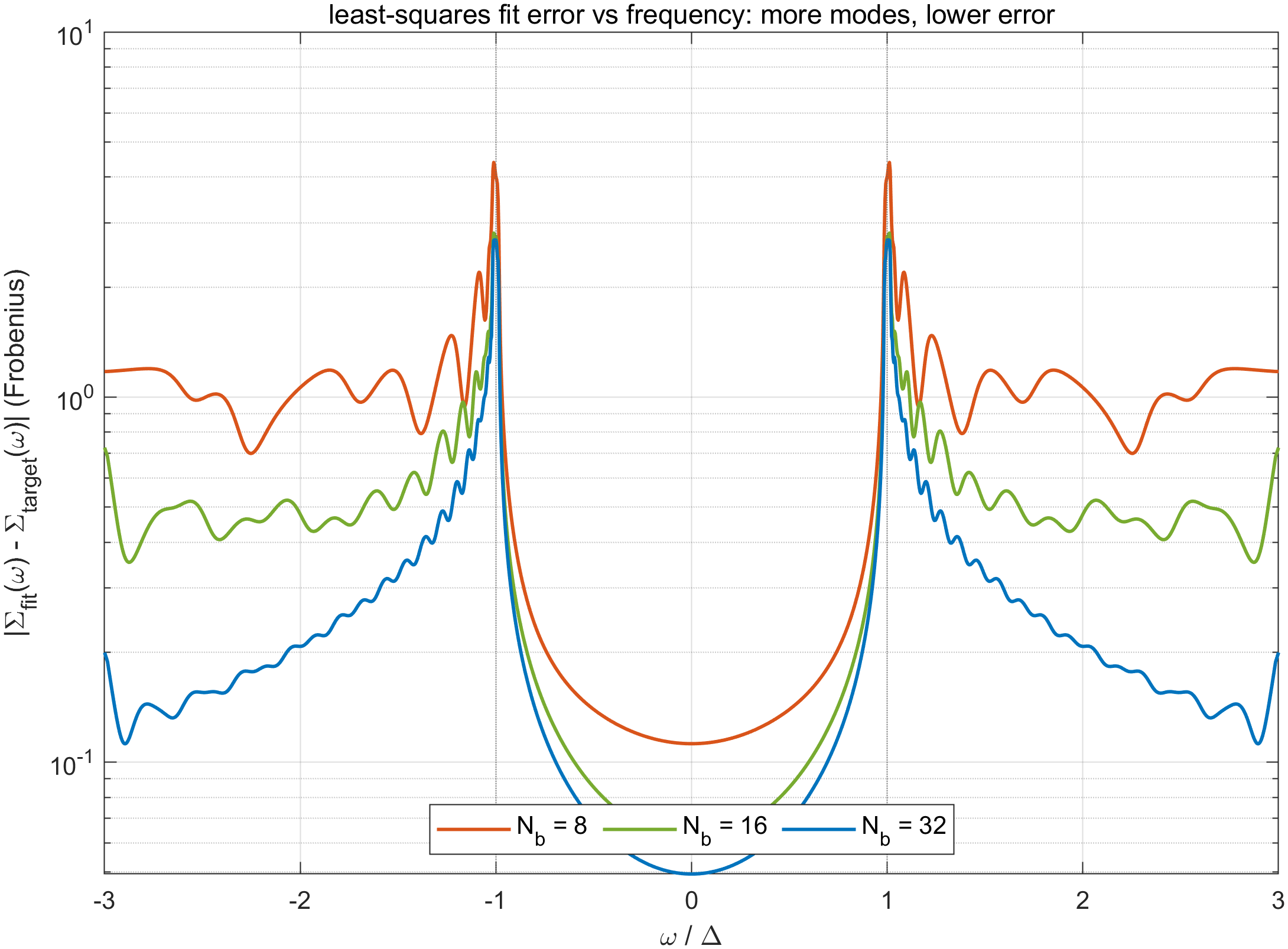}
\caption{Self-energy fit error versus frequency for least-squares baths
of increasing size ($N_b = 8, 16, 32$). The pointwise error
$|\Sigma_{\mathrm{fit}}(\omega) - \Sigma_{\mathrm{target}}(\omega)|$
(Frobenius over the Nambu components) drops across the spectrum as the
mode count grows; the residual peaks at the gap edges $\omega =
\pm\Delta$, where the finite-bandwidth BCS self-energy is
near-singular.}
\label{fig:sm_conv}
\end{figure}

\emph{Mode-resolved damping and fit convergence.} Before the
sensitivity test, we record two properties of the production
least-squares fit. Figure~\ref{fig:sm_gamma} shows its per-mode damping
structure $\gamma_j$ versus mode energy $E_j$: gap-edge modes take
small dampings (protecting long-lived subgap physics), in-band modes
take large ones (doing the absorbing)---the design principle behind the
gap-clean weighting. Figure~\ref{fig:sm_conv} shows the fit
convergence: the pointwise self-energy error $|\Sigma_{\mathrm{fit}} -
\Sigma_{\mathrm{target}}|$ decreases across the spectrum as the mode
count grows ($N_b = 8, 16, 32$), the residual peaking at the gap edges
$\omega = \pm\Delta$ where the finite-bandwidth self-energy is
near-singular. These underlie the ansatz residual floor quantified
below.

\emph{Test design.} The main-line protocol ($N_b = 32$, window $[-3,
3]$, gap weight 30, $\gamma_{\max} = 0.25$, 3 starts) is fitted
separately against targets evaluated at $\eta_0 = 10^{-3}$
($N_{\mathrm{int}} = 40001$) and $\eta_0 = 10^{-4}$ ($N_{\mathrm{int}}
= 400001$; the integration resolution is refined in step to keep the
poles resolved). The $\eta_0 = 10^{-3}$ side reproduced the
production bath digit by digit ($\mathrm{rel}\,L_2 = 0.15246$---the
fitter's \emph{unweighted} in-window relative $L_2$---gap
leakage 0.1486), which doubles as a protocol regression test.

\emph{Results} (all in the \emph{weighted} objective norm of the
main text, evaluated on the full fitting grid including margins and
gap weight, and quoted as dimensionless residuals---a different
convention from the in-window rel~$L_2$ above, which is why the
ansatz floor below sits beneath the production in-window value
0.152): the two targets differ by 0.059,
below the $N_b = 32$ ansatz residual floor of 0.10--0.12;
cross-residuals (bath A against target B, 0.1232, versus bath B
against its own target, 0.1197; and conversely) differ by at most
0.024; deep-gap leakage ($|\omega| < 0.5$) is 0.0470 versus 0.0487.
The comb parameters of the two baths differ by 19.6\%---but each fits
the other's target equally well: this is the \emph{fitting null
space} (at the 10--12\% residual level, $N_b = 32$ admits a large
family of equivalent baths; compare the $N_b = 48$ warm/cold-start
refit archive: different local minima, equal fit quality, MAR
verdicts unchanged digit by digit). Parameter-level nonuniqueness
thus does not propagate to the transport observables---the stability
test relevant for the physics.

\emph{Regulator--resolution coupling.} The $\eta_0$ sensitivity test
requires the integration resolution to be rescaled together with
$\eta_0$: at $\eta_0 = 10^{-4}$ the pole width is smaller than a
$2\times 10^{-4}$ integration spacing, which would render the target
curve numerical noise. Whenever a regulator parameter is changed, the
accompanying numerical resolution must be rescaled with it.

\emph{Gap-weight sensitivity.} The other structural hyperparameter of
the objective, the in-gap weight (production value 30), was varied
$\times 3$ down and up with everything else fixed ($N_b = 32$, window
$[-3,3]$, three starts; five-bias-point MAR spot check against the
Gauss-110 reference values, production spot value 2.9\%). Weight 10:
the in-gap Im leakage rises from 0.15 to 0.21 and the spot check
degrades to 7.5\%---the ohmic-shunt error mode that motivated the
gap-clean constraint in the first place reappears, i.e., the
parameter fails in exactly the direction and for exactly the reason
the design states. Weight 100: leakage 0.11, spot check 1.8\%,
a change within the reference-floor scatter. The production value
sits on the flat (large-weight) side of this trade-off; what matters
is the constraint ``suppress in-gap leakage hard,'' not a tuned
magic number.

\section{Edge-beat artifact and the density-floor bath}\label{sec:s4}

\begin{figure}
\includegraphics[width=\columnwidth]{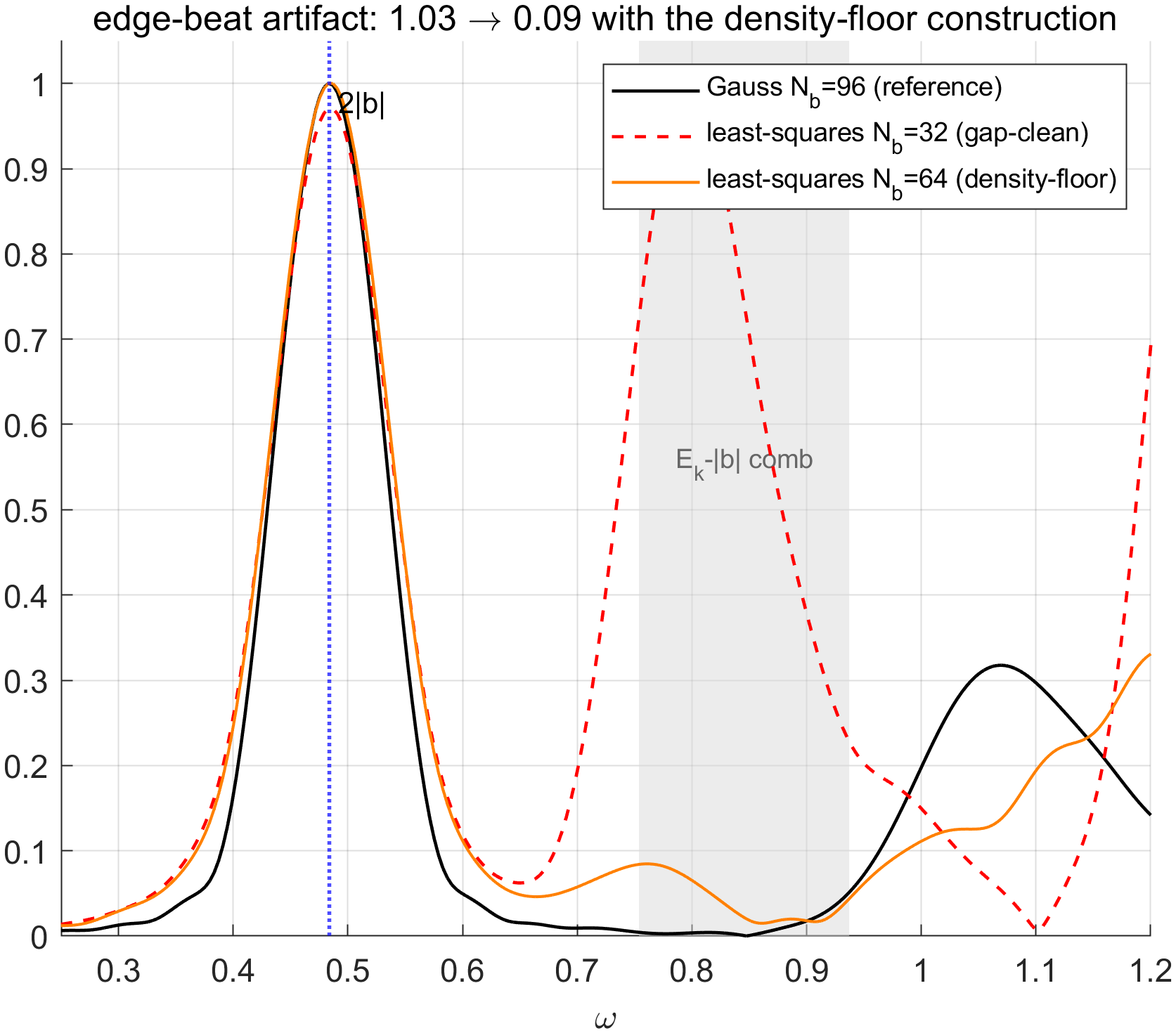}
\caption{\label{fig:edgebeat}Late-window spectra of the turn-off
current $I_A + I_B$ for three baths: Gauss $N_b = 96$ (reference,
black), gap-clean least-squares $N_b = 32$ (red dashed), density-floor
least-squares $N_b = 64$ (orange). The gray shading is the $E_j -
|\varepsilon_b|$ difference-frequency band predicted a priori from
the mode table; the in-band artifact peak of the gap-clean bath
(artifact/main = 1.03) is pushed to 0.09 by the density-floor
construction; the reference floor is 0.03. Normalization: maximum
within the display window (Sec.~\ref{sec:s4}).}
\end{figure}

\emph{Phenomenon and mechanism.} Gap-clean fitting assigns gap-edge
modes very small $\gamma$ (0.004--0.014, Table~\ref{tab:bath32});
their difference-frequency
comb with the ABS, at $E_j - |\varepsilon_b|$ (decay $\sim$0.023),
outlives the main line ($2|\varepsilon_b|$, decay $\sim$0.042) and
dominates the late-time window; the long-time spectrum of LS-32 shows
an artifact-to-main-peak ratio of 1.03 inside the assessment band
[Fig.~\ref{fig:edgebeat}]. The same comb sets the estimator boundary
behind the LS-32 native point of the main-text $\kappa$ panel [main
Fig.~7(b)], and for \emph{envelope-decay fits} the frequency condition
alone is not the discriminator (the in-band beats are spectrally
unresolvable from the main line at the window resolution): the
relevant a priori quantity is the \emph{takeover time} set by the
decay-rate difference between comb ($\sim$0.023) and main line
(0.042), both read off the $\{\varepsilon_j, \gamma_j\}$ tables. The
native envelope fit window (sliding windows out to $t \approx 520$)
reaches beyond the takeover: the measured slope is
$\kappa_{\mathrm{spec}}$ to 4\% over $t \in [80, 210]$ and collapses
toward the beat rate at late times (0.018 over $[330, 490]$),
producing the mixed-slope
$\kappa_{\mathrm{dyn}}/\kappa_{\mathrm{spec}} = 0.78$ of the $\kappa$
panel. Halving all dampings (the $\times 0.5$ variant) halves the
decay-rate difference and pushes the takeover beyond the same fit
window: its late-window slope stays at its own
$\kappa_{\mathrm{spec}}$ (0.021), which is why it remains a clean
test point while the native bath does not. \emph{A priori criterion
(spectral lines)}: the
contamination condition can be read directly off the mode table---a
difference frequency $E_j - |\varepsilon_b|$ falling into the $\pm
15\%$ assessment band around the target frequency contaminates it (in
the quench dispersion scan the two phases $\phi = 0.3\pi$ and
$0.45\pi$ are shown hollow and excluded from the quantitative assessment for this reason,
with 9--13\% deviations).

\emph{Density-floor construction (design rationale).} Three simpler
constructions are inadequate, which fixes the smooth-partition choice.
Equal-$E$ binning: the quadratic $\xi \leftrightarrow E$ map piles
$\Delta\xi \approx 0.2$ of weight into single modes, gap leakage
0.35. Uniform-$\xi$ equal weights: ineffective. Hard $\xi$
split: a positive-weight comb \emph{cannot subtract} the negative
in-gap Lorentzian tails of the core region (wing residual stuck at
0.45), and the cut point manufactures an artificial $1/\sqrt{\;}$
band edge. The smooth partition (the prescription of
Sec.~\ref{sec:s1}) resolves these: artifact $1.03 \to 0.085$,
$2|\varepsilon_b|$ deviation 0.36\%, MAR 10\% (the wing-fit residual
0.42 remains high). The LS-64 density-floor variant is therefore used
\emph{only} for long-time spectroscopy---peak assignment and
edge-beat screening---and \emph{does not enter} the MAR quantitative
error statistics; the quantitative MAR scheme remains gap-clean LS-32
(3.4\%, main-text Sec.~III.B). The two baths are selected by
observable and do not substitute for each other.

\emph{Fit-objective blindness to time-domain artifacts.} An
$N_b = 48$ cold-start refit lands in a worse local minimum (resolved
by warm starting), and baths of equal fit
quality have identical edge clusters and identical artifacts: the
spectral resolution of the late-time beat ($\sim 1/t$) is far below
the smoothing scale of the fit, so \emph{the fitting objective is
inherently blind to time-domain artifacts}. This is why the density
floor must be a constructive discipline rather than a fitting option.

\section{Normal-lead discipline: occupation pinning and the hybrid
bath}\label{sec:s5}

In DLvN the normal-lead occupations are pinned to the reference, so
the spectral shape \emph{and} the occupations inside the transport
window must be pointwise correct: purely fitted baths produce
occupation-staircase artifacts in the window (archive: NESS current
staircases of a purely fitted normal bath). The hybrid construction
(main-text Sec.~II.C, constraint 4) keeps an equally spaced dense
core with uniform small damping and analytic weights $\Gamma_\alpha
h/\pi$ ($\Gamma_\alpha$ = that normal lead's own hybridization,
Sec.~\ref{sec:s1} convention note), and compresses only the wings by
fitting the residual hybridization (target minus core comb); lead
strengths rescale exactly by $\sqrt{\Gamma_\alpha}$, so a single fit
serves all leads.

The production instance used in this work is the \emph{three-terminal
Shapiro hybrid-77} (\texttt{run\_n1d\_hybrid.m}; the bath loaded
verbatim by the Shapiro production runs): core $|\varepsilon| \le
1.6$ with $h = \gamma = 0.05$ and weights $\Gamma h/\pi$ at the
fitting hybridization $\Gamma = 0.1$ (the core was widened from 1.2
to 1.6 to push the core-edge transition band beyond the transport
reach of the protocol, $V \le 1.5$), plus 12 wing modes fitted to the
residual over the accuracy window $[-2, 2]$ with $\gamma \le 0.3$ and
positions constrained outside the core ($1.7 \le |\varepsilon| \le
6.0$; the constraint repairs a failure mode in which unconstrained
wings wandered into the window and polluted the hybridization
there)---$65 + 12 = 77$ modes, replacing a purely uniform bath of
about 400 modes; the full mode table is Table~\ref{tab:bath77}.
Superconducting leads are exempt from the dense
core: at $T \ll \Delta$ the in-band quasiparticle filling is flat
(the Fermi edge falls in the zero-DOS gap region), so occupation
pinning is harmless---this is the spectral-structure reason
superconducting leads compress more efficiently.

\section{Numerical archive of the gauge discipline}\label{sec:s6}

Numerical verification of gauge equivalence without damping
(benchmark cases as noted): the pairing gauge and the hopping-phase
(Peierls) gauge give $I(t)$ identical point by point---a two-terminal
S-QD-S dc-biased ac-Josephson case (unitary evolution, finite bath,
full $I(t)$) at rmse $6.6\times 10^{-7}$; a three-terminal case (two
superconducting leads plus a normal lead) under multifrequency drive
at $1.9\times 10^{-6}$, with the continuity identity on the same case
$\max|\sum_\alpha I_\alpha - dn/dt| = 1.3\times 10^{-4}$ ($n$ = dot
occupation, the same symbol as main-text Sec.~II.A). With damping
on, the recorded failure of the pairing gauge: the bias winds
$\Delta_j(t)$, the damped bath orbitals become explicitly
time-dependent, and damping toward a static $C_{\mathrm{ref}}$ drags
the system to a wrong steady state (long-time current drift, no
NESS); under the hopping gauge (bath blocks frozen) the same
protocol converges cleanly. This is the numerical basis for
the main-text rule ``no rotating term may sit on a damped site.''

\section{Protocol verification and numerical consistency
checks}\label{sec:s7}

Three protocol-design checks that fix, respectively, the observable
combination, the reproduction setup, and the sampling density:

\begin{enumerate}
\item \emph{Observable combination.} The $2|\varepsilon_b|$
resonance amplitude for the check on Fig.~5 of
Ref.~\cite{cheng2024quasiparticle} must be measured on $n(t)$,
$I_A + I_B$, or the single-side $I_A$, not on the symmetrized
combination $I = \tfrac{1}{2}(I_A - I_B)$: there the continuity
identity makes the $dn/dt$ part cancel exactly, forcing an algebraic
null independent of the physics. This constrains the choice of
transient observable combinations through the main-text Sec.~II.A
continuity identity $\sum_\alpha I_\alpha = dn/dt$ (in a two-terminal
system $I_A + I_B = dn/dt$), which every transient combination must
respect.
\item \emph{Reproduction setup.} The reproduction follows the Fig.~5
protocol of Ref.~\cite{cheng2024quasiparticle} (a turn-on transient).
One setup sensitivity is worth recording: with the pre-drive duration
$t_{\mathrm{pre}} = 60$ close to the relaxation timescale 43,
St\"uckelberg-type fringes appear; they vanish for $t_{\mathrm{pre}}
= 150$. The preparation stage of a transient protocol must be much
longer than the system memory. (The production runs keep
$t_{\mathrm{pre}} = 60$ as the verbatim reproduction of
Ref.~\cite{cheng2024quasiparticle}; the audit in Sec.~\ref{sec:s1}
shows the three quoted observables are insensitive to the choice at
the $10^{-3}$ level.)
\item \emph{Four-point $\phi_0$ aliasing.} The archived 4-$\phi_0$
locking-band data at finite $\Delta$ cannot resolve $A_3$ (it
aliases into the $A_1$ channel, and $A_2$ retains only its cosine
projection); the $m = 3$ assessment uses an 8-$\phi_0$
scan (Sec.~\ref{sec:s9}). The number of $\phi_0$ samples must
exceed $2 m_{\max}$.
\end{enumerate}

\section{MAR supplement}\label{sec:s8}

\begin{figure}[!t]
\includegraphics[width=\columnwidth]{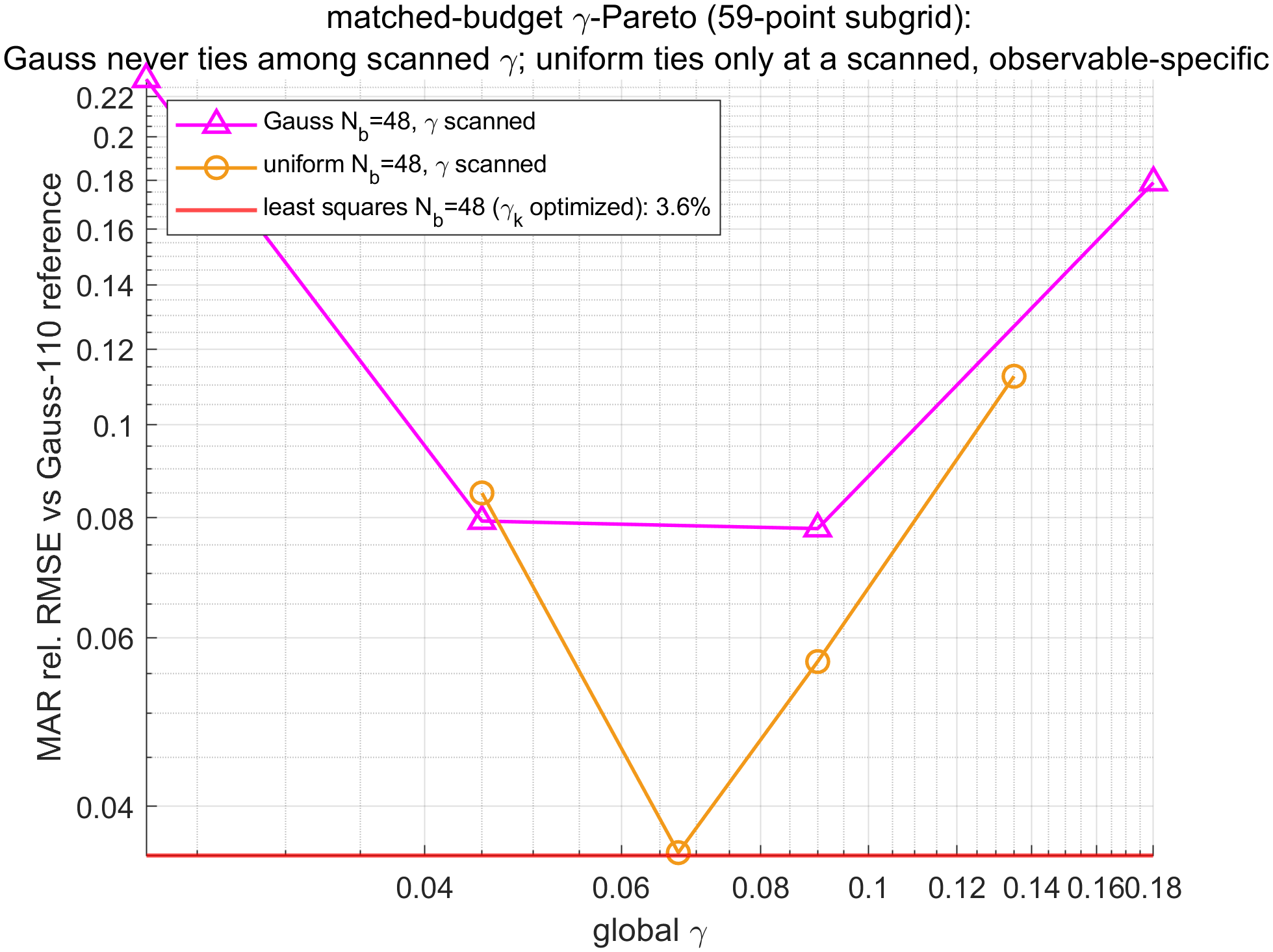}
\caption{\label{fig:pareto}Matched-budget $\gamma$-Pareto (59-point
subgrid): MAR error, relative to the Gauss-110 reference, of the
$\gamma$-scanned Gauss-48 and uniform-48 baths; the least-squares-48
result is the horizontal line (its own error against the same
Gauss-110 reference). The
uniform grid can be hand-tuned to tie on this single observable at
$\gamma = 0.0675$; that $\gamma$ forfeits the slow-dynamics window
($\kappa = 0.026$) and must be found by scanning against a
pre-computed high-resolution reference.}
\end{figure}

\emph{$\gamma$-Pareto completion of the matched-budget comparison
(Fig.~\ref{fig:pareto}).} The matched-budget comparison of main
Sec.~II.C pits an optimized bath against fixed-recipe baselines; the
fairness question ``could a scanned global $\gamma$ close the gap?''
is answered by scanning it. Protocol: the matched-budget MAR run
(48 modes per lead) on the anchor grid subsampled to every other
point (59 of 117; all numbers below, including the archived
full-grid configurations, are evaluated on this same subgrid).
Gauss-48: 22.9\% ($\gamma = 0.0225$, recurrence-contaminated),
7.9\% (0.045), 7.8\% (0.09), 17.9\% (0.18)---a U-shaped trade-off
whose optimum stays a factor 2.2 above the least-squares result.
Uniform-48: 8.5\% (0.045), \emph{3.6\%} (0.0675), 5.7\% (0.09),
11.2\% (0.135)---a sharp minimum that \emph{matches} the
least-squares 3.6\% on this observable (the tie point is
confirmed on the full 117-point grid: 3.60\% full-curve
vs 3.58\% subgrid, and is overlaid in main Fig.~4(b)). We report this
tie as found;
it sharpens rather than weakens the paper's claim, in two respects.
First, the winning $\gamma$ is located only by scanning full
$I$-$V$ curves against a pre-computed high-resolution reference---the very
resource the optimization is meant to replace---and it is
observable-specific (MAR, the most discretization-forgiving
observable of the set, in line with the stratified-assessment
discussion of the main text). Second, the same $\gamma = 0.0675$
fixes $\kappa = 0.386\gamma = 0.026$ for \emph{every} slow mode:
46$\times$ the resolution scale of the starred configuration of main
Fig.~7(b), i.e., the slow-dynamics window of validation 3 is
forfeited wholesale, whereas the least-squares bath reaches the same
MAR error with per-mode $\gamma_j$ and no tuning. The Gaussian
quadrature never ties among the scanned $\gamma$ values (its node
positions are locked by the rule; the four-point U-shape brackets its
optimum, but we make no claim beyond the scanned set).

\emph{Reference mode-number refinement.} To locate the
3--4\% plateau, the reference itself was refined in mode number at 14
representative biases (MAR peak tops, valleys, step edges, high-bias
tail; Gauss-140 on a 10-point subset), at the reference's own $\gamma
= 0.045$ and the production protocol verbatim. Result, relative to
Gauss-110: Gauss-140 deviates by $\le 0.4\%$ per point across the
subgap MAR window ($V \le 1.3$) but by 1.4\% and 1.9\% at the two
gap-edge biases $V = 1.6, 2.0$---the $n = 1$ threshold region, where
the quadrature comb is sparsest against the singular BCS edge---for
an aggregate relative $\ell_2$ of 1.2\% (Gauss-96: 1.4\%). The
pre-specified convergence gate ($<0.2\times$ the plateau) is
therefore not met at the gap-edge points, and the attribution is
worded accordingly in the main text: the reference's own discretization
dispersion is a real, subdominant ($\sim$1/3 of the plateau in
aggregate) reference-side component, the shared $\gamma$ broadening
remains the leading one, and the residual cannot be separated further
by these probes. The full 117-point refinement chain
($\sim(140/110)^3$ the 7.7-h reference cost per curve) remains
unrun.

The full $(V, \varepsilon_d)$ color map is $41 \times 86$ (removed
from the main figures as visually saturated) with the
tracked ridges overlaid. Ridge tracking: per column (fixed
$\varepsilon_d$), peaks of $dI/dV$ are found under a continuity
constraint (adjacent-column peak positions within a grid-spacing
tolerance) with subgrid parabolic refinement; the $k = 0$ family
linear fit gives $V = (2.031 \pm 0.045) - (2.069 \pm
0.100)\,\varepsilon_d$ (95\% CI from ordinary least squares under an
independent-residual assumption; the stated error bars are the local
bias-grid spacing, i.e., a resolution statement, and the CIs should
be read as descriptive rather than strictly probabilistic); the
dominant $n = 3$ resonance has ridge slope $+0.210 \pm
0.041$. The independent benchmark is reported in both error measures
to expose the division artifact: the relative deviation diverges as
$I \to 0$ (lesson: small-denominator observables are always reported
as absolute plus conditional relative deviations).

\emph{Peak position versus MAR threshold ($n = 3$).} The kinematic
voltages $eV = 2\Delta/n$ mark where the $n$-th multiple-Andreev
channel opens; the $dI/dV$ maximum is a finite-width proxy for that
onset and sits slightly above it. At $\varepsilon_d = 0$ the extracted
$n = 3$ maxima are $0.680$ (Floquet), $0.677$ (least squares) and
$0.666$ (Gauss-110) under a common central-difference scheme---all
within $\sim 2\%$ of $2\Delta/3 = 0.667$, whereas the $n = 5, 7$ maxima
fall on $2\Delta/5, 2\Delta/7$ to the grid resolution. The Floquet peak
is robust to the differentiation template ($0.680$ to within
$2\times 10^{-3}$ across smoothing windows), and its rising edge---the
maximum-curvature point $0.662$ and the half-maximum $0.657$---brackets
$2\Delta/3$, confirming that the underlying threshold sits at
$2\Delta/n$ while the derivative peak lies $\sim 2\%$ above it. The
finite-bath $dI/dV$ carries the discretization wiggle (main
Sec.~III.B), so the extracted discrete-bath peak acquires an additional
few-percent scatter under smoothing; we therefore anchor the threshold
statement on the wiggle-free Floquet onset and do not over-interpret
the exact discrete-bath peak location. This resolves the small $n = 3$
displacement visible in main Fig.~4(a): a lineshape and extraction
effect on the derivative maximum, not a shift of the MAR onset.

\emph{$\gamma$ scaling of the conductance peaks (main Fig.~5): model
and fit.} Writing $G \equiv dI/dV$, the model is a one-parameter
convolution,
\begin{equation}
G_{\mathrm{model}}(V;\gamma,n)
= \int dV'\, G_{\mathrm{Flq}}(V')\,
L_{R}\!\left(V - V';\, c_n\gamma\right),
\label{eq:gofmodel}
\end{equation}
where $G_{\mathrm{Flq}}$ is the exact Floquet conductance and $L_R(x;
w)$ is a Lorentzian of FWHM $w$ truncated at $|x| \le R
= 0.275$ and renormalized to unit mass, and the Floquet curve enters
on its full computed support ($V \in [0.15, 2.05]$). The Lorentzian
form is not ad hoc: a Lindblad-damped mode carries an exactly
Lorentzian spectral function, extended-reservoir transport reduces to
the Landauer expression with Lorentzian-broadened reservoir spectral
functions~\cite{gruss2016landauer,elenewski2017markovian,wojtowicz2021dual},
and Lorentzian lifetime broadening of a superconducting density of
states is the classic Dynes operation~\cite{dynes1978direct}. The
principle:
each damped mode gives the quasiparticle an energy-space lifetime
width $c_E\gamma$; the MAR resonance condition $n\,eV \approx 2\Delta$
maps energies onto the voltage axis with an $n$-fold lever arm, so the
voltage-space kernel is $c_n = c_E/n$. The single constant $c_3 =
1.345$ is fitted once, by scalar least squares over the pointwise
lineshapes of all five $n = 3$ rungs simultaneously; $c_E = 3c_3 =
4.04$, and $c_5 = c_E/5 = 0.807$ is used with no further fitting (the
$n = 5$ rows below are a parameter-free prediction). \emph{Kernel
convention}: the truncation radius matters because Lorentzian tails
are fat---widening $R$ to 0.41 (0.60) drags the refit constant to 1.24
(1.17) and degrades every rung (e.g., $R^2$ at $\gamma = 0.18$ falls
$0.91 \to 0.57 \to 0.24$): the data support a Lorentzian \emph{core}
with compact tails, not the full fat-tailed profile. $R = 0.275$
($\approx 2.3\times$ the largest kernel HWHM used) is therefore part
of the model definition; a refit under the standardized convention
reproduces $c_3$ to 0.1\%. Because $R$ was fixed after inspecting the
data, the kernel is a \emph{semi-empirical} broadening model with one
fitted constant and one data-selected truncation scale; a
leave-rung-out stability check at fixed, data-selected $R$ refits
$c_3$ on the rung subsets $\{0.18, 0.09\}$ and $\{0.0675, 0.045\}$ gives $c_3
= 1.340$ and $1.353$ (1.0\% spread around the production 1.345), and
the held-out rungs and the parameter-free $n = 5$ rows reproduce the
production-table metrics essentially unchanged (e.g., held-out $n =
3$ peak deviations $-1.9$ to $-4.8\%$; $n = 5$ peak deviations
$-11.4$ to $-16.2\%$)---the kernel scale is data-determined, not
rung-selected. (All model curves and quoted $n = 5$ deviations use the
standardized assessment convention above---the interval-averaged,
non-edge-truncated kernel [Table~\ref{tab:gof}].)

\emph{Bidirectional $R$ scan.} Scanning the truncation radius in both
directions (refitting $c_3$ jointly at each $R$) shows a well-defined
quality valley rather than a plateau. Narrowing below about twice the
largest kernel HWHM eats genuine core mass and forces $c_3$ into
compensation ($R = 0.15$: $c_3 = 2.05$, $R^2$ at $\gamma = 0.18$
falls to $-7.7$; $R = 0.10$: $c_3$ hits the optimizer bound);
widening lets the fat tails import structure from outside the
resonance neighborhood ($R = 0.35/0.41/0.60$: $R^2 = 0.63/0.57/0.24$).
The valley floor $R \in [0.275, 0.30]$ ($c_3$ stable to $\pm 2\%$,
mean lineshape deviation $0.059/0.060$) coincides with the two a
priori scales---full coverage of the Lorentzian core
($2.3$--$2.5\times$ the largest HWHM) and the adjacent-resonance
spacing $0.667 - 0.400 = 0.267$, the locality bound beyond which the
$n = 3$ lever-arm mapping should not propagate energy widths.
The $n = 5$ family, refit independently with its own free constant,
selects its own shallow valley at $R \approx 0.114$--$0.125$---again
at its own adjacent-resonance spacing ($0.400 - 0.286 = 0.114$)---
though with the shallow $n = 5$ lineshapes and the $\sim$5\% ripple
floor the valley depth (12\%) makes this consistent rather than
demonstrated. The freely fitted $c_5$ (five rungs, overall scale
fixed to the reference) comes out $0.68$--$0.71$ depending on
estimator convention---$12$--$16\%$ below the lever-arm value
$0.807$---quantifying the systematically negative
model$-$measurement peak deviations of Table~\ref{tab:gof}:
higher-order resonances are slightly sharper than the leading-order
lever arm predicts.

\emph{Anchor recovery (circularity check).} The model's $\gamma \to
0$ member is the Floquet curve \emph{by construction}, and the
measured rungs reach only 23--54\% of the Floquet peak height, so we
checked whether the $\gamma \to 0$ statement is data-supported or
model-imposed. Freeing the overall scale $s$ of the underlying
unbroadened curve and fitting $(s, c_3)$ jointly on the five $n = 3$
rungs returns $s = 0.995$, $c_3 = 1.328$: the finite-$\gamma$ data
alone recover the Floquet anchor to 0.5\%. The $(s, c)$ degeneracy is
broken by the $4\times$ range in $\gamma$ (profiling: moving $c$ by
$\pm 10\%$ forces $s$ by about $\mp 4\%$ at a 6--28\% penalty in the
residual sum), and per-rung scales at fixed $c$ are $0.983$--$1.006$
(spread 2.3\%). A descriptive error budget for this anchor-free
recovery---rung scatter $\pm 1\%$, profile direction $\pm 4\%$,
kernel-form ($R$-valley) $\pm 1$--$2\%$---puts the blind recovery of
the $n = 3$ peak at about $\pm 5\%$ (conservative $\pm 10\%$). The $n
= 5$ family alone constrains the anchor only weakly: freeing $(s,
c_5)$ returns $s = 0.90$ with a soft profile ($s = 0.82$--$0.97$
within a $1.4\times$ residual ratio)---all five rungs sit in the
strong-smearing regime, so the amplitude--width degeneracy that the
$4\times$ range in $\gamma$ breaks for $n = 3$ is not broken here.
Locking the kernel to the cross-order value $c_5 = c_E/5$ and freeing
the amplitude alone, however, returns $s = 1.04$ (joint $R^2 = 0.87$),
with the per-rung amplitude approaching the anchor monotonically
($1.16 \to 1.01$) as $\gamma$ decreases: the $n = 3$-calibrated
kernel and the $n = 5$ data jointly recover the anchor to $4\%$
(symmetrized estimator, below), while the blind single-family anchor
statement remains exclusive to $n = 3$. As throughout, these
intervals are descriptive (comb-ripple-dominated residuals).

\emph{Fine-grid extension and estimator convention.} The $n = 5$
family comprises the same five dampings as the $n = 3$ ladder,
$\gamma = 0.045$--$0.18$, all on the Gauss-110 reference: $\gamma =
0.135, 0.0675$ and a fine regrid of $0.045$ are included
on the fine peak grid (production protocol otherwise verbatim;
per-point NESS drift $\le 5\times 10^{-4}$). Two disclosures. First,
an estimator-convention audit: the centered finite difference of
$I(V)$ \emph{exactly} equals the interval average of $dI/dV$ over
$[V_{j-1}, V_{j+1}]$, whereas the model was previously evaluated
pointwise---an asymmetry biasing sharp-peak comparisons by up to
$-(2h)^2/(3W^2)$. All cross-checks here therefore also average the
model over the same per-point intervals (exact, since the box
average is itself a convolution and the model family is closed under
it). Under this symmetrization every published $n = 3$ number is
reproduced within its quoted band ($c_3$: $1.341$ vs $1.345$; peak
deviations $-2.0$ to $-4.4\%$), and the $n = 5$ peak deviations read
$-11.6$ to $-16.4\%$ (the Table~\ref{tab:gof} values; the
pointwise-convention counterparts read $-8.0$ to $-15.9\%$---the two
conventions differ by the box-average correction, are both monotone
in smearing depth, and the interval-averaged one is the
like-for-like comparison for finite-difference data). Second, the
deviation of the cross-order
prediction is \emph{monotone in smearing depth} $c_5\gamma/W$:
$-11.6\%$ at kernel-to-intrinsic-width ratio $2.6$, rising to
$-16.4\%$ at ratio $10$---the one-sided band of
Table~\ref{tab:gof} is a strong-smearing effect, largest where the
kernel most exceeds the intrinsic width.

Goodness of fit per rung (over all measured grid points of that rung;
$G$ denotes the measured $dI/dV$ by centered finite differences):
relative deviation $\lVert G - G_{\mathrm{model}}\rVert / \lVert G
\rVert$, coefficient of determination $R^2 = 1 -
\sum(G-G_{\mathrm{model}})^2/\sum(G-\bar G)^2$, and the peak-height
deviation:
\begin{table}[h]
\caption{\label{tab:gof}Goodness of fit of the convolution model
Eq.~\eqref{eq:gofmodel}. The $n = 3$ rows share the single fitted $c_3
= 1.345$; the $n = 5$ rows are parameter-free. Estimator convention:
measured $G$ by centered finite differences throughout; for the sharp
$n = 5$ peaks the model is averaged over the same per-point intervals
(the like-for-like convention, see text), while the wide $n = 3$
peaks are convention-insensitive and retain the pointwise evaluation;
the pointwise $n = 5$ values are quoted in the text as estimator
sensitivity.}
\begin{ruledtabular}
\begin{tabular}{ccccc}
$n$ & $\gamma$ & rel.\ dev. & $R^2$ & peak dev. \\
\hline
3 & 0.180  & 0.022 & 0.91 & $-1.6\%$ \\
3 & 0.135  & 0.039 & 0.87 & $-4.4\%$ \\
3 & 0.090  & 0.061 & 0.86 & $-3.6\%$ \\
3 & 0.0675 & 0.071 & 0.89 & $-2.1\%$ \\
3 & 0.045  & 0.101 & 0.89 & $-2.5\%$ \\
5 & 0.180  & 0.152 & $-1.2$ & $-16.4\%$ \\
5 & 0.135  & 0.122 & 0.50 & $-14.7\%$ \\
5 & 0.090  & 0.124 & 0.84 & $-13.3\%$ \\
5 & 0.0675 & 0.148 & 0.87 & $-12.8\%$ \\
5 & 0.045  & 0.174 & 0.90 & $-11.6\%$ \\
\end{tabular}
\end{ruledtabular}
\end{table}
The fit residuals here are dominated by the deterministic comb ripple
of the finite baths, not by independent noise, so $R^2$ (and every CI
in this SM) is reported as a \emph{descriptive} summary rather than a
probabilistic statement (measured basis in the decomposition below).
$R^2$ measures explained \emph{shape variance} and collapses by
construction for near-flat lineshapes: on the $n = 5$, $\gamma = 0.18$
rung the kernel exceeds the intrinsic width (FWHM 0.014) by a factor
of ten, the smeared profile is nearly flat, and the residual is
dominated by the $\gamma$-independent differentiation ripple
($\sim$5\%, below), so $R^2$ turns negative there while the relative
deviation (15\%) and the peak row carry the verdict.

\emph{Residual-structure decomposition.} $R^2 < 1$ by itself does not
challenge the model---structured residuals would. Projecting each
rung's residual onto the three canonical misfit modes (amplitude
$\propto G_{\mathrm{model}}$, position $\propto
dG_{\mathrm{model}}/dV$, kernel width $\propto \partial
G_{\mathrm{model}}/\partial c$) shows that every structured component
maps onto an effect already quantified in this section: the $n = 5$
amplitude mode (72--84\% of $\lVert r\rVert^2$ at the flattest rungs)
\emph{is} the disclosed one-sided cross-order deficit; the $n = 3$
position mode (58--79\% at the middle rungs) is a sub-grid
peak-registration offset ($\le 0.004$, below the bias-grid spacing
and inside the quoted 0.673--0.680 pinning band); the width mode
grows only at the smearing extremes, consistent with the $R$ scan
above. The mode-removed remainder is strongly correlated across rungs
($n = 5$: 0.81--0.97; adjacent $n = 3$ rungs: 0.79--0.89)---the
deterministic, $\gamma$-smoothed comb ripple rather than independent
noise---which is the measured basis for the descriptive-statistics
convention used throughout. The growth of the relative deviation
toward small $\gamma$ likewise reflects sharper features projecting
registration offsets and ripple more strongly, not model
degradation.

Protocol: local bias grids $V = 0.62{:}0.01{:}0.74$ ($n = 3$ peak) and
$V = 0.36{:}0.01{:}0.44$ ($n = 5$ peak); reference bath Gauss-110, the
production MAR protocol otherwise verbatim ($dt = 0.05$,
$t_{\mathrm{settle}} = 75$, six-period NESS averages); dampings
$\gamma \in \{0.045, 0.0675, 0.09, 0.135, 0.18\}$ for both peaks (the
$n = 3$ 0.045 rung and the $n = 5$ 0.135, 0.0675, 0.045 rungs run on
the fine grids for grid consistency). Settledness is monitored per point by
comparing six- and three-period window averages (agreement $\sim
10^{-4}$ throughout). \emph{Estimator systematics}: raw
finite-difference peak heights carry $\sim$5\% scatter from the
residual comb ripple of the bath---a three-rung reciprocal-linear
extrapolation of the $n = 3$ peak height scattered between $+1\%$ and
$-7\%$ of the exact value depending on the rung subset---so the robust
assessment quoted in the main text is the direct lineshape and
peak-height comparison against the convolved Floquet curve.
\emph{Hypotheses tested and falsified along the way}: (i) a
linear-in-$\gamma$ peak-height extrapolation ($-32\%$; expected, since
$\gamma$ is not small against the intrinsic width); (ii) an
$n$-independent voltage kernel (transferring the $n = 3$ constant
unchanged to $n = 5$ overshoots the suppression by $\sim$75\%)---
replaced by the lever-arm scaling $c_n = c_E/n$ of the main text,
under which the $n = 5$ prediction is parameter-free. The intrinsic
peak widths extracted from the Floquet curve (FWHM 0.071 for $n = 3$,
0.014 for $n = 5$) confirm that higher-order MAR resonances are
intrinsically sharper, hence proportionally more sensitive to a fixed
$\gamma$.

\section{Shapiro supplement}\label{sec:s9}

\emph{Arnold-tongue cost discipline.} The full-map cost is controlled
by a fast Floquet kernel (row-wise Green-function solves: the $G^R$ row via
$G^{R\,\dagger}_{\mathrm{inv}} \backslash E_R$, and $G^<$ elements
as $\sum_j G^R(a,j)\, s_l(j)\, G^R(c,j)^*$), validated against the
reference implementation at $2.5\times 10^{-14}$, speedup $\times
1.8$; an $N_E$ downshift ($1401 \to 801$) consistent at $3\times
10^{-5}$; dimension cap 700, cells beyond the cap left blank and
marked gray (no extrapolation, no interpolation).

\emph{DLvN spot checks against the Floquet map.}
Table~\ref{tab:spots} lists the six spot checks (positions marked by
circles in main Fig.~8(c)): integer steps are quantitative
(2.0--3.5\%; the $n = 3$ step reaches 7.5\%), $m = 2$ subharmonics
are 11--28\% (small amplitudes under $\gamma$ broadening:
qualitative). The dense-petal locking amplitude comparison is $A_1 =
0.12775$ (Floquet) versus $0.13152$ (DLvN), a 2.95\% deviation [main
Fig.~8(a)].

\begin{table*}
\caption{\label{tab:spots}DLvN spot checks against the Floquet
Arnold-tongue map (low-energy effective junction; generated from the
archived \texttt{tongues\_spots.mat}). $A_m$ is the locking amplitude
of the resonant harmonic $m$ at the tongue position
$(V_{\mathrm{dc}}/\Omega, V_{\mathrm{ac}})$.}
\begin{ruledtabular}
\begin{tabular}{cccccc}
$V_{\mathrm{dc}}/\Omega$ & $V_{\mathrm{ac}}$ & $m$ &
$A_m$ (Floquet) & $A_m$ (DLvN) & dev. \\
\hline
0.500 & 0.30 & 1 & 0.1278 & 0.1315 & 2.9\% \\
0.500 & 0.90 & 1 & 0.0760 & 0.0787 & 3.5\% \\
1.000 & 0.30 & 1 & 0.0475 & 0.0466 & 2.0\% \\
1.500 & 0.50 & 1 & 0.0320 & 0.0296 & 7.5\% \\
0.250 & 0.50 & 2 & 0.0473 & 0.0420 & 11.1\% \\
0.250 & 0.90 & 2 & 0.0071 & 0.0051 & 28.4\% \\

\end{tabular}
\end{ruledtabular}
\end{table*}

The classic Shapiro staircase synthesized from these voltage-biased
data (overdamped current-bias dual, data-driven plateau labels $n/m$
of $V_{\mathrm{dc}}/(\Omega/2)$; locking bands below $3\times 10^{-3}$
not drawn) appears as main Fig.~9(a).

\emph{Finite-$\Delta$ rational-grid scan (LS-32 production bath, all $m
\le 3$ resonances at 8 $\phi_0$, alias-free).} Full locking-amplitude
table of the production run (main Fig.~10(a)); the display
threshold for a drawn plateau is $3\times 10^{-3}$. Integer steps:
$A_1 = 9.6\times 10^{-2}$ / $9.0\times 10^{-2}$ / $1.7\times 10^{-2}$
at $n = 1, 2, 3$. $m = 2$ family: $A_2 = 3.5\times 10^{-2}$ ($1/2$),
$3.6\times 10^{-3}$ ($3/2$), $4.4\times 10^{-4}$ ($5/2$, below
threshold; the archived Gaussian-recipe value at the same point was
$4.5\times 10^{-4}$---a two-discretization cross-check), $1.1\times
10^{-4}$ ($7/2$). $m = 3$ family: $A_3 = 9.0\times 10^{-3}$ ($1/3$),
$4.1\times 10^{-3}$ ($2/3$),
$1.0\times 10^{-3}$ ($4/3$, below threshold), then $\le 10^{-4}$ for
all higher members (fast decay within the family). On every
fractional step the non-resonant harmonics sit at the
$10^{-5}$--$10^{-4}$ averaging floor, one to three orders below the
resonant one (selection-rule fingerprint); $m \ge 4$ resonances
are not scanned (fed by the $m$-th CPR harmonic, below the display
threshold by extrapolation of the family decay). The archived
Gaussian-recipe rescan (1/3 member $A_3 = 1.09\times 10^{-2}$ vs the
LS-32 $9.0\times 10^{-3}$) agrees at the qualitative-amplitude level
declared for this channel.

\emph{Off-skeleton collapse.} On step points $A_1$ is independent of
the averaging window (0.129/0.047); between steps, four points show
$A_1$ smaller by a factor of 20--50 and $\propto
1/T_{\mathrm{avg}}$ (still shrinking as the window lengthens). The
rational-skeleton restriction of Floquet: the commensurate base
frequency is $w_b = \Omega/q$ with $q$ the reduced denominator of
$V_{\mathrm{dc}}/(2\Omega)$ [$= n/(4m)$ on the kinematic grid], and
the Floquet dimension grows $\propto 1/w_b$ (the first $m = 3$ member,
$n = 1$, has $w_b = \Omega/12$, beyond the dimension cap), so that
assessment is carried by the exact symmetry selection rule,
satisfied numerically to the DLvN averaging floor ($\sim 10^{-5}$).

\section{Static-baseline bookkeeping}\label{sec:s11}

The static baseline against which the discretizations are scored is
the noninteracting continuum current-phase relation obtained from the
(Matsubara) junction free energy differentiated in phase, $I(\phi) =
2\,\partial F/\partial\phi$ with $F = -(1/\beta)\sum_n \ln\det
G^{-1}(i\omega_n)$, on a finite band $D = 20$ at $\beta = 50$
(Matsubara cutoff $4000$); it carries no bath discretization and, for
the noninteracting ($U = 0$) junction, is numerically converged with
respect to the stated Matsubara cutoff (not analytically exact). This same free-energy CPR is evaluated by two
independent implementations (an external reference code and the
present pipeline), which agree at $3\times 10^{-5}$---an implementation
cross-check, not two distinct physical methods. As a further,
physically independent test we compare against a real-axis Keldysh
(nonequilibrium Green-function) current---the Andreev-bound-state plus
continuum Meir--Wingreen current---evaluated at the same working point
($D = 20$, $\beta = 50$, $\Gamma = 1$): it reproduces the same CPR at
the percent level ($0.3\%$ in the wide-band limit; the finite-band
residual is limited by real-axis $\eta$- and band-edge regularization),
up to an exact overall current-normalization convention between the
two engines.
``Free-energy CPR'' in the main text and in Table~S1 denotes this
benchmark. The finite-gap static benchmark of main Fig.~10(b) (a
different junction, $\Gamma = 0.8$, $D = 4$) is the same free-energy
CPR evaluated at $\beta = 400$ (with a $\beta = 200$ consistency
check), Matsubara cutoff $12000$, on a 33-point phase grid; both
discretized baths match it at $\le 3.5\times 10^{-4}$, comparable to
the combined temperature/cutoff convergence change of $3.1\times
10^{-4}$ (between $\beta = 200$ and $400$; the two are varied together
in the archive, so this is not a pure truncation floor).

Three bookkeeping conventions (all verified by numerical
cross-checks). (i) \emph{The hybridization factor of 2 is a
parameter-passing difference, not a definition difference} (verified
against the code): at the static benchmark
point the physical total hybridization is $\Gamma = 1$, per lead
$\Gamma_\alpha = 0.5$; the static pipeline's bath constructor is
passed $\Gamma_\alpha = 0.5$ \emph{per lead}, while the NEGF
pipeline is passed the \emph{total} $\Gamma = 1$ (the NEGF engine
splits $\Gamma_{\mathrm{total}}/2$ per lead internally; verified in
the code structure)---both sides share the same $\pi\rho V^2$
definition, so there is no $\pi\rho V^2$ versus $2\pi\rho V^2$
discrepancy between the two pipelines. (Passing the per-lead value $0.5$ gives
reconciliation ratio $-1.0000$ and
rel.\ $1.9\times 10^{-7}$; the total value would differ by
exactly a factor of 2. The NEGF engine belongs to the same formula family:
passed $\Gamma = 1.0$; if 0.5 is passed by mistake an 8\% deviation
and a spurious ratio of 1.52 appear---that cross-check exists for
this purpose.)
(ii) \emph{Current sign}: $-1$ (a convention difference, not
physics). (iii) \emph{Temperature}: all curves are compared at
matched temperature $\beta = 50$ ($T = 0.02\Delta$). Reconciliation
results: reference ($\beta = 50$)
versus the free-energy CPR $1.9\times 10^{-7}$; the second independent
implementation $3\times 10^{-5}$.

\section{Naming map}\label{sec:s12}

Table~\ref{tab:naming} maps the paper-facing names to the archive
code names, for traceability when re-auditing the raw data.

\begin{table*}
\caption{\label{tab:naming}Paper-facing names versus data-archive
code names.}
\begin{ruledtabular}
\begin{tabular}{p{0.16\textwidth}p{0.12\textwidth}p{0.16\textwidth}}
Paper name & Archive name & Definition \\
\hline
Gauss $N_b{=}110/96$ (reference) & anchor110 / anchor96 &
gap-mapped Gaussian quadrature, uniform $\gamma = 0.045/0.05$ \\
least squares $N_b{=}32$ & fitted32 / gap-clean &
real-axis LS, gap weight 30, per-mode $\gamma$ \\
least squares $N_b{=}64$ (density floor) & two-zone 64 &
gap-edge dense core + smooth partition + LS wings \\
least squares $N_b{=}77$ (normal lead) & hybrid77 &
transport-window dense core + 12 LS wings \\
LS (imag-axis) & imaginary-axis variant &
statics-only convergence study \\
uniform $N_b$, $\gamma$ & uniform &
equal spacing and weights, global $\gamma$ \\
dense mapped $N_b{=}384/768$ & dense mapped grid &
extreme small-$\kappa$ regime only \\
$\Gamma_S$ (effective-junction SC hybridization) & GS0 &
plays the induced pairing amplitude on the dot in the
$\Delta \to \infty$ limit (defined at first use, main Sec.~III.D) \\
\end{tabular}
\end{ruledtabular}
\end{table*}

\section{Derivation of the structure-preservation
proposition}\label{sec:s13}

\emph{Proposition conditions.} The damping matrix $\Lambda$ is
proportional to the identity within each bath mode's $2\times 2$
Nambu block---i.e., the particle and hole components of the same
mode take the same $\gamma_j$; every implementation in this work
does so, including the per-mode-$\gamma$ fitted baths (the engine
assigns $\Lambda$ pairwise over the particle and hole copies of each
mode). The reference correlation matrix is $C_{\mathrm{ref}} =
f(H_{\mathrm{ref}})$ with $H_{\mathrm{ref}}$ a BdG-type reference
Hamiltonian (decoupled leads, or the coupled equilibrium junction).

\emph{(a) Hermiticity.} The right-hand side $-i[H, C] -
\tfrac{1}{2}\{\Lambda, C - C_{\mathrm{ref}}\}$ is manifestly
Hermitian when $C$ and $C_{\mathrm{ref}}$ are Hermitian and
$\Lambda$ is real diagonal, so $C(t)$ stays Hermitian.

\emph{(b) Particle-hole mirror consistency.} Let $X$ denote the
Nambu exchange matrix (particle block $\leftrightarrow$ hole block).
The construction guarantees $X H^*(t) X = -H(t)$: the hole block is
$-h^*(t)$, the pairing amplitudes stay real, and the conjugate
orientation of the Peierls phases is carried automatically by the
Nambu structure (main-text Sec.~II.A); the proposition condition
reads $X \Lambda X = \Lambda$. For $C_{\mathrm{ref}} =
f(H_{\mathrm{ref}})$, the identities $f(-E) = 1 - f(E)$ and $X
H_{\mathrm{ref}}^* X = -H_{\mathrm{ref}}$ give
\begin{equation}
X\,\big(1 - C_{\mathrm{ref}}^{\mathsf T}\big)\,X = C_{\mathrm{ref}}.
\label{eq:refmirror}
\end{equation}
Now let $C(t)$ solve the DLvN equation and define $C'(t) \equiv X (1
- C^{\mathsf T}(t)) X$. Transposing and conjugating term by term and
using the three symmetries above: the unitary part maps as $-X\,
(i[H^{\mathsf T}, C^{\mathsf T}])\, X = -i[H, C']$, and the
dissipative part as $-X\, \big({-\tfrac{1}{2}}\{\Lambda, C^{\mathsf
T} - C_{\mathrm{ref}}^{\mathsf T}\}\big)\, X = -\tfrac{1}{2}
\{\Lambda, C' - C_{\mathrm{ref}}\}$---so $C'$ solves the \emph{same}
equation. Physically: the $(\uparrow$ particle, $\downarrow$ hole$)$
sector carried by $C$ and the $(\downarrow$ particle, $\uparrow$
hole$)$ sector obtained by the particle-hole map evolve
consistently, so taking both spin components of the main-text
Sec.~II.A bond currents from the single propagated $C$ is free of
contradiction.

\emph{(c) Exact quadratic Lindblad form for the decoupled
reference.} The decoupled $H_{\mathrm{ref}}$ is a direct sum of bath-mode
BdG blocks; mode $j$'s $2\times 2$ block $h_j = [[\varepsilon_j,
\Delta], [\Delta, -\varepsilon_j]]$ diagonalizes to $\pm E_j$ in the
quasiparticle basis. By the proposition condition, $\Lambda$
restricted to that block is $\gamma_j \openone_2$ and commutes
with any intra-block basis change, so in the quasiparticle basis
$\Lambda$ remains diagonal with $\gamma_j$ and $C_{\mathrm{ref}}$ is
diagonal with $f(\pm E_j)$. Then
\begin{equation}
-\tfrac{1}{2}\{\Lambda,\, C - C_{\mathrm{ref}}\}
= -\tfrac{1}{2}\{\Gamma_1 + \Gamma_2,\, C\} + \Gamma_2,
\label{eq:lindbladform}
\end{equation}
with $\Gamma_2 = \Lambda f \succeq 0$ (gain) and $\Gamma_1 =
\Lambda(1 - f) \succeq 0$ (loss)---precisely the covariance-matrix
equation of linear Lindblad jump operators
$\sqrt{\gamma_\nu (1 - f_\nu)}\, a_\nu$ and $\sqrt{\gamma_\nu
f_\nu}\, a_\nu^\dagger$ (the standard result for quadratic open
systems~\cite{prosen2008third,barthel2021solving}); complete
positivity and $0 \le C \le 1$ follow exactly.

\emph{(d) Coupled reference.} Parts (a) and (b) use only the
Hermiticity and particle-hole property of $C_{\mathrm{ref}}$ and
hold equally for the coupled reference (junction equilibrium
$f(H)$); but then $C_{\mathrm{ref}}$ no longer commutes with
$\Lambda$ in the damped-site basis, and canonical Lindblad form is
not claimed---consistent with the domain of the Lindblad
construction in the DLvN literature in the infinite-lead
limit~\cite{hod2016driven}.

\section*{Supplementary tables: production bath parameters}

Tables~\ref{tab:bath32}--\ref{tab:bath77} list the complete mode
tables $\{\varepsilon_j, w_j, \gamma_j\}$ of the three production
least-squares baths, generated directly from the archived
\texttt{.mat} caches by \texttt{sm\_bath\_tables.m} (no hand
copying). $\varepsilon_j$ is the normal-state energy ($E_j =
\sqrt{\varepsilon_j^2 + \Delta^2}$ for superconducting baths), $w_j
= |V_j|^2$ the squared coupling in the convention stated with each
table, and $\gamma_j$ the mode damping. Machine-readable copies are
part of the data availability package (main text).

\begin{table*}
\caption{\label{tab:bath32}LS-32 (gap-clean), the main-line
superconducting bath. Total-$\Gamma$ convention (main-text
Sec.~II.C): per lead $|V_{\alpha j}|^2 = w_j/2$. Source:
\texttt{s3b\_refit.mat}.}
\begin{ruledtabular}
\begin{tabular}{cccc@{\qquad}cccc@{\qquad}cccc}
$j$ & $\varepsilon_j$ & $w_j$ & $\gamma_j$ &
$j$ & $\varepsilon_j$ & $w_j$ & $\gamma_j$ &
$j$ & $\varepsilon_j$ & $w_j$ & $\gamma_j$ \\
\hline
1 & $-3.750000$ & $1.2358{\times}10^{-1}$ & 0.2500 & 12 & $-0.391316$ & $5.1662{\times}10^{-2}$ & 0.0321 & 23 & $+1.004586$ & $6.9875{\times}10^{-2}$ & 0.1190 \\
2 & $-3.302365$ & $1.1268{\times}10^{-1}$ & 0.2500 & 13 & $-0.362909$ & $2.9684{\times}10^{-13}$ & 0.2250 & 24 & $+1.284706$ & $7.2788{\times}10^{-2}$ & 0.1529 \\
3 & $-2.860822$ & $9.0321{\times}10^{-2}$ & 0.2344 & 14 & $-0.204330$ & $4.2769{\times}10^{-2}$ & 0.0137 & 25 & $+1.571224$ & $7.3795{\times}10^{-2}$ & 0.1812 \\
4 & $-2.572137$ & $7.5831{\times}10^{-2}$ & 0.2435 & 15 & $-0.120968$ & $2.9684{\times}10^{-13}$ & 0.2250 & 26 & $+1.860363$ & $7.5672{\times}10^{-2}$ & 0.2072 \\
5 & $-2.288676$ & $7.5468{\times}10^{-2}$ & 0.2321 & 16 & $-0.048997$ & $4.3594{\times}10^{-2}$ & 0.0037 & 27 & $+2.147344$ & $7.3457{\times}10^{-2}$ & 0.2201 \\
6 & $-2.003442$ & $7.3858{\times}10^{-2}$ & 0.2126 & 17 & $+0.120969$ & $2.9684{\times}10^{-13}$ & 0.2250 & 28 & $+2.430553$ & $7.4623{\times}10^{-2}$ & 0.2358 \\
7 & $-1.716022$ & $7.5152{\times}10^{-2}$ & 0.1952 & 18 & $+0.125048$ & $4.0587{\times}10^{-2}$ & 0.0082 & 29 & $+2.711377$ & $7.6104{\times}10^{-2}$ & 0.2425 \\
8 & $-1.427496$ & $7.3963{\times}10^{-2}$ & 0.1686 & 19 & $+0.292706$ & $4.7495{\times}10^{-2}$ & 0.0216 & 30 & $+3.035179$ & $9.2313{\times}10^{-2}$ & 0.2090 \\
9 & $-1.143193$ & $7.1713{\times}10^{-2}$ & 0.1365 & 20 & $+0.362907$ & $2.9684{\times}10^{-13}$ & 0.2250 & 31 & $+3.396945$ & $8.6008{\times}10^{-2}$ & 0.2120 \\
10 & $-0.869732$ & $6.7332{\times}10^{-2}$ & 0.0999 & 21 & $+0.499359$ & $5.6441{\times}10^{-2}$ & 0.0459 & 32 & $+3.750000$ & $1.1612{\times}10^{-1}$ & 0.2479 \\
11 & $-0.616076$ & $6.0261{\times}10^{-2}$ & 0.0621 & 22 & $+0.739780$ & $6.3910{\times}10^{-2}$ & 0.0804 &  & & &  \\

\end{tabular}
\end{ruledtabular}
\end{table*}

\begin{table*}
\caption{\label{tab:bath64}LS-64 (density floor), the long-time
spectroscopy variant (Sec.~\ref{sec:s4}). Total-$\Gamma$ convention
as in Table~\ref{tab:bath32}. Rows 49 and 64 ($\varepsilon = \pm 10$)
are the far Re-tail anchors disclosed in Sec.~S1: out-of-band
rational-approximation devices for the Kramers--Kronig tail of
$\mathrm{Re}\,\Sigma$, with in-window spectral density $\le 1.5\times
10^{-5}$. Source: \texttt{s3j\_edgezone\_v4.mat}.}
\begin{ruledtabular}
\begin{tabular}{cccc@{\qquad}cccc@{\qquad}cccc}
$j$ & $\varepsilon_j$ & $w_j$ & $\gamma_j$ &
$j$ & $\varepsilon_j$ & $w_j$ & $\gamma_j$ &
$j$ & $\varepsilon_j$ & $w_j$ & $\gamma_j$ \\
\hline
1 & $+0.025000$ & $1.2732{\times}10^{-2}$ & 0.0150 & 23 & $+1.125000$ & $3.1735{\times}10^{-4}$ & 0.0150 & 45 & $-1.025000$ & $1.1255{\times}10^{-3}$ & 0.0150 \\
2 & $+0.075000$ & $1.2732{\times}10^{-2}$ & 0.0150 & 24 & $+1.175000$ & $1.6493{\times}10^{-4}$ & 0.0150 & 46 & $-1.075000$ & $6.0384{\times}10^{-4}$ & 0.0150 \\
3 & $+0.125000$ & $1.2732{\times}10^{-2}$ & 0.0150 & 25 & $-0.025000$ & $1.2732{\times}10^{-2}$ & 0.0150 & 47 & $-1.125000$ & $3.1735{\times}10^{-4}$ & 0.0150 \\
4 & $+0.175000$ & $1.2731{\times}10^{-2}$ & 0.0150 & 26 & $-0.075000$ & $1.2732{\times}10^{-2}$ & 0.0150 & 48 & $-1.175000$ & $1.6493{\times}10^{-4}$ & 0.0150 \\
5 & $+0.225000$ & $1.2729{\times}10^{-2}$ & 0.0150 & 27 & $-0.125000$ & $1.2732{\times}10^{-2}$ & 0.0150 & 49 & $-10.000000$ & $3.6439{\times}10^{0}$ & 0.0010 \\
6 & $+0.275000$ & $1.2726{\times}10^{-2}$ & 0.0150 & 28 & $-0.175000$ & $1.2731{\times}10^{-2}$ & 0.0150 & 50 & $-8.666667$ & $3.4530{\times}10^{-12}$ & 0.2500 \\
7 & $+0.325000$ & $1.2721{\times}10^{-2}$ & 0.0150 & 29 & $-0.225000$ & $1.2729{\times}10^{-2}$ & 0.0150 & 51 & $-7.333333$ & $3.4530{\times}10^{-12}$ & 0.2500 \\
8 & $+0.375000$ & $1.2710{\times}10^{-2}$ & 0.0150 & 30 & $-0.275000$ & $1.2726{\times}10^{-2}$ & 0.0150 & 52 & $-6.000000$ & $3.4530{\times}10^{-12}$ & 0.2500 \\
9 & $+0.425000$ & $1.2689{\times}10^{-2}$ & 0.0150 & 31 & $-0.325000$ & $1.2721{\times}10^{-2}$ & 0.0150 & 53 & $-4.666663$ & $3.4530{\times}10^{-12}$ & 0.2500 \\
10 & $+0.475000$ & $1.2647{\times}10^{-2}$ & 0.0150 & 32 & $-0.375000$ & $1.2710{\times}10^{-2}$ & 0.0150 & 54 & $-2.375905$ & $1.4471{\times}10^{-1}$ & 0.2500 \\
11 & $+0.525000$ & $1.2567{\times}10^{-2}$ & 0.0150 & 33 & $-0.425000$ & $1.2689{\times}10^{-2}$ & 0.0150 & 55 & $-1.679566$ & $1.5420{\times}10^{-1}$ & 0.2500 \\
12 & $+0.575000$ & $1.2415{\times}10^{-2}$ & 0.0150 & 34 & $-0.475000$ & $1.2647{\times}10^{-2}$ & 0.0150 & 56 & $-1.065077$ & $1.4892{\times}10^{-1}$ & 0.2189 \\
13 & $+0.625000$ & $1.2129{\times}10^{-2}$ & 0.0150 & 35 & $-0.525000$ & $1.2567{\times}10^{-2}$ & 0.0150 & 57 & $+1.303133$ & $1.9049{\times}10^{-1}$ & 0.2500 \\
14 & $+0.675000$ & $1.1607{\times}10^{-2}$ & 0.0150 & 36 & $-0.575000$ & $1.2415{\times}10^{-2}$ & 0.0150 & 58 & $+1.992383$ & $1.5388{\times}10^{-1}$ & 0.2500 \\
15 & $+0.725000$ & $1.0710{\times}10^{-2}$ & 0.0150 & 37 & $-0.625000$ & $1.2129{\times}10^{-2}$ & 0.0150 & 59 & $+2.644745$ & $1.3992{\times}10^{-1}$ & 0.2500 \\
16 & $+0.775000$ & $9.3081{\times}10^{-3}$ & 0.0150 & 38 & $-0.675000$ & $1.1607{\times}10^{-2}$ & 0.0150 & 60 & $+4.666665$ & $3.4530{\times}10^{-12}$ & 0.2500 \\
17 & $+0.825000$ & $7.4175{\times}10^{-3}$ & 0.0150 & 39 & $-0.725000$ & $1.0710{\times}10^{-2}$ & 0.0150 & 61 & $+6.000000$ & $3.4530{\times}10^{-12}$ & 0.2500 \\
18 & $+0.875000$ & $5.3149{\times}10^{-3}$ & 0.0150 & 40 & $-0.775000$ & $9.3081{\times}10^{-3}$ & 0.0150 & 62 & $+7.333334$ & $3.4530{\times}10^{-12}$ & 0.2500 \\
19 & $+0.925000$ & $3.4243{\times}10^{-3}$ & 0.0150 & 41 & $-0.825000$ & $7.4175{\times}10^{-3}$ & 0.0150 & 63 & $+8.666667$ & $3.4530{\times}10^{-12}$ & 0.2500 \\
20 & $+0.975000$ & $2.0228{\times}10^{-3}$ & 0.0150 & 42 & $-0.875000$ & $5.3149{\times}10^{-3}$ & 0.0150 & 64 & $+10.000000$ & $3.5926{\times}10^{0}$ & 0.0010 \\
21 & $+1.025000$ & $1.1255{\times}10^{-3}$ & 0.0150 & 43 & $-0.925000$ & $3.4243{\times}10^{-3}$ & 0.0150 &  & & &  \\
22 & $+1.075000$ & $6.0384{\times}10^{-4}$ & 0.0150 & 44 & $-0.975000$ & $2.0228{\times}10^{-3}$ & 0.0150 &  & & &  \\

\end{tabular}
\end{ruledtabular}
\end{table*}

\begin{table*}
\caption{\label{tab:bath77}LS-77 (normal-lead hybrid,
Sec.~\ref{sec:s5}): modes 1--65 are the dense core ($|\varepsilon|
\le 1.6$, $h = \gamma = 0.05$, analytic weights $\Gamma h/\pi$),
modes 66--77 the fitted wings. Weights are quoted at the fitting
hybridization $\Gamma = 0.1$; for a lead with hybridization
$\Gamma_\alpha$ rescale $w_j \to w_j\,(\Gamma_\alpha/0.1)$. Source:
\texttt{n1d\_hybrid.mat}.}
\begin{ruledtabular}
\begin{tabular}{cccc@{\qquad}cccc@{\qquad}cccc}
$j$ & $\varepsilon_j$ & $w_j$ & $\gamma_j$ &
$j$ & $\varepsilon_j$ & $w_j$ & $\gamma_j$ &
$j$ & $\varepsilon_j$ & $w_j$ & $\gamma_j$ \\
\hline
1 & $-1.600000$ & $1.5915{\times}10^{-3}$ & 0.0500 & 27 & $-0.300000$ & $1.5915{\times}10^{-3}$ & 0.0500 & 53 & $+1.000000$ & $1.5915{\times}10^{-3}$ & 0.0500 \\
2 & $-1.550000$ & $1.5915{\times}10^{-3}$ & 0.0500 & 28 & $-0.250000$ & $1.5915{\times}10^{-3}$ & 0.0500 & 54 & $+1.050000$ & $1.5915{\times}10^{-3}$ & 0.0500 \\
3 & $-1.500000$ & $1.5915{\times}10^{-3}$ & 0.0500 & 29 & $-0.200000$ & $1.5915{\times}10^{-3}$ & 0.0500 & 55 & $+1.100000$ & $1.5915{\times}10^{-3}$ & 0.0500 \\
4 & $-1.450000$ & $1.5915{\times}10^{-3}$ & 0.0500 & 30 & $-0.150000$ & $1.5915{\times}10^{-3}$ & 0.0500 & 56 & $+1.150000$ & $1.5915{\times}10^{-3}$ & 0.0500 \\
5 & $-1.400000$ & $1.5915{\times}10^{-3}$ & 0.0500 & 31 & $-0.100000$ & $1.5915{\times}10^{-3}$ & 0.0500 & 57 & $+1.200000$ & $1.5915{\times}10^{-3}$ & 0.0500 \\
6 & $-1.350000$ & $1.5915{\times}10^{-3}$ & 0.0500 & 32 & $-0.050000$ & $1.5915{\times}10^{-3}$ & 0.0500 & 58 & $+1.250000$ & $1.5915{\times}10^{-3}$ & 0.0500 \\
7 & $-1.300000$ & $1.5915{\times}10^{-3}$ & 0.0500 & 33 & $+0.000000$ & $1.5915{\times}10^{-3}$ & 0.0500 & 59 & $+1.300000$ & $1.5915{\times}10^{-3}$ & 0.0500 \\
8 & $-1.250000$ & $1.5915{\times}10^{-3}$ & 0.0500 & 34 & $+0.050000$ & $1.5915{\times}10^{-3}$ & 0.0500 & 60 & $+1.350000$ & $1.5915{\times}10^{-3}$ & 0.0500 \\
9 & $-1.200000$ & $1.5915{\times}10^{-3}$ & 0.0500 & 35 & $+0.100000$ & $1.5915{\times}10^{-3}$ & 0.0500 & 61 & $+1.400000$ & $1.5915{\times}10^{-3}$ & 0.0500 \\
10 & $-1.150000$ & $1.5915{\times}10^{-3}$ & 0.0500 & 36 & $+0.150000$ & $1.5915{\times}10^{-3}$ & 0.0500 & 62 & $+1.450000$ & $1.5915{\times}10^{-3}$ & 0.0500 \\
11 & $-1.100000$ & $1.5915{\times}10^{-3}$ & 0.0500 & 37 & $+0.200000$ & $1.5915{\times}10^{-3}$ & 0.0500 & 63 & $+1.500000$ & $1.5915{\times}10^{-3}$ & 0.0500 \\
12 & $-1.050000$ & $1.5915{\times}10^{-3}$ & 0.0500 & 38 & $+0.250000$ & $1.5915{\times}10^{-3}$ & 0.0500 & 64 & $+1.550000$ & $1.5915{\times}10^{-3}$ & 0.0500 \\
13 & $-1.000000$ & $1.5915{\times}10^{-3}$ & 0.0500 & 39 & $+0.300000$ & $1.5915{\times}10^{-3}$ & 0.0500 & 65 & $+1.600000$ & $1.5915{\times}10^{-3}$ & 0.0500 \\
14 & $-0.950000$ & $1.5915{\times}10^{-3}$ & 0.0500 & 40 & $+0.350000$ & $1.5915{\times}10^{-3}$ & 0.0500 & 66 & $-5.704916$ & $2.0468{\times}10^{-2}$ & 0.3000 \\
15 & $-0.900000$ & $1.5915{\times}10^{-3}$ & 0.0500 & 41 & $+0.400000$ & $1.5915{\times}10^{-3}$ & 0.0500 & 67 & $-4.758102$ & $2.0199{\times}10^{-2}$ & 0.3000 \\
16 & $-0.850000$ & $1.5915{\times}10^{-3}$ & 0.0500 & 42 & $+0.450000$ & $1.5915{\times}10^{-3}$ & 0.0500 & 68 & $-3.831284$ & $2.0415{\times}10^{-2}$ & 0.3000 \\
17 & $-0.800000$ & $1.5915{\times}10^{-3}$ & 0.0500 & 43 & $+0.500000$ & $1.5915{\times}10^{-3}$ & 0.0500 & 69 & $-2.942462$ & $2.0838{\times}10^{-2}$ & 0.3000 \\
18 & $-0.750000$ & $1.5915{\times}10^{-3}$ & 0.0500 & 44 & $+0.550000$ & $1.5915{\times}10^{-3}$ & 0.0500 & 70 & $-2.131091$ & $2.2734{\times}10^{-2}$ & 0.3000 \\
19 & $-0.700000$ & $1.5915{\times}10^{-3}$ & 0.0500 & 45 & $+0.600000$ & $1.5915{\times}10^{-3}$ & 0.0500 & 71 & $-1.782733$ & $9.6900{\times}10^{-3}$ & 0.2356 \\
20 & $-0.650000$ & $1.5915{\times}10^{-3}$ & 0.0500 & 46 & $+0.650000$ & $1.5915{\times}10^{-3}$ & 0.0500 & 72 & $+1.782733$ & $9.6900{\times}10^{-3}$ & 0.2356 \\
21 & $-0.600000$ & $1.5915{\times}10^{-3}$ & 0.0500 & 47 & $+0.700000$ & $1.5915{\times}10^{-3}$ & 0.0500 & 73 & $+2.131091$ & $2.2734{\times}10^{-2}$ & 0.3000 \\
22 & $-0.550000$ & $1.5915{\times}10^{-3}$ & 0.0500 & 48 & $+0.750000$ & $1.5915{\times}10^{-3}$ & 0.0500 & 74 & $+2.942462$ & $2.0838{\times}10^{-2}$ & 0.3000 \\
23 & $-0.500000$ & $1.5915{\times}10^{-3}$ & 0.0500 & 49 & $+0.800000$ & $1.5915{\times}10^{-3}$ & 0.0500 & 75 & $+3.831284$ & $2.0415{\times}10^{-2}$ & 0.3000 \\
24 & $-0.450000$ & $1.5915{\times}10^{-3}$ & 0.0500 & 50 & $+0.850000$ & $1.5915{\times}10^{-3}$ & 0.0500 & 76 & $+4.758102$ & $2.0199{\times}10^{-2}$ & 0.3000 \\
25 & $-0.400000$ & $1.5915{\times}10^{-3}$ & 0.0500 & 51 & $+0.900000$ & $1.5915{\times}10^{-3}$ & 0.0500 & 77 & $+5.704916$ & $2.0468{\times}10^{-2}$ & 0.3000 \\
26 & $-0.350000$ & $1.5915{\times}10^{-3}$ & 0.0500 & 52 & $+0.950000$ & $1.5915{\times}10^{-3}$ & 0.0500 &  & & &  \\

\end{tabular}
\end{ruledtabular}
\end{table*}

\clearpage   